\documentclass[12pt]{article}
\def\a{\alpha}
\def\b{\beta}

\def\k{\kappa}

\def\m{\mu}
\def\n{\nu}

\def\be{\begin{equation}}
\def\ee{\end{equation}}
\def\bqn{\begin{eqnarray}}
\def\eqn{\end{eqnarray}}    
\def\nn{\nonumber}

\newtheorem{theorem}{Theorem}[subsubsection]

\newcommand{\tr}{\mbox{tr}}

\begin{document}

\begin{titlepage}
\begin{flushright}
IHES/P/02/06 \\
ULB-TH-02/01
\end{flushright}
\vskip 1.0cm

\begin{centering}

{\huge {\bf Kasner-like behaviour for 
subcritical Einstein-matter systems}}

\vspace{1cm}

{\Large Thibault Damour$^{a}$, Marc Henneaux$^{b,c}$, Alan D. Rendall$^{d}$ 
and \\
Marsha Weaver$^{b}$} \\
\vspace{.7cm}
$^a$ Institut des Hautes Etudes Scientifiques,  35, Route de
Chartres,  F-91440 Bures-sur-Yvette, France \\
\vspace{.2cm} 
$^b$ Physique Th\'eorique et Math\'ematique,  Universit\'e Libre
de Bruxelles,  C.P. 231, B-1050, Bruxelles, Belgium      \\
\vspace{.2cm}
$^c$ Centro de Estudios Cient\'{\i}ficos, Casilla 1469, Valdivia, Chile \\
\vspace{.2cm}

$^d $Max-Planck-Institut f\"ur Gravitationsphysik,
Albert-Einstein-Institut, 
Am M\"uhlenberg 1, D-14476 Golm, Germany

\vspace{1.5cm}

\end{centering}

\begin{abstract} 
Confirming previous heuristic analyses  \`a la Belinskii-Khalatnikov-Lifshitz,
it is  rigorously
proven that certain ``subcritical'' Einstein-matter systems  
exhibit a monotone, generalized Kasner behaviour in the vicinity of
a spacelike singularity. The $D-$dimensional coupled Einstein-dilaton-$p$-form
system is subcritical if the dilaton couplings of the $p$-forms
belong to some dimension dependent open neighbourhood of zero
\cite{dh1},
while pure gravity is subcritical if $D \geq 11$ \cite{DHS}. Our proof relies,
like the recent theorem \cite{AR} dealing with the
(always subcritical \cite{BK2}) Einstein-dilaton system, on the
use of Fuchsian techniques, which enable one to construct 
local, analytic solutions
to the full set of equations of motion.  The solutions constructed are
``general'' in the sense that they depend on the maximal expected number of
free functions.
\end{abstract}

\vfill
\end{titlepage}

\section{Introduction}
\setcounter{equation}{0}
\setcounter{theorem}{0}
\setcounter{lemma}{0}      
\subsection{The problem}
In recent papers \cite{dh1,dh2,tDmH}, the dynamics of the coupled
Einstein-dilaton-$p$-form system in $D$ spacetime dimensions,
with action (in units where $8 \pi G = 1$),
\begin{eqnarray}
S[g_{\alpha \beta}, \phi, A^{(j)}_{\gamma_1 \cdots \gamma_{n_j}}]
&=& S_E [g_{\alpha \beta}] + S_\phi[g_{\alpha \beta}, \phi]
+ \sum_{j=1}^k S_j[g_{\alpha \beta}, \phi, 
A^{(j)}_{\gamma_1 \cdots \gamma_{n_j}}]
+ \hbox{ ``more"},
\label{001}  \\
S_E [g_{\alpha \beta}] &=&{1 \over 2} \int R \sqrt{-g} \, d^D x,
\label{action2} \\
S_\phi[g_{\alpha \beta}, \phi]  &=& - {1 \over 2} \int 
\partial_\mu \phi \, \partial^\mu \phi  \sqrt{-g} \, d^D x, 
\label{action3} \\
S_j[g_{\alpha \beta}, \phi, A^{(j)}_{\gamma_1 \cdots \gamma_{n_j}}] &=&
- {1 \over 2 ( n_j + 1)!} \int F^{(j)}_{\mu_1 \cdots \mu_{n_j + 1}}
F^{(j) \, \mu_1 \cdots \mu_{n_j + 1}} e^{\lambda_j \phi} \sqrt{-g} \, d^D x,   
\label{action4}
\end{eqnarray}
was investigated in the vicinity of a spacelike (``cosmological")
singularity along the lines initiated
by Belinskii, Khalatnikov and Lifshitz (BKL)
\cite{BKL}. In (\ref{001}), $g_{\alpha \beta}$ is the spacetime
metric,
$\phi$ is a massless scalar field known as the
``dilaton", while the $A^{(j)}_{\gamma_1 \cdots \gamma_{n_j}}$ are
a collection of $k$ exterior form gauge fields
($j=1, \cdots, k$), with exponential couplings to the
dilaton, each coupling being characterized by an
individual constant
$\lambda_j$ (``dilaton coupling constant").  
The $F^{(j)}$'s are the exterior derivatives
$F^{(j)} = dA^{(j)}$, whereas ``more" stands for possible coupling
terms among the $p$-forms which can be either of the
Yang-Mills type ($1$-forms), Chern-Simons type \cite{jdt}
or Chapline-Manton type \cite{cm,pvnetal}.
The degrees of the $p$-forms are restricted to be smaller 
than or equal to $D-2$ since a $(D-1)$-form (or $D$-form)
gauge field
carries no local degree of freedom. 
In particular, scalars ($n_j=0$) are allowed among the
$A^{(j)}$'s but we then require that the corresponding dilaton
coupling $\lambda_j$ be non-zero, so that there is only one
``dilaton''.  Similarly, we require $\lambda_j \not=0$
for the $(D-2)$-forms (if any), since these are ``dual''\footnote{
We recall that the Hodge duality between a $(n_j+1)$-form and a
$(D - n_j -1)$-form allows one to replace (locally) a $n_j$-form potential
$A^{(j)}$ by a $(D - n_j -2)$-form potential $A^{(j')}$  (with dilaton
coupling $\lambda_j' = - \lambda_j$).}
to scalars.  This restriction to a single dilaton is mostly done for
notational convenience: if there
were other dilatons among the $0$-forms, then, these must
be explicitly treated on the same footing as $\phi$
and separated off from the $p$-forms because they play
a distinct r\^ole.  In particular, they would appear
explicitly in the generalized Kasner conditions
given below and in the determination of what we call the
subcritical domain.  The discussion would proceed otherwise 
in the same qualitative way. 

The main motivation for studying actions of the class
(\ref{001}) is that these arise as bosonic sectors of supergravity
theories related to superstring or M-theory.  In fact, in
view of various no-go theorems, $p$-form gauge fields appear
to be the only massless, higher spin fields that
can be consistently coupled to gravity.  Furthermore, there
can be only one type of graviton \cite{bdgh}.  With this observation
in mind, the action (\ref{001}) is actually quite general.  The only
restriction concerns the scalar sector: we assume the coupling
to the dilaton to be exponential because this corresponds to the tree-level
couplings of the dilaton field of string theory. Note, however, that 
string-loop effects are expected to generate more general couplings 
$\exp (\lambda \phi) \rightarrow B(\phi)$ which can exhibit interesting
``attractor'' behaviours \cite{tDaP}.  We also restrict
ourselves by not including scalar potentials; see, however, the end 
of the article for
some remarks on the addition of
a potential for the dilaton, which can be treated by our methods.

Two possible general, ``competing"
behaviours of the fields in the vicinity of
the spacelike singularity have been identified\footnote{For a
recent extension of these ideas to the  brane-worlds scenarios,
see \cite{jdBsH}.}:
\begin{enumerate}
\item \label{Kasnerbehaviour} The simplest is the {\it ``generalized Kasner
behaviour''}, in which the spatial scale factors and the field $\exp(\phi)$ behave
at each spatial point
in a monotone, power-law fashion in terms of the proper
time as one approaches the
singularity, while the effect of the $p$-form fields $A^{(j)}$'s on the
 evolution of $g_{\mu \nu}$ and $\phi$ can be asymptotically
neglected. In that regime the spatial curvature terms can be also
neglected with respect to the leading order part of the extrinsic
curvature terms. In other words, as emphasized by BKL, time derivatives
asymptotically dominate over space derivatives so that one sometimes uses
the terminology ``velocity-dominated" behaviour \cite{Eardley}, instead of
``generalized Kasner behaviour''.  We shall use both
terminologies indifferently in this paper, recalling that in the
presence of $p$-forms, which act as potentials for the evolution
of the spatial metric and the dilaton (as do the spatial curvature
terms),  ``velocity-dominance" means not only that the spatial
curvature terms can be neglected, but also that the $p$-forms
can be neglected in the Einstein-dilaton evolution equations.\footnote{
The Kasner solution is generalized in two ways: first, the
original Kasner exponents include a dilaton exponent (if there
is a dilaton),
which appears in the Kasner conditions; second, the exponents
are not assumed to be constant in space.
We shall shorten ``exhibits generalized Kasner behaviour"
to Kasner-like.  We stress that we do not use this term to indicate
that the solution becomes asymptotically homogeneous
in space.   }

\item \label{mixmasterbehaviour} The second regime, known
as {\it``oscillatory''} \cite{BKL}, or {\it``generalized mixmaster''} 
\cite{misner69} behaviour, is more complicated.  It can be described
as the succession of an infinite number of increasingly shorter 
Kasner regimes as one goes
to the singularity, one following the
other according to a well-defined ``collision" law.  
This asymptotic evolution is presumably strongly chaotic.
It is expected that, at each spatial point, the scale factors of a general
inhomogeneous solution essentially
behave as in certain homogeneous models. For instance, for $D=4$ pure gravity
this guiding homogeneous model is the Bianchi IX model \cite{BKL,misner69},
while for $D=11$ supergravity it is its naive one-dimensional reduction involving
space-independent metric and three-forms \cite{dh2}.
\end{enumerate}

Whether it is the first or the second behaviour that is relevant 
depends on: (i) the spacetime dimension $D$, (ii) the field content
(presence or absence of the dilaton, types of $p$-forms), and (iii)
the values of the various dilaton couplings $\lambda_j$. Previous 
work reached the following conclusions:
\begin{itemize}
\item The oscillatory behaviour is general for pure gravity 
in spacetime dimension $4$ \cite{BKL},
in fact, in all spacetime dimensions  $4 \leq D \leq 10$, but 
is replaced by a
Kasner-like behaviour in spacetime dimensions $ D \geq 11$ \cite{DHS}.
(The sense in which we use ``general" will be made
precise below.)
\item The Kasner-like behaviour is general for the
gravity-dilaton system in all spacetime dimensions
$D \geq 3$ (see \cite{BK2,AR} for $D=4$).
\item The oscillatory behaviour is general for gravity coupled to $p$-forms,
in absence of a dilaton or of a dual $(D-2)$-form
($0<p<D-2$) \cite{dh2}.  
In particular, the bosonic sector of $11$-dimensional supergravity
is oscillatory \cite{dh1}.  Particular instances of this case
have been studied in \cite{Jantzen,Leblanc,Weaver}.
\item The case of the gravity-dilaton-$p$-form system is more complicated
to discuss because its behaviour depends on a combination of several 
factors, namely the dimension $D$, the
menu of $p$-forms, and the numerical values of the dilaton couplings.
For a given $D$ and a given menu of $p$-forms there exists a ``subcritical''
domain $\cal D$ (an open neighbourhood of the origin $\lambda_j = 0$ for
all $j$'s) such that: (i) when the $\lambda_j$ belong to $\cal D$ the 
general behaviour is Kasner-like, but (ii) when the  $\lambda_j$ do not
 belong to $\cal D$ the behaviour
is oscillatory. Note that $\cal D$ is open. Indeed, the behaviour is
oscillatory when the $\lambda_j$ are on the boundary of $\cal D$,
as  happens for instance for the low-energy bosonic sectors of
type I or heterotic superstrings \cite{dh1}.
For a single $p$-form, the subcritical domain $\cal D$ takes
the simple form $\vert \lambda_j \vert <\lambda_j^c$,
where $\lambda_j^c$ depends on the form-degree and the spacetime
dimension.  ($\lambda_j^c$ can be infinite.)
For a collection of $p$-forms, $\cal D$ is more
complicated and not just given by the Cartesian product of the
subcritical intervals associated with each individual $p$-form.
\end{itemize}

The above statements were derived by adopting a line of
thought analogous to that followed
by BKL.  Now, as understood by BKL themselves,
these arguments, although quite convincing, are somewhat heuristic.
It is true that the original arguments have received since then 
a considerable amount of both numerical and analytical
support \cite{bgimw,BIW,U1,WIB,garf}.  Yet, they still await
a complete proof.  One notable exception is
the four-dimensional
gravity-dilaton system, which has been rigorously
demonstrated in \cite{AR} to be indeed Kasner-like, confirming the
original analysis \cite{BK2}.
  Using Fuchsian techniques, the authors of \cite{AR}
have proven the existence of a local (analytic) Kasner-like
 solution to the Einstein-dilaton equations
in four dimensions that contains as many arbitrary, physically
relevant functions of space as there are local degrees of freedom,
namely $6$ (counting $q$ and $\dot{q}$ independently).
To our knowledge, this was the first construction,
in a rigorous mathematical sense,
of a general singular  solution for a
coupled Einstein-matter system. 
Note in this respect several previous
works in which formal solutions had been constructed
near (Kasner-like) cosmological singularities
by explicit perturbative methods,
to all orders of perturbation theory \cite{GM93,BDV}.

The situation concerning the more
complicated (and in some sense more interesting) generalized mixmaster 
regime is unfortunately -- and perhaps not surprisingly --
not so well developed.  Rigorous results are scarce
(note \cite{mixinhom}) and even in the case
of the spatially homogeneous Bianchi IX model only partial results
exist in the literature \cite{mix}.
 
The purpose of this paper is to extend the Fuchsian approach
of \cite{AR} to the more complicated class of models described by the
action (\ref{001}) and to prove
that those among the above models that were predicted
in \cite{DHS,dh1,dh2} to be Kasner-like are indeed so.
This provides many new instances where one can 
rigorously construct
a general singular  solution for a
coupled Einstein-matter (or pure Einstein, in $D \geq 11$) system.
In fact, our (Fuchsian-system-based) results prove that the formal 
perturbative solutions that can be explicitly built for these 
models do converge to exact solutions.
This provides a further
confirmation of the general validity of the BKL ideas.
We shall also explicitly determine the subcritical 
domain $\cal D$ for a few illustrative models.
For all the relevant  systems, we 
construct local (near the singularity) analytic solutions, which
are ``general'' in the sense that they contain the right
number of freely adjustable arbitrary functions of
space (in particular, these solutions have generically
no isometries), and which exhibit the generalized (monotone) Kasner
time dependence.
 
\subsection{Strategy and outline of the paper}
Our approach is the same as in \cite{AR}, 
and results from that work will be used frequently here without restating
the arguments.  
Here is an
outline of the key steps.

A $d + 1$ decomposition is used, for $d$ spatial dimensions,
$d = D-1$.  A Gaussian time coordinate, $t$, is chosen
such that the singularity occurs at $t=0$.  The first step
in the argument consists of identifying the leading terms for all the
variables.  This is accomplished by writing down a set
of evolution equations which is obtained by truncating the full evolution
equations, and then solving this
simpler set of evolution equations.  This simpler evolution system
is called the Kasner-like\footnote{Note that
we use the terms ``Kasner-like solutions" to label both
exact solutions of the truncated system and solutions of the
full system that are asymptotic to such solutions.  Which
meaning is relevant should be clear from the context.}
evolution system (or, alternatively,
the velocity-dominated system). It is a system of
ordinary differential equations with respect to time (one
at each spatial point) which coincides with the system that arises
when investigating metric-dilaton solutions that depend only on
time. The precise truncation rules are given in
subsection \ref{stepone} below.
The second step is to write down constraint
equations for the Kasner-like system (called ``velocity-dominated"
constraints) and to show that these
constraints propagate, {\it i.e.,} that if they are satisfied
by a solution to the Kasner-like evolution equations
at some time $t_0>0$, then they are satisfied for all time $t>0$.
In the set of Kasner-like solutions, one expects that there is a subset,
denoted by $V$, of solutions which have the property of
being asymptotic to solutions of the complete
Einstein-dilaton-$p$-form equations as $t \rightarrow 0$,
i.e as  one goes to the singularity.
This subset is characterized by inequalities on some of
the initial data, which, however, are not always consistent. 
The existence of a non-empty
$V$ requires the dilaton couplings to belong to
some range, the ``subcritical range".  When $V$
is non-empty and open, the solutions in $V$ involve
as many arbitrary functions of
space as a ``general solution" of the full Einstein 
equations should.  On the other hand, if $V$ is empty
the construction given in this paper breaks down and
the dynamical system is expected to be not
Kasner-like but rather oscillatory.

To show that indeed, the solutions in $V$ (when it is non-empty) are
asymptotic to true solutions, the third step is to identify
decaying quantities such that these decaying quantities along
with the leading terms mentioned above uniquely determine
the variables, and to
write down a {\it Fuchsian} system for the decaying quantities
which is equivalent to the Einstein-matter evolution
system. As the use of Fuchsian systems is central to our work let us
briefly recall what a Fuchsian system is and how such a system is related to 
familiar iterative methods. 
For a more detailed introduction to Fuchsian techniques
see \cite{AR,KR,R00a,Rendall:2001ai}
and references therein. Note that we shall everywhere
restrict ourselves to the analytic case. We expect that our results
extend to the $C^{\infty}$ case, but it is a non-trivial task to prove
that they do. 

The general form of a Fuchsian
system for a vector-valued unknown function $u$ is
\begin{equation}
\label{fuchs0}
t \, \partial_t u + {\cal A}(x) \, u = f(t,x,u,u_x),
\end{equation}
where the matrix  ${\cal A}(x)$ is required to 
satisfy some positivity condition (see below),
while the ``source term'' $f$ on the right hand side is required
to be ``regular.'' (See \cite{AR} for precise criteria allowing
one to check when the positivity assumption on ${\cal A}(x)$
is satisfied and when $f$ is regular.)  A key point is that
$f$ is required to be bounded by terms of
order $O(t^{\delta})$ (with $ t \rightarrow 0, \delta > 0$) as soon
as $u $ and their space derivatives $ u_x$ are in a bounded set (a simple,
concrete example of a source term satisfying this condition is
$f = t^{\delta_1} + t^{\delta_2} u + t^{\delta_3} u_x$, with $\delta_i$'s
larger than $\delta$). A convenient form of positivity condition to be satisfied
by the matrix  ${\cal A}(x)$ is that the operator norm of $\tau^{{\cal A}(x)}$
be bounded when $ 0< \tau < 1$ (and when $x$ varies in any open set).
Essentially this condition restricts the eigenvalues of the matrix
${\cal A}(x)$ to have positive real parts.
The basic property of Fuchsian systems that we shall use is
that there is a unique solution
to the Fuchsian equation which vanishes as $t$ tends to zero \cite{KR}.
One can understand this theorem as a mathematically rigorous version of
the recursive method for solving the equation (\ref{fuchs0}).
Indeed, when confronted with equation (\ref{fuchs0}), it is natural to 
construct a solution by an iterative process, starting with the 
zeroth order approximation $u_0 =0$ (which is the
unique solution of (\ref{fuchs0}) with $f \equiv 0$ that tends
to zero as $t  \rightarrow 0$), and solving a sequence of equations
of the form $t \partial_t u^{(n)} + {\cal A}(x)  u^{(n)} =
f(t,x,u^{(n-1)},u^{(n-1)}_x)$. At each step in this iterative process the 
source term is a known function which essentially behaves (modulo logarithms)
like a sum of powers of $t$ (with space-dependent coefficients). The crucial 
step in the iteration is then to solve equations of the type 
$t \partial_t u + {\cal A}(x)  u = C(x) t^{\delta(x)}$.
The positivity condition on ${\cal A}(x)$ guarantees the absence of
homogeneous solutions remaining bounded as $ t \rightarrow 0$, and ensures
the absence of ``small denominators'' in the (unique bounded) inhomogeneous 
solution generated by each partial source term:
$ u_{\rm inhom} = (\delta + {\cal A})^{-1} C t^{\delta}$.
(See, {\it e.g.,} \cite{BDV}
for a concrete iterative construction of a Kasner-like solution and the 
proof that it extends to all orders.) This link between Fuchsian
systems and ``good systems'' that can be solved to all orders in a 
formal iteration makes it a priori probable that all cases which the
heuristic approach \`a la BKL has shown to be asymptotic to a Kasner-like
solution (by checking that the leading ``post-Kasner'' contribution is
asymptotically sub-dominant) can be cast in a Fuchsian form. The main
technical burden of the present work will indeed be to show in detail
how this can be carried out for the evolution systems corresponding
to all the sub-critical ({\it i.e.,} non-oscillatory) Einstein-matter
systems.
Our Fuchsian formulation proves that (in the analytic case) the formal
all-orders iterative solutions for 
the models we consider do actually converge to the unique, exact solution
having a given leading Kasner asymptotic behaviour as $ t \to 0$.

Finally, the fourth step of our strategy is to prove that the constructed
solution does satisfy also all the Einstein and Gauss-like constraints 
so that it is a solution
of the full set of Einstein-matter equations. We shall deal successively with
the matter (Gauss-like) constraints, and the Einstein constraints.

Our paper is organized as follows.
In section~\ref{scaMax4D}, we first consider the paradigmatic example of
gravity coupled to a
massless scalar field and to a Maxwell field in 4 spacetime
dimensions.  The action (\ref{001}) reads in this case
\begin{equation}
\label{action1}
S[g_{\alpha \beta}, \phi, A_\gamma] =
{1 \over 2} \int \{ R - \partial_\mu \phi \, \partial^\mu \phi
- {1 \over 2} F_{\mu \nu} F^{\mu \nu} e^{\lambda \phi} \}
\sqrt{-g} \, d^4 x.
\end{equation}
For this simple example, we shall explicitly determine the subcritical 
domain $\cal D$, {\it i.e.,} the critical value  $\lambda_c$ such that
 the system is Kasner-like when $-\lambda_c < \lambda < \lambda_c$.
Because this case is exemplary
of the general situation, while still being technically rather simple to handle,
we shall describe in some detail the explicit steps of the
Fuchsian approach.

In section~\ref{vacuum},
vacuum solutions governed by the pure Einstein action (\ref{action2}) with 
$D \geq 11$ are considered.   This system was argued in \cite{DHS}
to be Kasner-like and we show here how this rigorously 
follows from the Fuchsian approach. Note that, contrary to what
happens when a dilaton is present, Fuchsian techniques apply here even though
not all Kasner exponents can be positive.

In sections~\ref{scalar} --
\ref{yangmills}, the results of the previous sections
are generalized to the wider class of systems (\ref{001}).   First,
in section~\ref{scalar}, to solutions of Einstein's
equation with spacetime dimension $D \geq 3$ and
a matter source consisting of a massless scalar field,
governed by the action 
$S_E [g_{\alpha \beta}] + S_\phi[g_{\alpha \beta}, \phi]$.
This is the generalization to any $D \geq 3$ of the case $D = 4$
 treated in \cite{AR}.
In section~\ref{nform}, we turn to the general
situation described by the action (\ref{001}), without, however, including the
additional terms represented there by ``more''.
We then give some general rules for computing the
subcritical domain of the dilaton couplings guaranteeing
velocity-dominance (section~\ref{subcritical}).
The inclusion of interaction terms is considered in the last sections.
It is shown that they do not affect the asymptotic
analysis.  This is done first for the Chern-Simons and Chapline-Manton
interactions in section~\ref{couple}, and next, in section~\ref{yangmills},
for the Yang-Mills couplings (for some gauge group $G$),
 for which the action reads
\begin{equation}
\label{actionym}
S[g_{\alpha \beta}, \phi, A_\gamma] =
{1 \over 2} \int \{ R - \partial_\mu \phi \, \partial^\mu \phi
- {1 \over 2} F_{\mu \nu} \cdot F^{\mu \nu} e^{\lambda \phi} \}
\sqrt{-g} \, d^D x.
\end{equation}
Here the dot product,
$F \cdot F$, is a time-independent, Ad-invariant, non-degenerate
scalar product on the Lie algebra of $G$ (such a scalar product exists
if the algebra is
compact, or semi-simple).  Contrary to what is done in
sections~\ref{scaMax4D}, \ref{nform} and \ref{couple},
we must work now with the vector potential (and not just with
the field strength), since it appears explicitly in the coupling
terms.  

In section~\ref{nonlinear} we show that self-interactions of
a rather general type for the scalar field can be included 
without changing the asymptotics of the solutions.  Explicitly,
we add a (nonlinear) potential term,
\begin{equation}
\label{actionnl}
S_{\rm NL}[g_{\alpha \beta}, \phi]  =
- \int V(\phi) \sqrt{-g} \, d^D x,
\end{equation}
to the action (\ref{001}), where $V(\phi)$ must fulfill some
assumptions given in section~\ref{nonlinear}.  $V(\phi)$ may,
for example, be an exponential function of $\phi$, a constant,
or a suitable power of $\phi$.  Similar forms for $V(\phi)$ were 
considered with $D=4$ in \cite{R00b}. 

{}Finally, in section~\ref{conclusions}, we state two
theorems that summarize the main results of the
paper and give concluding remarks.

\subsection{On the generality of our construction}
As we shall see the number of arbitrary functions contained in solutions
to the velocity-dominated constraint equations is equal
to the number of arbitrary functions for solutions
to the Einstein-matter constraints.  In this 
function-counting sense, our construction describes 
what is customarily called a ``general'' solution of the system.
Intuitively speaking, our construction concerns some ``open set'' of the
set of all solutions (indeed, the Kasner-like behaviour of
the solution is unchanged under arbitrary, small perturbations
of the initial data, because this simply amounts to changing
the integration functions). Note that, in the physics literature, such a 
``general'' solution is often referred to as being a ``generic'' solution.
However, in the mathematics literature the word ``generic'' is restricted
to describing either an open dense subset of the set of all
solutions, or (when this can be defined) a subset of measure unity of the
set of all solutions. In this work we shall stick to the 
mathematical terminology. We shall have nothing rigorous to say about whether
our general solution is also generic. However, we wish to emphasize the
following points.

{}First, let us mention that
the set $V$ of solutions to the
velocity-dominated equations that are asymptotic to
solutions of the complete equations
is not identical to the set $U$ of all solutions
to the velocity-dominated constraint equations. The subset $V \subset U$
is defined by imposing some inequalities on the free data. These inequalities
do not change the number of free functions. Therefore the solutions in $V$
are still ``general''.
One can wonder whether there could be a co-existing
general behaviour, corresponding to initial data that
do not fulfill the inequalities.  For instance, 
could  such ``bad'' initial data lead to a
generalized mixmaster regime?  This is a difficult question and
we shall only 
summarize here
what is the existing evidence.  
There are heuristic arguments,
supported by numerical study, \cite{BK2,DHS2,Berger00,BG}
that suggest that if one starts with initial data that
do not fulfill the inequalities, one ends up,
after a finite transient period (with a finite number
of ``collisions" with potential walls), with
a solution that is asymptotically velocity-dominated,
for which the inequalities are fulfilled almost everywhere.  In that
sense, the inequalities would not
represent a real  restriction
since there is a dynamical mechanism that drives the solution to the
regime where they are satisfied.
For the subcritical values of the dilaton couplings
that make the inequalities defining $V$ consistent, there is thus no
evidence for
an alternative oscillatory regime corresponding to
a different (open) region in the space of initial data\footnote{The
oscillatory regime may however be present for peculiar
initial data, presumably forming a set
of zero measure.  For instance, gravity + dilaton
is generically Kasner-like, but exhibits an oscillatory
behaviour for initial data with $\phi = 0$ (in $D<11$
spacetime dimensions).}.  It has indeed been shown that the
inequalities defining $V$ are no restriction in a
large spatially homogeneous class \cite{mix}.
Such rigorous results
are, however,  lacking in the inhomogeneous case.
In fact, an interesting subtlety might take place in the inhomogeneous
case. The heuristic arguments
and numerical studies of \cite{Berger00,BG} 
suggest the possibility that
the mechanism driving the system to
$V$ may be suppressed at exceptional spatial points
in general spacetimes, with the result that the asymptotic data
at the exceptional spatial points are not consistent with the
inequalities we assume and lead to
so-called ``spikes".  This picture has been given a firm 
basis in a scalar field model with symmetries \cite{RW}
but the status of the spikes in a general context remains unclear.

{}Finally, since we only deal with spacelike
singularities, the classes of solutions
we consider do not contain all solutions governed by the
action (\ref{001}).  Other types of singularities
({\it e.g.} timelike or null ones)
are known to exist. Whether these other types of singularities
are general is, however, an open question.

\subsection{Billiard picture}

At each spatial point, the solution of the coupled Einstein-matter
system can be pictured, in the vicinity of a spacelike
singularity, as a billiard motion in a region of hyperbolic
space \cite{Chitre,Misnerb,tDmH,melni}.  Hyperbolic billiards are
chaotic when they have finite volume and non chaotic otherwise.
In this latter case, the ``billiard ball" generically
escapes freely to infinity after
a finite number of collisions with the bounding walls.  Subcritical
Einstein-matter systems define infinite-volume billiards.  The velocity-dominated
solutions correspond precisely to the last (as $t \rightarrow 0$) free
motion (after all collisions have taken place), in which the billiard
ball moves to infinity in hyperbolic space.

\subsection{Conventions}
We adopt a ``mostly plus'' signature ($-+++ \ldots$). 
The spacetime dimension is $ D \equiv d + 1$.
Greek indices range from $0$ to $d$,
while Latin
indices $\in \{1,\ldots,d\}$.
The spatial Ricci tensor is labeled $R$ and
the spacetime Ricci tensor is labeled $^{(D)}R$. Our curvature conventions
are such that the Ricci tensor of a sphere is positive definite.
Einstein's equations read $G_{\alpha \beta} = T_{\alpha \beta}$,
where  $G_{\alpha \beta} = R_{\alpha \beta} - R g_{\alpha \beta}/2$
denotes the Einstein tensor and $T_{\alpha \beta}$ denotes the matter
stress-energy tensor, $T_{\alpha \beta}
= - (2/ \sqrt{-g}) \delta S_{\rm matter}/ \delta g^{\alpha \beta}$,
and units such that $ 8 \pi G =1$.
The spatial metric compatible covariant derivative
is labeled $\nabla_a$ and the spacetime metric compatible
covariant derivative is labeled $^{(D)} \nabla_\alpha$.  The
velocity-dominated metric compatible covariant derivative
is labeled $^0 \nabla_a$.  According to the
context,  $g$ denotes the (positive) determinant
of $g_{ab}$ in $d +1$-decomposed expressions, and the (negative)
determinant of $g_{\mu \nu}$ in spacetime expressions.
Whenever $t^\delta$ or $t^{-\delta}$ appears,
$\delta$ denotes a strictly positive number, arbitrarily
small. We use Einstein's summation convention on repeated tensor
indices of different variances. (When the need arises to suspend the
summation conventions for some non-tensorial indices, we shall explicitly
mention it.)  In expressions
where there is a sum that the indices do not indicate, all
sums in the expression are indicated explicitly by a
summation symbol.  Indices on the velocity-dominated metric
and the velocity-dominated extrinsic curvature are raised
and lowered with the velocity-dominated metric.

\subsection{$d+1$ decomposition}
\label{stepzero}
Consider a solution to the Einstein's equations
following from (\ref{001}), consisting of a
Lorentz metric and matter fields on a $D$-dimensional manifold $M$
which is diffeomorphic to $(0,T) \times \Sigma$ for a $d$-dimensional
manifold, $\Sigma$, such that the metric induced on each $t =$ constant
hypersurface is Riemannian, for $t \in (0,T)$.  Here $D$ is an integer
strictly greater than two.  Furthermore, consider a $d+1$
decomposition of the Einstein tensor, $G_{\alpha \beta}$, and the
stress-energy tensor, $T_{\alpha \beta}$, with a Gaussian time
coordinate, $t \in (0,T)$, and a local frame $\{e_a\}$ on $\Sigma$.
Note that the frame $e_a = e_a^i(x) \partial_i$ is time-independent.
The spacetime metric reads $ds^2 = - dt^2 + g_{a b}(t,x) e^a e^b$,
where $e^a = e^a_i(x) d x^i$ (with $e^a_i e^i_b = \delta^a_b$) is the co-frame.
Let $\rho = T_{00}$, $j_a = - T_{0a}$ and $S_{ab} = T_{ab}$.
Define
\begin{eqnarray}
\label{hamiltonian1}
C & = & 2 G_{00} - 2 T_{0 0} \\
& = & -{k^a}_b \, {k^b}_a + ( \tr \, k )^2 + R - 2 \rho. \nonumber
\end{eqnarray}
$C=0$ is the Hamiltonian constraint.  Similarly, 
$C_a = 0$ is the momentum constraint, where
\begin{eqnarray}
\label{momentum1}
C_a & = & -G_{0a} + T_{0a} \\
& = & \nabla_b {k^b}_a - \nabla_a (\tr \, k) - j_a. \nonumber
\end{eqnarray}
In Gaussian coordinates, the relation 
between the metric and the extrinsic curvature is
\begin{equation}
\label{dtg}
\partial_t g_{ab} = -2 k_{ab}.
\end{equation}
The evolution equation for the extrinsic curvature is obtained
by setting ${E^a}_b=0$, with
\begin{eqnarray}
\label{eveq1}
{E^a}_b & = & ^{(D)}{R^a}_b - {T^a}_b + {1 \over(D- 2)} T \, {\delta^a}_b \\
\label{eveq}
\Rightarrow \; \partial_t {k^a}_b & = & 
 {R^a}_b + (\tr \, k) \, {k^a}_b - {M^a}_b.
\end{eqnarray}
Here
\begin{displaymath}
{M^a}_b = {S^a}_b - {1 \over D-2} ((\tr \, S) - \rho) {\delta^a}_b.
\end{displaymath}

\section{Scalar and Maxwell fields in four dimensions}
\setcounter{equation}{0}
\setcounter{theorem}{0}
\setcounter{lemma}{0}
\label{scaMax4D}

\subsection{Equations of motion}
As said above, let us start by considering in detail, as archetypal system,
the system defined by the action~(\ref{action1}),
{\it i.e.,} the spacetime dimension is $D=4$ and
the matter fields are a massless scalar field exponentially
coupled to a Maxwell field, with the magnitude of the dilaton
coupling constant smaller in magnitude
than some positive real number determined below,
$0 \leq |\lambda| < \lambda_c$.  The stress-energy
tensor of the matter fields is
\begin{displaymath}
T_{\mu \nu} = \, ^{(4)}\nabla_\mu \phi \,
^{(4)}\nabla_\nu \phi - {1 \over 2} g_{\mu \nu} \,
^{(4)}\nabla_\alpha \phi \,
^{(4)}\nabla^\alpha \phi + [F_{\mu \alpha} {F_\nu}^ \alpha
-{1 \over 4} g_{\mu \nu} F_{\alpha \beta} F^{\alpha \beta} ] e^{\lambda \phi}.
\end{displaymath}
The matter fields satisfy the following equations.
\begin{eqnarray}
^{(4)}\nabla_\alpha \, ^{(4)}\nabla^\alpha \phi & = & {\lambda \over 4}
 F_{\alpha \beta} F^{\alpha \beta}  e^{\lambda \phi}, \nonumber \\
^{(4)}\nabla_\mu (F^{\mu \nu} e^{\lambda \phi}) & = & 0, \nonumber \\
^{(4)}\nabla_{[ \alpha} F_{\beta \gamma ]} & = & 0. \nonumber
\end{eqnarray}
The 3+1 decomposition of the stress-energy tensor is best expressed
in terms of the electric spatial vector density
${\cal E}^a = \sqrt{g} \, F^{0a} e^{\lambda \phi}$
and the magnetic antisymmetric spatial tensor $F_{ab}$.
\begin{eqnarray}
\rho & = & {1 \over 2} \{ (\partial_t \phi)^2 + g^{ab} e_a(\phi) e_b(\phi)
+{1 \over g} g_{ab} {\cal E}^a {\cal E}^b e^{-\lambda \phi} +
{1 \over 2} g^{ab} g^{ch} F_{ac} F_{bh} e^{\lambda \phi} \}, \nonumber \\
j_a & = & - \partial_t \phi \, e_a(\phi) +
{1 \over \sqrt{g} } {\cal E}^b \, F_{ab}, \nonumber \\
\label{calcM}
{M^a}_b & = & g^{ac} e_b(\phi) \, e_c(\phi)
- {1 \over g} \{ g_{bc} {\cal E}^a {\cal E}^c
- {1 \over 2} {\delta^a}_b g_{ch} {\cal E}^c {\cal E}^h \}
e^{-\lambda \phi} \nonumber \\ & & \; \;
+ \{ g^{ac} g^{hi} F_{ch} F_{bi} - {1 \over 4} {\delta^a}_b
g^{ch} g^{ij} F_{ci} F_{hj} \} e^{\lambda \phi}.
\end{eqnarray}
The matter constraint equations are
\begin{eqnarray}
\label{elconstraint}
e_a ( {\cal E}^a) + f^b_{ba} \, {\cal E}^a & = & 0 \\
\label{magconstraint}
e_{[a}(F_{bc]}) + f^h_{[a b} F_{c] h}& =  & 0.
\end{eqnarray}
Here $f^c_{ab}$ are the (time-independent) structure functions of
the frame, $[e_a, e_b] = f^c_{ab} e_c$.  The matter evolution
equations are
\begin{eqnarray}
\partial^2_t \phi - (\tr k) \partial_t \phi & = & g^{ab} \nabla_a \nabla_b \phi
+ {\lambda \over 2 g } g_{ab} {\cal E}^a {\cal E}^b e^{-\lambda \phi}
- {\lambda \over 4 } g^{ab} g^{ch} F_{ac} F_{bh} e^{\lambda \phi}, 
\label{evoldil}\\
\partial_t {\cal E}^a & = & e_b( \sqrt{g} g^{ac} g^{bh} F_{ch}
e^{\lambda \phi})
+ (f^i_{ib} g^{ac} + {1 \over 2} f^a_{bi} g^{ic})
\sqrt{g} g^{bh} F_{ch} e^{\lambda \phi},  \label{evole}\\
\partial_t F_{ab} & = & -2 e_{[a} ( {1 \over \sqrt{g}}
g_{b] c} {\cal E}^c e^{-\lambda \phi}) + f^c_{ab}
{1 \over \sqrt{g}} g_{c h} {\cal E}^h e^{-\lambda \phi}.
\label{evolm}
\end{eqnarray}

\subsection{Velocity-dominated evolution equations and
solution}\label{stepone}
The Kasner-like, or velocity-dominated, evolution equations are obtained
from
the full evolution equations by: (i) dropping
the spatial derivatives from the right hand sides of
(\ref{eveq}), (\ref{evoldil}), (\ref{evole}) and (\ref{evolm})
(note that $f^c_{ab}$-terms count as derivatives and
that we keep the time derivatives of the magnetic field
in (\ref{evolm}) even though
$F_{a b} = \partial_a A_b -\partial_b A_a $);
 and (ii) dropping the $p$-form terms in
both the Einstein and dilaton evolution equations.  This is a general rule
and yields in this case
\begin{eqnarray}
\label{dtg0}
\partial_t \, ^0g_{ab} & = & -2 \, ^0k_{ab}, \\
\label{dtk0}
\partial_t \, ^0{k^a}_b & = & (\tr \, ^0k) \, ^0{k^a}_b, \\
\label{dtphi0}
\partial^2_t \, ^0 \phi - (\tr \, ^0k) \, \partial_t \, ^0 \phi & = & 0, \\
\label{dtE0}
\partial_t \, {}^0 {\cal E}^a & = & 0, \\
\label{dtF0}
\partial_t \, {}^0 F_{ab} & = & 0.
\end{eqnarray}
(As we shall see below, interaction terms of Yang-Mills or other types
-- if any -- should also be dropped.)  

It is easy to find the general analytic solution of the evolution system 
(\ref{dtg0}) -- (\ref{dtF0})
 since the equations are the same as for ``Bianchi type I"
homogeneous models (one such set of equations per spatial
point).  Taking the trace of (\ref{dtk0}) shows that
$ -1/ \tr \, ^0k = t + C(x)$. By a suitable redefinition of
the time variable one can set $C(x)$ to zero.  Then (\ref{dtk0})
shows that  $- t \,^0{k^a}_b \equiv {K^a}_b$ is a constant matrix
(which must satisfy $\tr K = {K^a}_a = 1$, and be such that
$^0g_{a c }(t_0) {K^c}_b$ is symmetric in $a$ and $b$),
\begin{equation}
\label{0kab1}
^0 {k^a}_b(t)  =  - t^{-1} \, {K^a}_b . 
\end{equation}
Injecting this information into (\ref{dtg0})
leads to a linear evolution system for $ ^0g_{a b}$: 
$ t \, \partial_t \, ^0 g_{a b} = 2 \;  ^0 g_{a c } {K^c}_b$,
which is solved by exponentiation,
\begin{equation}\label{exp}
\label{0gab1}
^0 g_{ab}(t)  =  \, ^0 g_{ac}(t_0) {{\left[ {\left(
{t \over t_0} \right)}^{2 K} \right]}^c}_b.
\end{equation}
The other evolution equations are also easy to solve,
\begin{eqnarray}
\label{0phi1}
^0 \phi(t) & = & A \ln t + B, \\
\label{0E1}
^0 {\cal E}^a(t) & = & ^0 {\cal E}^a \\
\label{0F1}
^0 F_{ab}(t) & = & ^0 F_{ab}.
\end{eqnarray}
In (\ref{0gab1}) $(t/t_0)^{2 K}$ denotes the exponentiation of a matrix.
Quantities on the left hand side of (\ref{0kab1}) -- (\ref{0F1})
may be functions of both time and space, while all the time dependence of 
the right hand side
is made explicit.  For instance, (\ref{0F1}) is saying that 
the spacetime dependence of the general magnetic field $^0 F_{ab}(t,x)$ (solution
of the velocity-dominated evolution system) is reduced to a simple space
dependence $^0 F_{ab}(x)$ (where $^0F_{ab}$ is an antisymmetric spatial tensor).
Let $p_a$ denote the eigenvalues
 of ${K^a}_b$, ordered such that $p_1 \leq p_2 \leq p_3$.
 Since $\tr K=1$, we have the constraint 
\be
p_1 + p_2 + p_3 =1.
\label{Kasn1a}
\ee
In the works of BKL the matrix solution (\ref{0gab1})
is simplified by using a special frame $\{e_a\}$ with respect to
which the matrices $^0 g_{ab}(t_0)$ and ${K^a}_b$ are diagonal.
However, as emphasized in \cite{AR}, this choice can not
necessarily be made analytically on neighbourhoods where
the number of distinct eigenvalues of ${K^a}_b$ is not constant.
To obtain an analytic solution, while still controlling
the relation of the solution to the eigenvalues
of ${K^a}_b$, a special construction was introduced
in \cite{AR}.  This construction is based on some (possibly small)
neighbourhood $U_0$ of an arbitrary spatial point $x_0 \in \Sigma$ and uses
a set of auxiliary exponents $q_a(x)$. These auxiliary exponents remain
numerically close to the exact ``Kasner exponents'' $p_a(x)$, are  
analytic and enable one to define an analytic frame (see below).
To construct the auxiliary
exponents $q_a(x)$ one distinguishes three cases:

Case I (near isotropic):  \, If all three eigenvalues are equal at $x_0$,
choose a number $\epsilon > 0$ so that for $x \in U_0$,
$\max_{a,b} |p_a(x) - p_b(x)| < \epsilon/2$.
In this case define $q_a = 1/3$ on $U_0$, $a = 1,2,3$.

Case II (near double eigenvalue):  \, If the number of 
distinct eigenvalues at $x_0$
is two,  choose $\epsilon > 0$ so that for
$x \in U_0$, $\max_{a,b} |p_a - p_b| > \epsilon/2$, and
$|p_{a'} - p_{b'}| < \epsilon/2$ for some pair, $a'$,
$b'$, $a' \neq b'$, shrinking $U_0$ if necessary.
Denote by $p_\perp$ the
distinguished exponent not equal to $p_{a'}$, $p_{b'}$.
In this case define $q_\perp = p_\perp$
and $q_{a'} = q_{b'} = (1 - q_\perp)/2$ on $U_0$.

Case III (near diagonalizable): \, If all eigenvalues are distinct at $x_0$,
choose $\epsilon > 0$ so that for $x \in U_0$,
$\min_{\stackrel{a,b}{a \neq b}} |p_a(x) - p_b(x)| > \epsilon/2$,
shrinking $U_0$ if necessary.
In this case define $q_a = p_a$ on $U_0$.

The frame $\{ e_a \}$, called the adapted frame, is required to be
such that the related (time-dependent) frame 
$\{ \tilde e_a (t) \equiv t^{-q_a} e_a \}$ is orthonormal with
respect to the velocity-dominated metric at some time $t_0 > 0$,
{\it i.e.,} such that $ ^0g_{a b}(t_0) = t_0^{2 q_a} \delta_{a b}$.
(Here and in the rest of the paper, the Einstein summation convention
does not apply to indices on $q_a$ and $p_a$.
These indices should be ignored when determining sums. Furthermore,
quantities with a tilde will refer to the frame $\{ \tilde e_a (t)\}$.)

In addition, in Case II it is required that $e_\perp$ be
an eigenvector of $K$ corresponding to $q_\perp$ and
that $e_{a'}$, $e_{b'}$ span the eigenspace of $K$
corresponding to the eigenvalues $p_{a'}$, $p_{b'}$.
In case III it is required that the $e_{a}$ be
eigenvectors of $K$ corresponding to the eigenvalues $q_a (\equiv p_a)$.
In all cases it is required that $\{ e_a \}$ be analytic.
The auxiliary exponents, $q_a$, are analytic,
satisfy the Kasner relation $\sum q_a = 1$, are ordered
($q_1 \leq q_2 \leq q_3$), and satisfy $q_1 \geq p_1$,
$q_3 \leq p_3$ and $\max_a |q_a - p_a| < \epsilon/2$.
If $q_a \neq q_b$, then $^0g_{ab}$, $^0g^{ab}$,
$^0 \tilde g_{ab}$ and $^0 \tilde g^{ab}$ all vanish,
and the same is true with $g$ replaced by $k$. 

Equations~(\ref{0kab1}) -- (\ref{0F1}), with the form of
$g_{ab}(t_0)$ and ${K^a}_b$ specialized as given just above, are
the general analytic solution to the velocity-dominated evolution
equations in the sense that any analytic solution to the
velocity-dominated evolution equations takes this form
near any $x_0 \in \Sigma$ by choice of
(global) time coordinate and (local) spatial frame.

\subsection{Velocity-dominated constraint equations}\label{steptwo}

When written in terms of the velocity-dominated variables, the
velocity-dominated constraints take the same form as the full constraint equations,
except the Hamiltonian constraint, which is obtained by dropping
spatial gradients and electromagnetic contributions to the energy-density.
This is a general rule, valid also for the more general models
considered below.
Thus, if we define
\begin{eqnarray}
^0 \rho & = & {1 \over 2} (\partial_t \, ^0 \phi)^2, \nonumber \\
^0 j_a & = & - \partial_t \, ^0 \phi \, e_a(\, ^0 \phi) +
{1 \over \sqrt{ \, ^0 g} } \, ^0 {\cal E}^b \, \, ^0 F_{ab}, \nonumber
\end{eqnarray}
we get $^0C=0$ and $^0C_a = 0$
for the velocity-dominated constraints corresponding to the Hamiltonian
and momentum constraints, with
\begin{eqnarray}
\label{hamiltonian0}
^0C & = & -\,{^0k^a}_b \, {^0k^b}_a + ( \tr \, ^0k )^2 - 2 \, ^0\rho, \\
\label{momentum0}
^0C_a  & = & \, ^0\nabla_b \, ^0{k^b}_a - e_a (\tr \, ^0k) - \, ^0j_a.
\end{eqnarray}                                
The velocity-dominated matter constraint equations read
\begin{eqnarray}
e_a ( \, ^0 {\cal E}^a) + f^b_{ba} \, ^0 {\cal E}^a & = & 0, \nonumber \\
e_{[a}( \, ^0 F_{bc]}) + f^h_{[a b} \, ^0 F_{c] h} & = & 0. \nonumber
\end{eqnarray}

For the solution (\ref{0kab1})  -- (\ref{0phi1})
the velocity-dominated Hamiltonian constraint
equation is equivalent to 
\be
\sum {p_a}^2 + A^2 = 1.
\label{Kasn1b}
\ee
The conditions (\ref{Kasn1a}) and (\ref{Kasn1b}) are the famous
Kasner conditions when the dilaton is present.  While $p_1$
is necessarily non-positive when $A =0$, this is no longer the
case when the dilaton is nontrivial ($A \not=0$): all 
$p_a$'s can then  be positive.  This is the major
feature associated with the presence of the dilaton, which turns
the mixmaster behaviour of (4-dimensional) vacuum gravity into the
velocity-dominated behaviour.  We shall call $(p_a,A)$ the Kasner
exponents (because they are the exponents of the
proper time in the solution for the scale factors and
$\exp \phi$) and refer to (\ref{Kasn1a}) and (\ref{Kasn1b})
as the Kasner conditions (note that $A$ is often denoted $p_{\phi}$ to emphasize 
its relation to the kinetic energy of $\phi$, and its similarity with the other exponents).

A straightforward calculation shows that
\begin{eqnarray}
\label{dt0C}
\partial_t \, {}^0 C - 2 (\tr \, {}^0 k)  \, {}^0 C & = & 0 ,\\
\label{dt0Ca}
\partial_t \, {}^0 C_a - (\tr \, {}^0 k) \, {}^0 C_a & = &
-{1 \over 2} e_a( \, {}^0 C ).
\end{eqnarray}
Thus if the velocity-dominated Hamiltonian and
momentum constraints are satisfied at some
$t_0 > 0$, then they are satisfied for all $t>0$.
Similarly, since $^0 {\cal E}^a$ and $^0 F_{ab}$ are independent
of time, if the matter constraints are satisfied at
some time $t_0 > 0$, then they are clearly satisfied for all time.

\subsection{Critical value of dilaton coupling $\lambda_c$}
\label{criticalvalue}
Our ultimate goal is to show that the velocity-dominated solutions
asymptotically approach (as $t \rightarrow 0$) solutions
of the original system of equations.  We shall prove that this is the
case provided the Kasner exponents $p_i, A$, subject
to the Kasner conditions
\be
\sum {p_a}^2 - (\sum {p_a})^2 + A^2 = 0, \; \; \; \sum {p_a} = 1
\label{Kasner1}
\ee    
 obey additional restrictions. 
These restrictions are inequalities on the Kasner exponents
and read explicitly
\begin{equation}
\label{inequalities1}
2 p_1 - \lambda A > 0,  \hspace{20pt} p_1 >0, 
\hspace{20pt} 2 p_1 + \lambda A > 0.
\end{equation}
As explained in \cite{dh1}, and rigorously checked below,
these restrictions are necessary and sufficient to ensure that the terms
that are dropped when replacing the full Einstein-dilaton-Maxwell equations
by the velocity-dominated equations become indeed negligible as $t \to 0$.
More precisely, the first condition (respectively the third)
among (\ref{inequalities1})
guarantees that one can neglect the electric (respectively, magnetic)
part of the
energy-momentum tensor of the electromagnetic field
in the Einstein equations, whereas
the condition $p_1>0$ is necessary for the spatial curvature terms
to be asymptotically negligible.  The conditions (\ref{inequalities1}) define
the set $V$ of velocity-dominated solutions referred to in the introduction.

It is clear that if $\vert \lambda \vert$ is small enough -- in particular,
if $\lambda = 0$ -- the inequalities (\ref{inequalities1}) can 
be fulfilled since the Kasner exponents can be all positive when the
dilaton is included.  But if $\vert \lambda \vert$ is greater that some critical
value $\lambda_c$, it is impossible to fulfill simultaneously
the Kasner conditions (\ref{Kasner1}) and the 
inequalities (\ref{inequalities1}), because one of the
terms $\pm \lambda A$ becomes more negative than $2p_1$ is positive. 
In that case, the set
$V$ is empty and our construction breaks down.  For $\vert \lambda \vert
<\lambda_c$, however, the set  $V$ is non-empty and, in fact,
stable under small perturbations of the Kasner exponents
since (\ref{inequalities1}) defines an open region on the
Kasner sphere.  
We determine in this subsection the critical
value $\lambda_c$ such that (\ref{Kasner1}) and (\ref{inequalities1}) are
compatible whenever $\vert \lambda \vert
<\lambda_c$. 

To that end, we follow the geometric approach of \cite{tDmH,DHJN}.
In the 4-dimensional space of the Kasner exponents $(p_a,A)$,
we consider the ``wall chamber'' ${\cal W}$ defined to be the conical domain where 
\be
p_1 \leq p_2 \leq p_3, \; \; 
2 p_a - \lambda A \geq 0, \; \;
p_a \geq 0, \; \;
2 p_a + \lambda A \geq 0.
\ee
These inequalities are not all independent since the four conditions
\be
p_1 \leq p_2 \leq p_3, \; \; 2 p_1  - \lambda A \geq 0, \; \;
2 p_1 + \lambda A \geq 0
\label{wallcone}
\ee
imply all others.  The quadratic Kasner condition (\ref{Kasner1}) can be rewritten
\be
G_{\m \n} p^\m p^\n = 0, \hspace{20pt} (p^\m) \equiv  (p_a, A)
\label{light}
\ee
where $G_{\m \n}$ defines a metric in ``Kasner-exponent space''
\be
dS^2 = G_{\m \n} dp^\m dp^\n = 
\sum {dp_a}^2 - (\sum {dp_a})^2 + (dA)^2
\label{Kmetric}
\ee
The metric (\ref{Kmetric}) has Minkowskian signature
$(-,+,+,+)$.  An example of timelike direction is given
by $p_1 = p_2 = p_3$, $A=0$.  
Inside or on the light cone,
the function $\sum {p_a}$ does not vanish.  The upper light cone 
(in the space of the Kasner exponents) is conventionally
defined by (\ref{light}) and the extra condition
$\sum {p_a} >0$.
It is  clear from our discussion
that the Kasner conditions (\ref{Kasner1}) and the
inequalities (\ref{inequalities1}) are compatible if and only if
there are lightlike directions in the interior of the wall chamber
${\cal W}$
(by rescaling $p^\m \rightarrow \a p^\m$, $\a >0$, one can always
make $\sum p_a = 1$ for any point in the interior of the wall chamber so
that this condition does not bring a restriction).
The problem amounts accordingly to determining the relative position
of the light cone (\ref{light}) and the wall chamber (\ref{wallcone}).  

This is most easily done by computing the edges of (\ref{wallcone}),
{\it i.e.,} the one-dimensional intersections of three faces among
the four faces (\ref{wallcone}) of ${\cal W}$.  There
are four of them: (i) $p_1 = p_2 = A = 0$, $p_3 = \a$; (ii) $p_1 = A = 0; p_2
=p_3 = \a$; (iii) $2p_1 = 2p_2 = 2p_3 = \lambda A = \a$; and (iv)
$2p_1 = 2p_2 = 2p_3 = -\lambda A = \a$, where in each case, 
$\a \geq 0$ is a parameter along the edge ($\a = 0$ being the origin).  
The vectors $e^\m_A$ ($A=1,2,3,4$)
along the edges corresponding to $\a = 1$, namely
$(0,0,1,0)$, $(0,1,1,0)$, $(1/2, 1/2, 1/2, 1/\lambda)$ and
$(1/2, 1/2, 1/2, -1/\lambda)$ form
a basis in Kasner-exponent space.  Any vector $v^\m$ can thus be expanded
along the $e^\m_A$, $v^\m = v^A e^\m_A$. 
A point $P$ in Kasner-exponent space is on or inside the wall chamber
${\cal W}$ if and only if its coordinates $p^A$ in this basis
fulfill $p^A \geq 0$ with $P$ inside when $p^A >0$ for all $A's$. 
Thus, if all the edge vectors $e^\m_A$ are timelike or lightlike,
the Kasner conditions
are incompatible with the inequalities (\ref{inequalities1}) since
any linear combination of causal vectors with non-negative coefficients
is on or inside the forward light cone (the $e^\m_A$'s are future-directed
since $p_1 + p_2 + p_3 >0$ for all of them).  If, however,
one (or more) of the edge vectors lies outside the light cone, then, the
Kasner conditions and the inequalities (\ref{inequalities1}) are compatible.
The nature of some of the edge vectors depends 
on the value of the dilaton coupling $\lambda$: 
while the first one is always lightlike and  the 
second one always timelike,  the squared norm
of the last two
is $ - 3/2 + 1/\lambda^2 = (2 - 3 \lambda^2) /( 2 \lambda^2)$.
This determines the critical value 
\be
\lambda_c = \sqrt{2 \over 3}
\label{critical}
\ee 
such that the edge vectors are timelike or null (incompatible inequalities) 
if $ \vert \lambda \vert \geq \lambda_c$, but spacelike (compatible
inequalities) if $  \vert \lambda \vert < \lambda_c$.  Note that the
value of $\lambda$ that arises from dimensionally reducing $5$-dimensional
vacuum gravity down to $4$ dimensions is $\lambda = \sqrt{6}$ and
exceeds the critical value.  This ``explains'' the conclusion reached
in \cite{BK2} that the gravity-dilaton-Maxwell system obtained
by Kaluza-Klein reduction of 5-dimensional gravity is oscillatory.

We shall assume from now on that $\vert \lambda \vert
<\lambda_c$ and that the Kasner exponents fulfill the
above inequalities.
For later use, we choose a number $\sigma > 0$ so that, for all $x \in U_0$,
$\sigma < 2 p_1 - \lambda A $,
$\sigma < 2 p_1 + \lambda A $ and
$\sigma < p_1/2 $.  Reduce $\epsilon$ if necessary so that
$\epsilon < \sigma/7 $.  If $\epsilon$ is reduced, it
may be necessary to shrink $U_0$ so that the conditions imposed in
section \ref{stepone} remain satisfied.  In section \ref{stepthree}
it is assumed that $\epsilon$ and $U_0$ are such that the conditions
imposed in section \ref{stepone} and the conditions imposed
in this paragraph are all satisfied.         

\subsection{Fuchsian system which is equivalent to the
Einstein-matter evolution equations} \label{stepthree}
\subsubsection{Rewriting of equations}
Theorem 3 in \cite{AR} (theorem 4.2 in preprint 
version), on which we rely for our result, states
that a Fuchsian equation ({\it i.e.,} as we mentioned above, an
equation of the form (\ref{fuchs0}) where $\cal A$ satisfies a
positivity condition and $f$ is regular, which includes a boundedness
property) has a unique solution $u$ that vanishes as $t \downarrow 0$,
and furthermore spatial derivatives of $u$ of any order
vanish as $t \downarrow 0$, as shown in \cite{KR}.  Our goal is to
recast the Einstein-matter evolution equations as a Fuchsian equation
for the deviations from the velocity-dominated solutions.  Thus,
we denote the unknown vector $u$ as
\begin{equation}\label{ucomponents}
u = ({\gamma^a}_b, {\lambda^a}_{bc},
{\kappa^a}_b, \psi, \omega_a, \chi,\xi^a,\varphi_{ab})
\end{equation}
where the variables ${\gamma^a}_b$ {\it etc.} are related to
the Einstein-matter variables by
\begin{eqnarray}
\label{defgamma}
g_{ab} & = & \, ^0 g_{ab} + \, ^0 g_{ac} t^{{\alpha^c}_b}
{\gamma^c}_b, \\
\label{deflambda}
e_c({\gamma^a}_b) & = & t^{-\zeta} {\lambda^a}_{bc}, \\
\label{defkappa}
k_{ab} & = & g_{ac}( \, ^0{k^c}_b + t^{-1 + {\alpha^c}_b}
{\kappa^c}_b), \\
\label{defpsi}
\phi & = & \,^0 \phi + t^\beta \psi, \\
\label{defomega}
e_a(\psi) & = & t^{-\zeta} \omega_a, \\
\label{defchi}
t \, \partial_t \psi + \beta \, \psi & = & \chi, \\
\label{defxi}
{\cal E}^a & = & \, ^0 {\cal E}^a + t^\beta \xi^a, \\
\label{defvarphi}
F_{ab} & = & \, ^0 F_{ab} + t^\beta \varphi_{ab}.
\end{eqnarray}
In the first of these equations $t^{{\alpha^c}_b}$ is {\it not} the 
exponentiation of a matrix with components ${\alpha^c}_b$ such as occurs 
in (\ref{exp}). The
expression $t^{{\alpha^c}_b}$ is for each fixed value of $c$ and $b$ the
number which is $t$ raised to the power given by the number ${\alpha^c}_b$
(defined below). 
In equations~(\ref{defgamma}) and (\ref{defkappa}) there is no summation
on the  index $b$ (but there is a summation on $c$). In
(\ref{defvarphi}) $\varphi_{ab}$ is a totally antisymmetric spatial
tensor, which contributes three independent components to $u$.
This assumption is consistent with the form of the evolution
equation for $\varphi_{ab}$, equation (\ref{fuchvarphi}) below. 
The exponents appearing in (\ref{defgamma}) -- (\ref{defvarphi})
are as follows.  Define $\alpha_0 = 4 \epsilon$,
$\beta = \epsilon/100$ and $\zeta = \epsilon/200$ (where $\epsilon$ is
the same (small) quantity which entered the definition of the
auxiliary exponents $q_a$ in section~\ref{stepone} and which
was further restricted at the end of section~\ref{criticalvalue}).
All of these quantities are independent of $t$ and $x$.  Finally define
\begin{displaymath}
{\alpha^a}_b = 2 \max(q_b - q_a, 0) + \alpha_0 =
 2 q_{\max \{ a,b \} } - 2 q_a + \alpha_0.
\end{displaymath}
Note that the numbers ${\alpha^a}_b$ are all strictly positive. In the
second definition of ${\alpha^a}_b$ we have used the fact that the $q_a$'s
are ordered. The role of ${\alpha^a}_b$ is to shift the spectrum of the 
Fuchsian-system matrix $\cal A$, in equation (\ref{fuchs0}), to be
positive. It is not clear to what extent the choice of ${\alpha^a}_b$ is
fixed by the requirement of getting a Fuchsian system. It seems that the
(triangle-like) inequality (42) of \cite{AR} 
(inequality (5.9) in preprint version)
is a key property of these coefficients.
We shall further comment below on the specific choice of
${\alpha^a}_b$ and its link with the BKL-type approach
to the cosmological behaviour near $t = 0$.

When writing the first-order evolution system for $u$ we momentarily
abandon the restriction that $g_{ab}$ and $k_{ab}$ be symmetric,
as in \cite{AR}.
Thus we need to define $g^{ab}$, and we do so by requiring that
$g_{ab} g^{bc} = {\delta_a}^c$.  This implies that
$g^{ab} g_{bc} = {\delta^a}_c$.  We lower indices on tensors
by contraction with the second index of $g_{ab}$ and also raise
indices on tensors by contraction with the second index of $g^{ab}$.
This choice is so that raising and then lowering a given
index results in the original tensor, and the same for lowering
and then raising an index.  The position of the indices
on quantities appearing in $u$ and other
such quantities is fixed.  Repeated indices on these quantities
imply a summation.  On the other hand, as we already mentioned above,
one qualifies the summation convention by insisting that indices repeated only 
because of their occurrence on
$p_a$, $q_a$, ${\alpha^a}_b$ and other such non-tensorial quantities should be
ignored when determining sums.

Substituting (\ref{defgamma}) -- (\ref{defvarphi})
in the evolution equations yields equations of motion for $u$
of the form (\ref{fuchs0})
\begin{eqnarray}
\label{fuchgamma}
&&\hspace{-40pt}
t \, \partial_t {\gamma^a}_b  + {\alpha^a}_b {\gamma^a}_b
+ 2 {\kappa^a}_b -2 (t \, ^0{k^a}_c) {\gamma^c}_b
+ 2 {\gamma^a}_c (t \, ^0{k^c}_b) \; = \;
-2 \, t^{{\alpha^a}_c + {\alpha^c}_b - {\alpha^a}_b}
{\gamma^a}_c {\kappa^c}_b \\
\label{fuchlambda}
&&\hspace{-40pt}
t \, \partial_t {\lambda^a}_{bc} \; = \;
t^\zeta e_c(t \, \partial_t {\gamma^a}_b)
+ \zeta \, t^\zeta e_c({\gamma^a}_b) \\
\label{fuchkappa}
&&\hspace{-40pt}
t \, \partial_t {\kappa^a}_b + {\alpha^a}_b {\kappa^a}_b
 -(t \, ^0{k^a}_b) (\tr \kappa) \; = \; t^{\alpha_0} (\tr \kappa) {\kappa^a}_b
+ t^{2 - {\alpha^a}_b} (\, ^S {R^a}_b - {M^a}_b) \\
\label{fuchpsi}
&&\hspace{-40pt}
t \, \partial_t \psi + \beta \psi - \chi \; = \; 0 \\
\label{fuchomega}
&&\hspace{-40pt}
t \, \partial_t \omega_a \; = \; t^\zeta \{ e_a(\chi)
+ (\zeta - \beta) e_a(\psi) \} \\
\label{fuchchi}
&&\hspace{-40pt}
t \, \partial_t \chi + \beta \chi \; = \; t^{\alpha_0 - \beta}
(\tr \, \kappa) ( A + t ^\beta \chi ) + t^{2 - \beta}
g^{ab} \nabla_a \nabla_b \, ^0 \phi + t^{2 - \zeta} \nabla^a \omega_a
\nonumber \\ && \hspace{60pt}
+t^{2 - \beta} \{ {\lambda \over 2 g}
g_{ab} {\cal E}^a {\cal E}^b e^{-\lambda \phi}
- {\lambda \over 4} g^{ab} g^{ch} F_{ac} F_{bh}
e^{\lambda \phi} \} \\
\label{fuchxi}
&&\hspace{-40pt}
t \, \partial_t \xi^a + \beta \xi^a \; = \;
t^{1 - \beta} \{ 
e_b( \sqrt{g} g^{ac} g^{bh} F_{ch} e^{\lambda \phi})  \nonumber \\
&& \hspace{60pt} + (f^i_{ib} g^{ac}
+ {1 \over 2} f^a_{bi} g^{ic})
\sqrt{g} g^{bh} F_{ch} e^{\lambda \phi} \}  \\
\label{fuchvarphi}
&&\hspace{-40pt}
t \, \partial_t \varphi_{ab} + \beta \varphi_{ab} \; = \;
t^{1 - \beta} \{ -2 e_{[a} ( {1 \over \sqrt{g}} 
g_{b] c} {\cal E}^c e^{-\lambda \phi})
+ f^c_{ab} {1 \over \sqrt{g}} g_{c h} {\cal E}^h e^{-\lambda \phi} \} 
\end{eqnarray}
All the quantities entering these equations have been defined, except
$^S {R^a}_b$.  This is done by taking the Ricci tensor of
the  symmetric part $ g_{(a b)}$ of  $g_{a b}$ \cite{AR}. More explicitly,  
$^S {R^a}_b = g^{ac} \, ^S R_{cb}$, with
\begin{equation}
\label{curvature}
^S R_{ab} = 
t^{q_a + q_b} \, ^S \tilde R_{ahb}^{\; \; \; \; \; \; h}
= t^{q_a + q_b} \{ \tilde e_h ( \, ^S
\tilde \Gamma^h_{ab}) - \tilde e_a ( \, ^S \tilde \Gamma^h_{hb})
+ \, ^S \tilde \Gamma^i_{ab} \, ^S \tilde \Gamma^h_{hi}
- \, ^S \tilde \Gamma^i_{hb} \, ^S \tilde \Gamma^h_{ai}
+ \tilde f^i_{ah} \, ^S \tilde \Gamma^h_{ib} \},
\end{equation}
and the connection coefficients in the frame $\{ \tilde e_a \}$,
\begin{equation}
\label{christoffel}
^S \tilde \Gamma^c_{a b} = 
{1 \over 2} \, ^S \tilde g^{ch} \left\{
\tilde e_a (\tilde g_{(bh)}) + \tilde e_b (\tilde g_{(ha)})
- \tilde e_h (\tilde g_{(ab)})
- \tilde g_{(ia)} \tilde f^i_{bh}
- \tilde g_{(bi)} \tilde f^i_{ah} \right\}
+ {1 \over 2} \tilde f^c_{ab}.
\end{equation}
Here, $^S \tilde{g}^{ab}$ is defined as the inverse of $\tilde{g}_{(ab)}$.
Once it is shown that the tensor $g_{ab}$ in equation (\ref{defgamma})
is symmetric, then it follows that
$^S {R^a}_b = {R^a}_b$  and that equations
(\ref{fuchgamma}) -- (\ref{fuchvarphi}) are equivalent to the
Einstein-matter equations.

\subsubsection{The system (\ref{fuchgamma})--(\ref{fuchvarphi})
is Fuchsian}

A good deal of the work needed to show that equation (\ref{fuchs0})
(as written out in equations (\ref{fuchgamma}) -- (\ref{fuchvarphi}))
is Fuchsian was done in \cite{AR}, in the massless scalar field case
considered there.  The form of the velocity-dominated evolution and the
form of the function $u$ are the same in the two cases except for
the crucial addition of new source terms and new 
evolution equations involving  the Maxwell field. The
presence of the new components does not alter already existing parts of
the matrix $\cal A$, nor already existing terms in $f$. 
The difference between $\cal A$ here
and $\cal A$ in the massless scalar field case considered
in \cite{AR} is that here there are additional rows and
columns, such that the only non-vanishing new entries
are on the diagonal and strictly positive.  Therefore
the argument in \cite{AR} that their $\cal A$ satisfies
the appropriate positivity condition implies that our
$\cal A$ satisfies the appropriate positivity condition.

On the other hand, it is crucial to control in detail the new source
terms in $f$, connected to the Maxwell field, which were absent
in \cite{AR}.  It is for the study of these terms that the results
of \cite{dh1}, and in particular the inequalities (\ref{inequalities1})
which were shown there to guarantee that Maxwell source terms become
asymptotically subdominant near the singularity, become important.
Recall that the crucial criterion for the source $f(t,x,u,u_x)$ is
that it be $O(t^\delta)$, for some strictly positive $\delta$.
In regard to this estimate, we use the notation ``big $O$,''
``$\preceq$'' and ``small $o$'' as follows.
Given two functions $F(t,x,u,u_x)$ and $G(t,x,u,u_x)$ we use the
notation $F \preceq G $, to denote that, for every compact set
$K$, there exists a  constant $C$ and a number $t_0 > 0$ such that
$ \vert F(t,x,u,u_x) \vert \leq C \vert G(t,x,u,u_x) \vert$ when
$(x,u,u_x) \in K$ and $0< t \leq t_0$
(see Definition 1 in \cite{AR}).  If $G$ is a function only
of $t$ (e.g. a power of $t$), then we replace $F \preceq G $ with
$F = O(G)$.  If $f(t,x,u,u_x) = O(t^\delta)$,
then by reducing the value of $\delta$ (keeping it
positive) we have that $f(t,x,u,u_x) = o(t^\delta)$
with a ``small o'' which denotes that $f/t^\delta$ tends
to zero uniformly on compact sets $K$ as $t \rightarrow 0$.

The new source terms involving the Maxwell field are:
the last four terms in ${M^a}_b$ (see equation~(\ref{calcM})),
the last two terms on the right hand side of equation~(\ref{fuchchi})
and the terms of the right hand sides of equations~(\ref{fuchxi})
and (\ref{fuchvarphi}).

The calculation of the estimates starts in the frame,
$\{ \tilde e_a \}$, defined in section~\ref{stepone}.
For more details concerning the basic estimates, we
refer the reader to \cite{AR}.  In the frame $\{ \tilde e_a \}$
the Kasner-like metric is ({\it cf.} (\ref{0gab1}))
\begin{eqnarray}
\label{0tildegab1} 
^0 \tilde g_{ab} & = & ^0 \tilde g_{ac}(t_0) {{\left[ {\left(
{t \over t_0} \right)}^{2 ( K - Q)}\right]}^c}_b, \\
\label{0tildegab2}
^0 \tilde g^{ab} & = &
{{\left[ {\left(
{t \over t_0} \right)}^{-2 ( K - Q)}\right]}^a}_c
 \, ^0 \tilde g^{cb}(t_0),
\end{eqnarray}
where the matrix $Q$ is the diagonal matrix
$ {Q^a}_b \equiv q_a {\delta^a}_b$
which commutes with $K$.  With our choice of frame,
$^0 \tilde g_{ab}(t_0) = \delta_{ab}$
and $^0 \tilde g^{ab}(t_0) = \delta^{ab}$.
In Lemma 2 in \cite{AR} (lemma 5.1 in preprint
version), the form of (\ref{0tildegab1}) and
(\ref{0tildegab2}) is considered and it is shown that
$^0\tilde g_{ab} = O(t^{-\epsilon})$ and
$^0\tilde g^{ab} = O(t^{-\epsilon})$.
It is useful to write down expressions for the proposed
metric and extrinsic curvature in the frame $\{\tilde e_a\}$.
The components in terms of this frame are 
\begin{eqnarray}
\tilde g_{ab} & = & \, ^0 \tilde g_{ab} +
\, ^0 \tilde g_{ac} t^{\tilde \alpha^c_{\; \; b}}
{\gamma^c}_b,  \nonumber \\
\tilde k_{ab} & = & \tilde g_{ac}( \, ^0 \tilde k^c_{\; \; b} +
t^{-1 + \tilde \alpha^c_{\; \; b}} {\kappa^c}_b). \nonumber
\end{eqnarray}
Here, $\tilde \alpha^a_{\; \; b} = \alpha^a_{\; \; b} + q_a - q_b =
|q_a - q_b| + \alpha_0$ is symmetric in $a$, $b$,
$\tilde \alpha^a_{\; \; b} = \tilde \alpha^b_{\; \; a}$.
To get an estimate for the inverse metric, we note
first that the inverse of $g_{ac}{}^0 g^{cb}$ is given by 
$g^{ca}{}^0 g_{cb}$. Thus it is possible to express the latter 
quantity algebraically in terms of ${}^0 g_{ab}$ and ${\gamma^a}_b$.
Now define 
\begin{equation}
\bar{\gamma^a}_b=-t^{-{\tilde\alpha^a}_b}(\delta^a_b-\tilde{g}^{ac}
{}^0\tilde{g}_{cb}),
\end{equation}
which, from what we just observed, can be expressed algebraically in
terms of known quantities and ${\gamma^a}_b$.
Then one has
\begin{equation}
\label{tildeinvg}
\tilde g^{ab}  =  ^0 \tilde g^{ab} +
t^{\tilde \alpha^a_{\; \; c}} \bar \gamma^a_{\; \; c} \, ^0 \tilde g^{cb}.
\end{equation}                  
As a consequence of an
argument given in \cite{AR} which uses the (triangle-like)
inequality (42) 
of that paper ((5.9) in preprint version)
and the matrix identity preceding it, this
exhibits $\bar{\gamma^a}_b$ as a regular function of ${\gamma^a}_b$.  
In particular, if it is known that ${\gamma^a}_b$ is $o(1)$ then
the same is true of $\bar{\gamma^a}_b$.

To better grasp the usefulness of the introduction of the exponents
$\alpha^a_{\; \; b}$ and $\tilde \alpha^a_{\; \; b}$, and the link of
the Fuchsian estimates with the approximate estimates used in the
BKL-like works, let us consider more closely the simple case where
all the Kasner exponents are distinct (Case III). In this case
$p_a = q_a$ and one can diagonalize the Kasner-metric,
 so that, in the rescaled
frame $\tilde e_a$, we have simply (for all $t \leq t_0$)
$\,^0 \tilde g_{ab}(t)  = \delta_{ab}$. In such a case, the BKL-type
estimates would be obtained (in the time-dependent
 rescaled frame ${\tilde e_a}$)
by approximating the exact metric by its Kasner limit, i.e. simply
$ \tilde g_{ab}^{\rm BKL}(t)  = \delta_{ab}$. By contrast, the estimates
of the Fuchsian analysis are made with the exact metric,
$\tilde g_{ab}(t)  = \delta_{ab} 
+ t^{\tilde \alpha^a_{\; \; b}}  \gamma^a_{\; \; b}$, 
in which  $\gamma^a_{\; \; b}$, being part of $u$, is considered
to be in a compact set and hence is
bounded. As the diagonal $\tilde \alpha^a_{\; \; a} = \alpha_0 >0$,
we see that (in the frame ${\tilde e_a}$) 
the diagonal components of the ``Fuchsian'' metric
asymptote those of the ``BKL'' metric, and that both are close to one.
Concerning the non-diagonal components (in the frame ${\tilde e_a}$) 
of the ``Fuchsian'' metric we see
that they are constrained, by construction (i.e. by the choice
$\tilde \alpha^a_{\; \; b} = |q_a - q_b| + \alpha_0$),
 to tend to zero faster than
$ t^{|q_a - q_b|}$. This closeness between the metrics used in the
two types of estimates explains the parallelism between the rigorous
results derived here and the heuristic estimates used in BKL-type
works. If we come back to the general case where the Kasner metric
cannot be diagonalized in an analytic fashion, the optimal estimates
become worse by a negative power of $t$ (coming from the estimate 
of the matrix difference $ 2 ( K - Q) $ in equations
(\ref{0tildegab1}), (\ref{0tildegab2}) above). 
The proposed metric in the frame $\{\tilde e_a\}$ satisfies then
\begin{displaymath}
\tilde g_{ab}  \preceq t^{|q_a - q_b| - \epsilon}
\hspace{20pt} \mbox{and} \hspace{20pt}
\tilde g^{ab}  \preceq t^{|q_a - q_b| - \epsilon}.
\end{displaymath}
The proposed inverse metric in the adapted frame is
\begin{displaymath}
g^{ab} = \,^0 g^{ab} + t^{{\alpha^a}_c} \bar \gamma^a_{\; \; c} \, ^0 g^{cb}.
\end{displaymath}
The proposed metric in the adapted frame satisfies
\begin{equation}
\label{estmetric}
g_{ab} \preceq t^{ 2 q_{\max \{ a,b \}} - \epsilon}
\hspace{20pt} \mbox{and} \hspace{20pt}
g^{ab} \preceq t^{ -2 q_{\min \{ a,b \}} - \epsilon}.
\end{equation}
Estimates of spatial derivatives of the proposed
metric are also needed.
\begin{eqnarray}
e_c(\tilde g_{ab}) \preceq t^{|q_a - q_b| - \delta - \epsilon}
& \hspace{20pt} \mbox{and} \hspace{20pt} &
e_c(\tilde g^{ab}) \preceq t^{|q_a - q_b| - \delta - \epsilon},
\nonumber \\
\label{estdgab}
e_c(g_{ab}) \preceq t^{ 2 q_{\max \{ a,b \}} - \delta - \epsilon}
& \hspace{20pt} \mbox{and} \hspace{20pt} &
e_c(g^{ab}) \preceq t^{ -2 q_{\min \{ a,b \}} - \delta - \epsilon}
\end{eqnarray}
for some strictly positive $\delta$.

The determinant of the proposed metric also appears in some of
the new source terms.  From (\ref{0gab1}), the form of
$^0g_{ab}(t_0)$ and $\tr \, K = 1$, one gets
$^0 g = t^2$.  From (\ref{0tildegab1}) and
$^0 \tilde g_{ab}(t_0) = \delta_{ab}$ one gets
$^0 \tilde g = 1$.  The expression for the determinant
is a sum of terms of the form $g_{ab} g_{cd} g_{ef}$,
such that in each term, each index, 1, 2, 3, occurs exactly
twice.  {}From the Kasner relation for the $q_a$'s
and the relation between the two
frames, it follows that $g = t^2 \tilde g$.  Considering the
form of the various expressions, one then obtains $1/g = O(t^{-2})$,
$\sqrt{g} = O(t)$, $1/\sqrt{g} = O(t^{-1})$, and
$1/\sqrt{g} - 1/\sqrt{\, ^0g} = O(t^{-1 + \alpha_0 - 3 \epsilon})
= O(t^{-1 + \epsilon})$.
Spatial derivatives of the determinant also appear in $f$.
Considering the form of $\tilde g - \, ^0 \tilde g$ and that
$e_a(\,^0 \tilde g) = 0$, it follows that
$e_a(\tilde g) = O(t^{\alpha_0 - \delta - 3 \epsilon})$, and
\begin{equation}
\label{estddetg1}
e_a(g) = O(t^{2 + \alpha_0 - \delta - 3 \epsilon}).
\end{equation}
Finally,
\begin{displaymath}
e_a(g^{-1/2}) = - {e_a(g) \over 2 g^{3/2}}
= O(t^{-1 + \alpha_0 - \delta - 3 \epsilon}).
\end{displaymath}

Let us now consider the new source terms in $f$,
beginning with the last four terms of $t^{2 - {\alpha^a}_b} {M^a}_b$.
To estimate the contributions of ${\cal E}^a$ and $F_{a b}$ it is sufficient
to note from (\ref{defxi}) and (\ref{defvarphi}) that 
${\cal E}^a = O(1)$ and $F_{a b} = O(1)$. Then we get,
using the definition of ${\alpha^a}_b$
and (\ref{estmetric}),
\begin{eqnarray}
& & t^{2 - {\alpha^a}_b} {1 \over g} \{g_{bc} {\cal E}^a {\cal E}^c
-{1 \over 2} {\delta^a}_b g_{ch}  {\cal E}^c {\cal E}^h\}
e^{-\lambda \phi} \nonumber \\ & & \hspace{40pt} \preceq \,
\sum_c t^{-2 q_{\max\{a,b\}} + 2 q_a + 2 q_{\max\{b,c\}}
- \lambda A - \alpha_0 - \epsilon} +
\sum_{c,h} t^{+ 2 q_{\max\{c,h\}} - \lambda A - \alpha_0 - 
\epsilon} \nonumber \\  & & \hspace{40pt} \preceq \,
t^{2 q_1 - \lambda A - \alpha_0 - \epsilon}
\, = \, O(t^{- \alpha_0 - \epsilon + \sigma})
\, = \, O(t^{\delta}), \nonumber \\
& & t^{2 - {\alpha^a}_b} \{g^{ac}
g^{hi} F_{ch} F_{bi} -{1 \over 4}{\delta^a}_b
g^{ch} g^{ij} F_{ci} F_{hj}\} e^{\lambda \phi}
\nonumber \\
&& \hspace{40pt} \preceq \,
\sum_{c,h \neq c,i \neq b} t^{2-2 q_{\max\{a,b\}} + 2 q_a - 2 q_{\min\{a,c\}}
- 2 q_{\min\{h,i\}} + \lambda A - \alpha_0 - 2 \epsilon}
\nonumber \\ & & \hspace{60pt}
+ \sum_{c,h,i \neq c,j\neq h}
t^{2- 2 q_{\min\{c,h\}} - 2 q_{\min\{i,j\}}
+ \lambda A - \alpha_0 - 2 \epsilon} \nonumber \\
&& \hspace{40pt}  \preceq \, t^{2 q_1 + \lambda A - \alpha_0 - 2 \epsilon}
\, = \, O(t^{- \alpha_0 - 2 \epsilon + \sigma})
\, = \, O(t^{\delta})  \nonumber
\end{eqnarray}
for some strictly positive $\delta$.
The crucial inputs in getting these estimates are the inequalities
(\ref{inequalities1}).  We recall also that the quantity $\sigma$ (linked
to (\ref{inequalities1}) being satisfied) was introduced at the end
of subsection \ref{criticalvalue}.  The estimate of the last
two terms on the right hand side of (\ref{fuchchi}) is
\begin{eqnarray}
t^{2 - \beta} {1 \over g}
g_{ab} {\cal E}^a {\cal E}^b e^{-\lambda \phi}
& = & O(t^{- \beta - \epsilon + \sigma}) \; = \; O(t^{\delta}),
\nonumber \\
t^{2-\beta}  g^{ab} g^{ch} F_{ac} F_{bh} e^{\lambda \phi}
& = & O(t^{- \beta - 2 \epsilon + \sigma}) \; = \; O(t^{\delta}) \nonumber
\end{eqnarray}
The right hand side of (\ref{fuchxi}) is
$O(t^{\alpha_0- \beta - \delta - 5\epsilon + \sigma})=O(t^{\delta})$.
The right hand side of (\ref{fuchvarphi}) is
$O(t^{\alpha_0- \beta - \delta - 4\epsilon + \sigma})=O(t^{\delta})$.
The other terms which occur in $f$ were estimated in \cite{AR}, resulting
in $f = O(t^\delta)$.

To show that we indeed have a Fuchsian equation, we need to check not only
 that $f = O(t^\delta)$, but also that $\partial_u f= O(t^\delta)$
and $\partial_{u_x} f = O(t^\delta)$, along with other
regularity conditions \cite{AR,KR}.  In \cite{AR} it is shown
that $f$ is regular with equation (31) in that paper and the
remarks following equation (31).  In our case there is a
factor involving the determinant of the metric in various of
the terms in $f$ which are not present in the case considered
in \cite{AR}.  The discussion surrounding equation (31) in
\cite{AR} applies to our case as well, even for terms in $f$
containing $g^{\pm 1/2}$.  The Kasner-like contribution
is the leading term, and this function of $t$ and $x$ can
be factored out.  What is left is of the form
$w(t,x,u,u_x) (1 + h(t,x,u))^{\pm 1/2}$, which is analytic in $h$
at $h = 0$.  The conditions listed following equation (31) hold.
Thus we conclude that (\ref{fuchs0}) as written out in 
(\ref{fuchgamma}) -- (\ref{fuchvarphi}) is a Fuchsian equation.

\subsubsection{Symmetry of metric}
\label{symmetryofg}

It remains to show that $g_{ab}$ is symmetric, so that 
equation~(\ref{fuchs0}) as written out in
(\ref{fuchgamma}) -- (\ref{fuchvarphi}) is equivalent to the Einstein-matter 
evolution equations. The structure of the argument is the same in any
dimension and so it will be written down for general $d$\footnote{The 
argument for the symmetry of the metric in
\cite{AR} is not valid as written since some terms were omitted in the 
evolution equation for the antisymmetric part of the extrinsic
curvature. The correct equation is
\begin{displaymath}
\partial_t(k_{ab}-k_{ba})=(\tr k)(k_{ab}-k_{ba})
-2(k_{ac}k^c{}_b-k_{bc}k^c{}_a).
\end{displaymath}
In the following a proof of the symmetry of the tensors $g_{ab}$ and 
$k_{ab}$ is supplied with the help of a different method.}.
The number of distinct eigenvalues of ${K^a}_b$ is maximal
almost everywhere.  Thus it is enough to show that
$g_{[ab]}$ and $k_{[ab]}$ vanish in the case that
the Kasner-like metric is diagonal, since then by analytic
continuation they vanish on the entire domain.  We therefore
consider the case that the Kasner-like metric is
diagonal.

The redefinitions (\ref{defgamma}), (\ref{defkappa})
from the variables $g_{ab}$, $k_{ab}$ to the variables 
$\gamma^a{}_b$, ${\kappa^a}_b$
were viewed in the previous subsections as a change between
variables with no particular symmetry properties in
their indices (18 on each side).  One can, however enforce $g_{[ab]} = 0$
by assuming that $\gamma^a{}_b$ is symmetric and vice-versa.
Indeed, under our diagonality assumption for the
Kasner-like metric, $\tilde g_{ab}(t)  = \delta_{ab}
+ t^{\tilde \alpha^a_{\; \; b}}  \gamma^a_{\; \; b}$ where
${\tilde \alpha^a_{\; \; b}} = |q_a - q_b| + \alpha_0$ is
{\it symmetric} in $(a,b)$.  Accordingly, imposing the symmetry
$\gamma^a{}_b = \gamma^b{}_a$ algebraically ensures the symmetry of $g_{ab}$.
Similarly, one can enforce $k_{[ab]}$ to vanish by imposing
consistent constraints on $\kappa^a{}_b$:
inserting
(\ref{defgamma}) into (\ref{defkappa}) (with the velocity-dominated
solution diagonal) and writing out the 
constraint $k_{ab} - k_{ba} = 0$ gives the following condition 
on  $\kappa^a{}_b$
\begin{equation}
\label{symkappa}
{\kappa^a}_b - {\kappa^b}_a - {\gamma^a}_b p_b
+ {\gamma^b}_a p_a + t^{\alpha_{(ab) c}}
({\gamma^a}_c {\kappa^c}_b - {\gamma^b}_c {\kappa^c}_a) = 0,
\end{equation}
with $\alpha_{(ab) c} = 2 p_{\max \{a,c\}} + 2 p_{\max \{b,c\}}
- 2 p_{\max \{a,b\}} -2 p_c + \alpha_0$.  
These conditions show that there are only six
independent components among the $\kappa^a{}_b$, which can
be taken to be those with $a \leq b$.
This is because, the relation (\ref{symkappa}) can be solved uniquely for
the components $\kappa^a{}_b$ with $a>b$, given the other ones,
at least for $t$ small.
That this is true can be seen as follows. Rearrange the equations
(\ref{symkappa}) so that the terms containing $\kappa^a{}_b$ with 
$a>b$ are on the left hand side and all other terms are on the 
right hand side. The result is an inhomogeneous linear system of
the form $A(t,x)v(t,x)=w(t,x)$ where $A(t,x)$ and $w(t,x)$ are known
quantities and $v$ denotes the components $\kappa^a{}_b$ with $a>b$
which we want to determine. Furthermore $A(t,x)=I+o(1)$, where $I$
denotes the identity matrix. It follows that $A(t)$ 
is invertible for $t$ small, which is what we wanted to show.
The solution $\kappa^a{}_b$ ($a>b$) remains moreover 
bounded
when ${\gamma^b}_a$ and $\kappa^a{}_b$ are in
a compact set.
We shall assume from now on that $\gamma^a{}_b$ is symmetric
and $\kappa^a{}_b$ constrained by (\ref{symkappa}), so that 
symmetry of the metric is automatic.
The redefinitions (\ref{defgamma}), (\ref{defkappa})
from $g_{ab}$, $k_{ab}$ to $\gamma^a{}_b$, ${\kappa^a}_b$
can now be viewed as an invertible change of variables,
from 12 independent variables to 12 independent variables.
We can also clearly
assume $\lambda^a{}_{bc}$ in (\ref{deflambda}) to be
symmetric in $a$, $b$.

With these conventions, there are less components in $u$ than in the previous
subsections.  The independent components can be taken to be
$\gamma^a{}_b$, ${\kappa^a}_b$ and $\lambda^a{}_{bc}$
with $a \leq b$, together with the matter variables.
An independent system of evolution equations 
is given by (\ref{fuchgamma}) -- (\ref{fuchkappa}) 
with $a\le b$ for the gravitational variables, and the
same evolution equations as before for the matter
variables.  These evolution equations are equivalent to all
the original evolution equations, since
the equations (\ref{fuchgamma}) -- (\ref{fuchkappa})
with $a>b$ are then automatically fulfilled, as can be shown using the
fact that the Einstein tensor and the stress-energy tensor
are symmetric for symmetric metrics. To see
this it must be shown that given a symmetric tensor $S_{ab}$,
the vanishing of ${S^a}_b = g^{ac} S_{cb}$ for $a \leq b$
implies that $S_{ab} = 0$.  Consider the linear map which
takes a symmetric tensor $S_{ab}$, raises an index, and keeps
the components of the result with $a \leq b$.  This is a mapping
between vector spaces of dimension $d (d+1)/2$ and can be shown
to be an isomorphism by elementary linear algebra.  This proves
the desired result.

Now, this reduced evolution system is also Fuchsian.  This
follows from the same reasoning as above, which still holds because
all components of $u$, including the non-independent ones, can still be
assumed to be bounded.   Therefore,  there is a unique $u$
that goes to zero, which must be equal to the one considered
in the previous subsections.
The metric considered previously is thus indeed symmetric.

\subsubsection{Unique solution on a neighbourhood of
the singularity}

Given an analytic solution to the velocity-dominated evolution
equations on $(0,\infty) \times \Sigma$,  such that
inequalities~(\ref{inequalities1}) are satisfied, we now
have a solution
$u$ to a Fuchsian equation (and a corresponding solution to the
Einstein-matter evolution equations) in the intersection of
a neighbourhood of the singularity with $(0,\infty) \times U_0$
where $U_0$ is a neighbourhood of an arbitrary point on $\Sigma$.
These local solutions can
be patched together to get a solution to the Einstein-matter
evolution equations everywhere in space near the singularity. It
may seem like there could be a problem patching together the
solutions obtained on distinct neighborhoods with non-empty
intersection because
the Fuchsian equation is not the same for different allowed
choices of $\epsilon$ and adapted local frame.
The construction is possible because different allowed
choices of $\epsilon$ and local frame result in a well-defined
relationship between the different solutions $u$ which are
obtained, such that the corresponding Einstein-matter
variables agree on the intersection (up to change of basis).
It therefore follows that
given an analytic solution to the velocity-dominated evolution
equations on $(0,\infty) \times \Sigma$,  such that
inequalities~(\ref{inequalities1}) are satisfied,
our construction uniquely determines a solution to the
Einstein-matter evolution equations
everywhere in space, near the singularity.

\subsection{Einstein-matter constraints}\label{stepfour}

\subsubsection{Matter constraints}

The time derivative of the matter constraint quantities
(the left hand side of equations~(\ref{elconstraint})
and (\ref{magconstraint})) vanishes.
If the velocity-dominated matter constraints
are satisfied, the matter constraint quantities are $o(1)$.
A quantity which is constant in time and $o(1)$ must vanish.
Therefore the matter constraints are satisfied.

\subsubsection{Diagonal Kasner metrics}

It remains to show that the
Hamiltonian and momentum constraints are satisfied,
that $C$ and $C_a$, defined in (\ref{hamiltonian1})
and (\ref{momentum1}), vanish.  Since we now have a
metric, $g_{\mu \nu}$, it follows that
$\nabla_\mu G^{\mu \nu} = 0$.
Since the matter evolution and constraint
equations are satisfied, it follows that
$\nabla_\mu T^{\mu \nu} = 0$.
{}From the vanishing of the right hand side of (\ref{eveq1})
and the vanishing of the covariant divergence of
both the Einstein tensor and the stress-energy tensor,
it follows that
\begin{eqnarray}
\label{evC}
\partial_t C & = & 2 (\tr k) C - 2 \nabla^a C_a \\
\label{evCa}
\partial_t C_a & = & (\tr k) \, C_a - {1 \over 2} \nabla_a C.
\end{eqnarray}  
Now define $\bar C = t^{2 - \eta_1} C$ and
$\bar C_a = t^{1-\eta_2} C_a $, with
$0 < \eta_2 < \eta_1 < \beta$.
\begin{eqnarray}
\label{fuchsianC}
t \, \partial_t \bar C + \eta_1 \bar C & = & 2 (1 + t \, \tr \, k)
\bar C - 2 t^{2 - \eta_1 + \eta_2} \nabla^a \bar C_a \\
\label{fuchsianCa}
t \, \partial_t \bar C_a + \eta_2 \bar C_a & = &
(1 + t \, \tr \, k) \bar C_a -
{1 \over 2} t^{\eta_1 - \eta_2} \nabla_a \bar C
\end{eqnarray}
On the right hand side of (\ref{fuchsianC}) and
(\ref{fuchsianCa}) $\bar C$ and
$\bar C_a$ are to be considered as components
of $u = (\bar C, \bar C_a)$.
If it is shown that (\ref{fuchsianC}) and (\ref{fuchsianCa})
is a Fuchsian system, then there is a unique solution
$u$ such that $u = o(1)$.  It is clear
that $u = 0$ is a solution to (\ref{fuchsianC}) and
(\ref{fuchsianCa}).  If it is shown that
$\bar C = o(1)$ and $\bar C_a=o(1)$, (i.e. that
$C = o(t^{-2 + \eta_1})$ and $C_a =  o(t^{-1 + \eta_2})$),
then they must
be this unique solution.  Furthermore, it is sufficient
to consider the case that the Kasner-like metric
is diagonal, since the number of distinct eigenvalues
of ${K^a}_b$ is maximal on an open set of $\Sigma$.  If
the constraints vanish on an open set of their domain,
then by analytic continuation they vanish everywhere
on their domain.

Therefore we consider the case that the Kasner-like
metric is diagonal and show first that
\begin{eqnarray}
\label{estconstraints3}
1 + t \, \tr \, k & = & O(t^\delta) \\
\label{estconstraints4}
\nabla^a \bar C_a & = & O(t^{-2 + \delta + \eta_1 - \eta_2})
\end{eqnarray}
(when $\bar C_a$ is bounded) so that the system (\ref{fuchsianC}),
(\ref{fuchsianCa}) is Fuchsian
(the complete regularity of $f(t,x,u, u_x)$ defined by (\ref{fuchsianC})
and (\ref{fuchsianCa}) can be easily verified); and second, that
\begin{eqnarray}
\label{estconstraints1}
C & = & o(t^{-2 + \eta_1}) \\
\label{estconstraints2}
C_a & = & o(t^{-1 + \eta_2}) .
\end{eqnarray}

Some facts which will be used to show this follow.
Consider indices $a \in \{1,2,3\}$.
The following inequalities hold for
some positive integer $n$ and
for real numbers, $q_a$, ordered such that
if $a < b$, then $q_a \leq q_b$.
(In later sections we define ordered auxiliary exponents,
$q_a$, for $a \in \{1,...,d\}$, for arbitrary fixed $d \geq 2$.
Then (\ref{makeest1}) -- (\ref{makeest3}) hold more generally
for indices in $\{1,...,d\}$.)
\begin{equation}
\label{makeest1}
q_{a_0} + \sum_{i=1}^n|q_{a_{i-1}} - q_{a_i}| - q_{a_n} \geq 0
\end{equation}
\begin{equation}
\label{makeest2}
q_{a_0} + \sum_{i=1}^n|q_{a_{i-1}} - q_{a_i}| + q_{a_n} \geq
2 q_{\max\{a_k,a_j\}}
\end{equation}
\begin{equation}
\label{makeest3}
-q_{a_0} + \sum_{i=1}^n|q_{a_{i-1}} - q_{a_i}| - q_{a_n} \geq
-2q_{\min\{a_k,a_j\}}
\end{equation}
The latter two inequalities hold for any $k$, $j$ in $\{0,\ldots, n \}$.

In the case that the Kasner-like metric
is diagonal, $q_a = p_a$.
The metric in the frame $\{\tilde e_a \}$ is
$^0 \tilde g_{ab} = \delta_{ab}$,
\begin{eqnarray}
\tilde g_{ab} & = & \delta_{ab} +
t^{\tilde \alpha^a_{\; \; b}} {\gamma^a}_b
\; \preceq \; t^{|p_a - p_b|}, \nonumber \\
\tilde g^{ab} & = & \delta^{ab} +
t^{\tilde \alpha^a_{\; \; b}} \bar \gamma^a_{\; \; b}
\; \preceq \; t^{|p_a - p_b|}.  \nonumber
\end{eqnarray}
The extrinsic curvature satisfies
$ t \; ^0{k^a}_b = -{\delta^a}_b \, p_b $,
\begin{eqnarray}
t \, {k^a}_b & = & -{\delta^a}_b \, p_b +
t^{{\alpha^a}_b} {\kappa^a}_b, \nonumber \\
t \, (\tilde k^a_{\; \; b} - \, ^0\tilde k^a_{\; \; b})
& = & t^{\tilde \alpha^a_{\; \; b}} {\kappa^a}_b, \nonumber
\end{eqnarray}
and $t \, \tr \; ^0 k =  -1 $,
\begin{eqnarray}
\label{estttrk}
t \, \tr \, k & = & -1 + t^{\alpha_0} \, \tr \kappa, \\
\label{esttrksq}
t^2 \{ (\tr \, k)^2 - (\tr \, ^0 k)^2 \} & = & O(t^{\alpha_0}).
\end{eqnarray}
The following estimates will also be useful.
\begin{eqnarray}
- {k^a}_b \,{k^b}_a + \, ^0 {k^a}_b \, ^0 {k^b}_a & = & -
2 t^{-2 + \alpha_0} {\kappa^a}_a \, p_a - t^{-2 + {\alpha^a}_b
+ {\alpha^b}_a} {\kappa^a}_b \,{\kappa^b}_a \nonumber \\
\label{estksq}
& = & O(t^{-2 + \alpha_0}),
\end{eqnarray}
and
\begin{equation}
\label{estebtrk}
e_a ( \tr \, k - \tr \, ^0k) =
e_a(t^{-1 + \alpha_0} \tr \, \kappa) = O (t^{-1 + \alpha_0}).
\end{equation}
The structure functions of the frame $\{ \tilde e_a \}$ are
\begin{eqnarray}
\tilde f^c_{ab} & = & t^{p_c - p_a - p_b} f^c_{ab}
- \ln t \, e_a(p_b) \, t^{-p_a} {\delta^c}_b
+ \ln t \, e_b(p_a) \, t^{-p_b} {\delta^c}_a \nonumber \\
& \preceq & t^{p_c - p_a - p_b -\delta}. \nonumber
\end{eqnarray}
It is convenient to have an estimate of $\tilde \Gamma^c_{ab}$,
the connection coefficients (\ref{christoffel})
in the frame $\{ \tilde e_a \}$,
term by term.
\begin{eqnarray}
\label{christoffela}
&&\mbox{Term A:} \hspace{40pt}
\tilde g^{ch} \tilde e_a (\tilde g_{bh})
\preceq \sum_h t^{|p_c - p_h | -p_a + |p_b - p_h| - \delta}, \\
\label{christoffelb}
&&\mbox{Term B:} \hspace{40pt}
\tilde g^{ch} \tilde e_b (\tilde g_{ha})
\preceq \sum_h t^{|p_c - p_h | -p_b + |p_h - p_a| - \delta}, \\
\label{christoffelc}
&&\mbox{Term C:} \hspace{40pt}
\tilde g^{ch} \tilde e_h (\tilde g_{ab})
\preceq \sum_h t^{|p_c - p_h | -p_h + |p_a - p_b| - \delta}, \\
\label{christoffeld}
&&\mbox{Term D:} \hspace{40pt}
\tilde g^{ci} \tilde g_{ha} \tilde f^h_{bi}
\preceq \sum_{h,i \neq b} t^{|p_c - p_i | + |p_h - p_a| + p_h - p_b - p_i - \delta}, \\
\label{christoffele}
&&\mbox{Term E:} \hspace{40pt}
\tilde g^{ci} \tilde g_{bh} \tilde f^h_{ai}
\preceq \sum_{h,i \neq a} t^{|p_c - p_i | + |p_b - p_h| + p_h - p_a - p_i - \delta}, \\
\label{christoffelf}
&&\mbox{Term F:} \hspace{40pt}
\tilde f^c_{ab} \preceq t^{ p_c - p_a - p_b - \delta}.
\end{eqnarray}
The difference between the connection coefficients for the metric
$\tilde g_{ab}$  and those for the Kasner-like metric is
$\Delta \tilde \Gamma^c_{ab} =
\tilde \Gamma^c_{ab} - \, ^0\tilde \Gamma^c_{ab}$.
It is useful to have the estimates
\begin{displaymath}
\tilde \Gamma^a_{ac} = {1 \over 2} \tilde g^{ab} 
\tilde e_c(\tilde g_{ab}) + \tilde f^a_{ac} \preceq t^{-p_c - \delta},
\end{displaymath}
and
\begin{equation}
\label{deltatracetildegamma}
\Delta \tilde \Gamma^a_{ac} =
{1 \over 2} \tilde g^{ab} \tilde e_c (\tilde g_{ab})
\preceq t^{-p_c + \alpha_0 - \delta}.
\end{equation}

\subsubsection{Momentum and Hamiltonian constraints}

First, we show (\ref{estconstraints3}) and
(\ref{estconstraints4}). 
{}From equation (\ref{estttrk}),
$1 + t \, \tr \, k = O(t^{\alpha_0})$. 
Similarly, we can estimate $\nabla^a \bar C_a$,
\begin{eqnarray}
\nabla^a \bar C_a & = &
\tilde g^{ab} \, \tilde \nabla_a \tilde {\bar C}_b \nonumber \\
& = & \tilde g^{ab}
\{ t^{-p_a} \, e_a (\bar C_b \, t^{-p_b})
-\tilde \Gamma^c_{ab} \, \bar C_c \, t^{-p_c} \}  \nonumber
\end{eqnarray}
The first term is
\begin{equation}
\label{divmom1}
\tilde g^{ab} t^{-p_a} \, e_a ( \bar C_b \, t^{-p_b} )
\preceq t^{|p_a - p_b| - p_a - p_b - \delta}
\preceq t^{-2 p_{\min\{a,b\}} - \delta}.
\end{equation}
{}From (\ref{christoffela}) -- (\ref{christoffele})
the second term is
\begin{equation}
\label{divmom2}
\tilde g^{ab} \tilde \Gamma^c_{ab} \, \bar C_c \, t^{-p_c}
\preceq t^{-2 p_3 - \delta}.
\end{equation}
{}From (\ref{divmom1}) and (\ref{divmom2}), the desired estimate,
$\nabla^a \bar C_a = O(t^{-2 + \eta_1 - \eta_2})$ is obtained.     
Thus, the system (\ref{fuchsianC}),
(\ref{fuchsianCa}) is Fuchsian.

Next we turn to (\ref{estconstraints1}) and
(\ref{estconstraints2}).
A term that appears in the momentum constraint
is $\nabla_a {k^a}_b$.  The estimate is needed in
the adapted frame, and the covariant derivative
is calculated in the frame $\{\tilde e_a \}$.
This adds a factor of $t^{p_b}$,
\begin{displaymath}
\nabla_a {k^a}_b =
\{ \tilde e_a (\tilde k^a_{\; \; b}) +
\tilde \Gamma^a_{a c} \tilde k^c_{\; \; b}
- \tilde \Gamma^c_{a b} \tilde k^a_{\; \; c} \}  \, t^{p_b}
\end{displaymath}
Furthermore, the quantity whose estimate will be
required is the difference between this term and
the corresponding term in the velocity-dominated
constraint,
\begin{eqnarray}
\label{estderk1a}
\nabla_a {k^a}_b - \, ^0 \nabla_a \, {^0 k^a}_b & = &
\{ \tilde e_a (\tilde k^a_{\; \; b}
- \, ^0\tilde k^a_{\; \; b})
+ \Delta \tilde \Gamma^a_{ac}
\, ^0 \tilde k^c_{\; \; b}
+ \tilde \Gamma^a_{a c} (\tilde k^c_{\; \; b}
- \, ^0\tilde k^c_{\; \; b}) \} t^{p_b} \\
\label{estderk1b} & &
- \tilde \Gamma^c_{a b} \tilde k^a_{\; \; c} t^{p_b}
+ \, ^0 \tilde \Gamma^c_{ab}
\, ^0 \tilde k^a_{\; \; c} \, t^{p_b}
\end{eqnarray}

The right hand side of (\ref{estderk1a}) is
$O(t^{-1 + \alpha_0 - \delta})$.
The terms in line (\ref{estderk1b}) originating from Term E
of the connection coefficients (see (\ref{christoffele}))
are cancelled in the sum, due to the antisymmetry of
$\tilde f^h_{ai}$ and the symmetry of $\tilde k^{ai}$ and
$^0 \tilde k^{ai}$.  For estimating the rest of the terms in
line (\ref{estderk1b}), it is convenient to rewrite this line as,
\begin{equation}
\label{estderk1c}
- \tilde \Gamma^c_{a b} \tilde k^a_{\; \; c} t^{p_b}
+ \, ^0 \tilde \Gamma^c_{ab}
\, ^0 \tilde k^a_{\; \; c} \, t^{p_b}
= - \Delta \tilde \Gamma^c_{ab}
\, ^0 \tilde k^a_{\; \; c} \, t^{p_b}
- \tilde \Gamma^c_{a b} (\tilde k^a_{\; \; c}
- \, ^0\tilde k^a_{\; \; c})  \, t^{p_b},
\end{equation}
with
\begin{eqnarray}
\label{deltagamma1}
\Delta \tilde \Gamma^c_{ab} & = & {1 \over 2} \{
\tilde e_a(t^{\tilde \alpha^b_{\; \; c}} {\gamma^b}_c)
+ \tilde e_b(t^{\tilde \alpha^c_{\; \; a}} {\gamma^c}_a)
- \tilde e_c(t^{\tilde \alpha^a_{\; \; b}} {\gamma^a}_b) \\
\label{deltagamma2} & & \; \;
+ \sum_h t^{\tilde \alpha^c_{\; \; h}} \bar \gamma^c_{\; \; h} [
\tilde e_a(t^{\tilde \alpha^b_{\; \; h}} {\gamma^b}_h )
+ \tilde e_b(t^{\tilde \alpha^h_{\; \; a}} {\gamma^h}_a )
- \tilde e_h(t^{\tilde \alpha^a_{\; \; b}} {\gamma^a}_b ) ] \\
\label{deltagamma3} & & \; \;
- \sum_i t^{\tilde \alpha^i_{\; \; a}}{\gamma^i}_a \tilde f^i_{bc}
- \sum_h t^{\tilde \alpha^c_{\; \; h}} \bar \gamma^c_{\; \; h}
\tilde f^a_{bh}
- \sum_{hi} t^{\tilde \alpha^c_{\; \; h}} \bar \gamma^c_{\; \; h}
t^{\tilde \alpha^i_{\; \; a}}{\gamma^i}_a \tilde f^i_{bh} \\
\label{deltagamma4} & & \; \;
- t^{\tilde \alpha^b_{\; \; i}}{\gamma^b}_i \tilde f^i_{ac}
- \sum_h t^{\tilde \alpha^c_{\; \; h}} \bar \gamma^c_{\; \; h}
\tilde f^b_{ah}
- \sum_{hi} t^{\tilde \alpha^c_{\; \; h}} \bar \gamma^c_{\; \; h}
t^{\tilde \alpha^b_{\; \; i}}{\gamma^b}_i \tilde f^i_{ah} \}.
\end{eqnarray}
The terms in line (\ref{deltagamma4}) need not be considered
since they originate from Term E of the connection coefficients
and as stated above the contribution from this term is cancelled
by terms in
$\Lambda = \tilde \Gamma^c_{a b} \, ^0\tilde k^a_{\; \; c} \, t^{p_b}$.
So considering only lines (\ref{deltagamma1}) -- (\ref{deltagamma3}),
the first term on the right hand side of (\ref{estderk1c}) is
\begin{equation}
\label{estimate4}
\Delta \tilde \Gamma^a_{ab} \, p_a \, t^{-1 + p_b}
= O(t^{-1 + \alpha_0 - \delta}) + \, \mbox{terms which are
cancelled by $\Lambda$.}
\end{equation}
Since the terms in the sum come with different weights, $p_a$,
(\ref{deltatracetildegamma}) cannot be used
in~(\ref{estimate4}).  But the estimate is straightforward.
For example, the term in (\ref{estimate4}) originating from
the 3rd term in line (\ref{deltagamma3}) $\preceq \sum_{a,h,i}
t^{-1 + |p_a - p_h| + |p_i - p_a| + p_i - p_h + 2 \alpha_0
-\delta} = O(t^{-1 + \alpha_0 - \delta})$.  Finally consider
the rest of the right hand side of (\ref{estderk1c}),
\begin{equation}
\label{dangerous}
- {1 \over 2} \left [ \tilde g^{ch} \{
\tilde e_a (\tilde g_{bh}) + \tilde e_b (\tilde g_{ha})
- \tilde e_h (\tilde g_{ab})
- \tilde g_{ia} \tilde f^i_{bh}
- \tilde g_{bi} \tilde f^i_{ah} \}
+  \tilde f^c_{ab} \right ] \, t^{\tilde \alpha^a_{\; \; c}}
{\kappa^a}_c \, t^{-1 + p_b}.
\end{equation}
For all terms except the 5th term in (\ref{dangerous}),
the estimate, $O(t^{-1 + \alpha_0 - \delta})$ can be
obtained from (\ref{christoffela}) -- (\ref{christoffelf}).
The fifth term originates from Term E of the connection
coefficients, and was already considered above.  Therefore
\begin{equation}
\label{estderk2}
\nabla_a {k^a}_b - \, ^0 \nabla_a \, {^0 k^a}_b = 
O(t^{-1 + \alpha_0 - \delta}).
\end{equation}

Next the matter terms are estimated.  For the Hamiltonian
constraint, an estimate of $\rho - \, ^0 \rho$ is needed.
\begin{eqnarray}
(\partial_t \phi)^2 - (\partial_t \, ^0 \phi)^2
& = & \{ 2 \, \partial_t \, ^0 \phi +
t^{-1 + \beta} (\beta \psi + t \partial_t \psi)\} \,
t^{-1 + \beta} (\beta \psi + t \partial_t \psi) \nonumber \\
& = & \{ 2 A + t^\beta (\beta \psi + t \partial_t \psi)\} \,
t^{-2 + \beta} (\beta \psi + t \partial_t \psi) \nonumber \\
& = & o(t^{-2 + \eta_1}), \nonumber \\
g^{ab} e_a(\phi) e_b(\phi) & \preceq & t^{-2p_3 - \delta - \epsilon}
\; = \; o(t^{-2 + \eta_1}), \nonumber \\
{ 1 \over g } g_{a b} {\cal E}^a {\cal E}^b e^{-\lambda_j \phi}
& = & O(t^{-2 - \epsilon + \sigma}) = o(t^{-2 + \eta_1}), \nonumber \\
g^{a b} g^{c h} F_{a c} F_{b h} e^{\lambda_j \phi}
& = & O(t^{-2 - 2 \epsilon + \sigma})
= o(t^{-2 + \eta_1}). \nonumber
\end{eqnarray}
Therefore,
\begin{equation}
\label{estrho}
\rho - \, ^0 \rho = o(t^{-2 + \eta_1}).
\end{equation}
The difference between the matter terms in the momentum constraint
and in $^0C_a$ is
\begin{eqnarray}
- \partial_t \phi \, e_a(\phi) + \partial_t \,
^0 \phi \,  e_a( \, ^0 \phi)
\; = \; - \partial_t \, ^0 \phi \, e_a(t^\beta \psi)
- \partial_t (t^\beta \psi) e_a(\phi) & = &
O(t^{-1 + \beta - \delta}), \nonumber \\
\label{estjb1f}
({1 \over \sqrt{g} } - {1 \over \sqrt{\, ^0 g} })
\, ^0{\cal E}^b \, ^0 F_{a b} +
{1 \over \sqrt{g} } ( {\cal E}^b \, F_{a b} -
\, ^0{\cal E}^b \, ^0 F_{a b} ) & = & o(t^{-1 + \eta_2}).
\end{eqnarray}
Estimates related to the determinant which are relevant to
(\ref{estjb1f}) can be found immediately preceding
equation~(\ref{estddetg1}).  {}From the estimates just obtained,
\begin{equation}
\label{estjb}
j_a - \, ^0 j_a = o(t^{-1 + \eta_2}).
\end{equation}

{}From $R = O(t^{-2 + \alpha_0})$ (shown in \cite{AR}) and from
$^0 C = 0$, (\ref{estksq}), (\ref{esttrksq}),
(\ref{estrho}) and the relative magnitude of the various
exponents, it follows that $C = o(t^{-2 + \eta_1})$.
{}From $^0C_a = 0$, (\ref{estderk2}),
(\ref{estebtrk}), (\ref{estjb}) and the relative
magnitude of the various exponents, it follows
that $C_a = o(t^{-1 + \eta_2})$.

Since (\ref{estconstraints1}) -- (\ref{estconstraints2}) are
satisfied, the Hamiltonian and momentum constraints are satisfied.

\subsection{Counting the number of arbitrary functions}
The number of degrees of freedom of the Einstein-Maxwell-dilaton system
in 4 spacetime dimensions is 5: 2 for the gravitational field,
2 for the electromagnetic field and 1 for the dilaton.  Hence,
a general solution to the equations of motion should contain
10 freely adjustable, physically relevant, functions of
space (each degree of freedom needs two initial data, $q$
and $\dot{q}$).  This is exactly the number that appears
in the above Kasner-like solutions.
\begin{itemize}
\item The metric carries four, physically relevant, distinct
functions of space.  This is the standard calculation
\cite{BKL}.
\item The scalar field carries two functions of space, $A$
and $B$.
\item The electromagnetic field carries six functions of
space, $\! ^0 {\cal E}^a$ and $\! ^0 F_{ab}$.  These are
physically relevant because they are gauge invariant, but
they are subject to two constraints, leaving four
independent functions.
\end{itemize}
A different way to arrive at the same conclusions is to
observe that the respective number of fields, dynamical equations
and (first class) constraints are the same for the
velocity-dominated system and the full system.
Hence,  a general solution of the velocity-dominated
system (in the sense of function counting) will contain
the same number of physically distinct, arbitrary functions
as a general solution of the full system.  This general argument
applies to all systems considered below and hence will
not be repeated.

In \cite{AR} a different way of assessing the generality of the solutions
constructed was used. This involved exhibiting a correspondence between
solutions of the velocity-dominated constraints and solutions of the full
constraints using the conformal method. That method starts with certain 
free data and shows the existence of a unique solution of the constraints
corresponding to each set of free data. It is a standard method for 
exploring the solution space of the full Einstein constraints \cite{choquet} 
and in \cite{AR} it was shown how to modify it to apply to the 
velocity-dominated constraints. While it is likely that the conformal 
method can be applied in some way to all the matter models considered in 
this paper, the details will only be worked out in two cases which suffice
to illustrate the main aspects of the procedure. These are the 
Einstein-Maxwell-dilaton system with $D=4$ (this section) and the Einstein 
vacuum equations with arbitrary $D\ge 4$ (next section).
Even in those cases no attempt will be made to give an
exhaustive treatment of all issues arising. It will, however, be shown
that the strategies presented for solving the velocity-dominated 
constraints are successful in some important situations.

The procedure presented in the following is slightly different from that
used in \cite{AR}. Even for the case of the Einstein-scalar field system 
with $D=4$ it gives results which are in principle stronger than
those in \cite{AR} since they are not confined to solutions which are
close to isotropic ones. In the presence of exponential dilaton couplings
a change of
method seems unavoidable. One part of the conformal method concerns the 
construction of symmetric second rank tensors which are traceless and have 
prescribed divergence from the truly free data. In this step there is no 
difference between the full constraints and the velocity-dominated ones.
An account of the methods applied to the full
constraints in the case $D=4$ can be found in 
\cite{choquet}. (These arguments generalize
in a straightforward way to other $D$. It is merely necessary to
find the correct conformal rescalings. For $D\ge 4$ and vacuum these are
written down in the next section.) In view of this
we say, with a slight abuse 
of terminology, that the free data consists of a collection $\tilde g_{ab}$, 
$\tilde k_{ab}$, $H$, $\phi$, $\tilde\phi_t$, ${\cal E}^a$, 
$F_{ab}$ where $\tilde g_{ab}$ is a Riemannian metric, $\tilde k_{ab}$ is 
a symmetric tensor with vanishing trace and prescribed divergence with 
respect to $\tilde g_{ab}$, $H$ is a non-zero constant, $\phi$ and 
$\tilde\phi_t$ are scalar functions and ${\cal E}^a$ and $F_{ab}$ are 
objects of the same kind as elsewhere in this section. All these objects 
are defined on a three-dimensional manifold. Next we introduce a positive 
real-valued function $\omega$ which is used to construct solutions of the
constraints from the free data. Define $g_{ab}=\omega^4 \tilde g_{ab}$, 
$k_{ab}=\omega^{-2}\tilde k_{ab}+Hg_{ab}$, $\phi_t=\omega^{-6}\tilde\phi_t$.
The objects $g_{ab}$, $k_{ab}$, $\phi$, $\phi_t$, ${\cal E}^a$ and  
$F_{ab}$ satisfy the constraints provided the divergence of $\tilde k_{ab}$
is prescribed as $\omega^6 j_a$ and $\omega$ satisfies a nonlinear equation
which in the case of the full Einstein equations is known as the 
Lichnerowicz equation. In the case of the velocity-dominated constraints 
it is an algebraic equation. The Lichnerowicz equation is of the form
\begin{equation}\label{lichno3}
\Delta_{\tilde g}\omega-\frac 18 R_{\tilde g}\omega+\sum_{i=1}^3 
a_i \omega^{\alpha_i}-\frac 34 H^2\omega^5=0
\end{equation} 
Here $\alpha_1=1$, $\alpha_2=-3$ and $\alpha_3=-7$. The functions $a_i$
depend on the free data and their exact form is unimportant. All that is 
of interest are that each $a_i$ is non-negative and that at any point of 
space $a_1=0$ iff $\nabla_a\phi=0$ , $a_2=0$ iff the electromagnetic data
vanish and $a_3=0$ iff $\phi_t$ and $\tilde k_{ab}$ vanish. Next consider
the velocity-dominated constraints for $d=3$. The analogue of the elliptic
equation (\ref{lichno3}) is the algebraic equation
\begin{equation}\label{vdlichno3}
b \omega^{-7}-\frac 34 H^2\omega^5=0
\end{equation} 
Here $b$ is a non-negative function which vanishes at a point of space iff 
$\phi_t$ and $\tilde k_{ab}$ vanish. This can be solved trivially for 
$\omega>0$ provided $b$ does not vanish at any point since the mean 
curvature $H$ is non-zero. For each choice of free data satisfying this 
non-vanishing condition there is a unique solution $\omega$ of (\ref{vdlichno3}).

In order to compare the sets of solutions of the full and
velocity-dominated
constraints in these two cases it remains to investigate the solvability of
the elliptic equation (\ref{lichno3}) for $\omega$. A discussion of this type 
of problem in any dimension can be found in \cite{isenberg}. We would like 
to show that for suitable metrics on a compact manifold the equation for 
$\omega$ always has a unique solution,  {\it i.e.,} the situation
is exactly as in the case of the velocity-dominated 
equations. The problem can be simplified by the use of the Yamabe theorem, 
which says that any metric can be conformally transformed to a metric of 
constant scalar curvature $-1$, $0$ or $1$. In the following only the
cases of negative and vanishing scalar curvature of the metric supplied 
by the Yamabe theorem will be considered. A key role in the existence and
uniqueness theorems for equation (\ref{lichno3}) is played by the positive 
zeros of the algebraic expressions 
$x+8\sum_{i=1}^3 a_i x^{\alpha_i}-6 H^2 x^5$ 
and $8\sum_{i=1}^3 a_i x^{\alpha_i}-6 H^2 x^5$. Provided 
$\sum_{a=1}^3 a_i$ does not vanish anywhere it is possible to show that 
each of the algebraic expressions has a unique positive zero for each value 
of the parameters.

The significance of the information which has been obtained concerning the
zeros of certain algebraic expressions is that it guarantees the existence 
of a positive solution of the corresponding elliptic equations for any set
of free data satisfying the inequalities already stated using the method
of sub- and supersolutions ({\it cf.} \cite{isenberg}).
In the case of equation
(\ref{lichno3}) uniqueness also holds. For in that case the equation has
a form considered in \cite{om} for which uniqueness is demonstrated in that
paper. The advantage of the three dimensional case is that there the problem 
reduces to the analysis of the roots of a cubic equation, a relatively 
simple task compared to the analysis of the zeros of the more complicated 
algebraic expressions occurring in higher dimensions.

For the purpose of analysing Kasner-like (monotone) singularities it is not 
enough to know about producing just any solutions of the constraints.
What we have shown is that (i) if the
Kasner constraints are satisfied at time $t_0$, then they are propagated
at all times by the velocity-dominated evolution equations; and (ii) if
the Kasner constraints are satisfied, the exact constraints are also
satisfied.  It is also necessary to verify, however, that one can satisfy
simultaneously the Kasner constraints {\em and} the inequalities
necessary for applying the Fuchsian arguments, i.e., 
we must make sure that we can produce a sufficiently large class of
solutions which satisfy the inequalities necessary to make them
consistent with Kasner behaviour. Because of the indirect nature of the 
way of solving the momentum constraint (which has not been explained
here) it is not easy to control the generalized Kasner exponents of the
resulting spacetime. There is however, one practical possibility. Choose
a spatially homogeneous solution with Abelian isometry group (for $d=3$
this means Bianchi type I) which satisfies the necessary inequalities.
Take the free data from that solution and deform it slightly. Then the
generalized Kasner exponents of the final solution of the
velocity-dominated
equations will also only be changed slightly. If the homogeneous solution
is defined on the torus $T^3$ then it is known that any other metric of
constant scalar curvature has non-positive scalar curvature. Therefore we
are in the case for which existence and uniqueness is discussed above. 
We could also start with a negatively curved Friedmann model.

\section{Vacuum solutions with $D \geq 11$}\label{vacuum}
\setcounter{equation}{0}
\setcounter{theorem}{0}
\setcounter{lemma}{0}      
The second class of solutions we consider is governed by the action
(\ref{action2}), with $D \geq 11$.  The $d+1$ decomposition is as
in section \ref{stepzero}, with the matter terms vanishing.
The Kasner-like evolution equations are (\ref{dtg0})
and (\ref{dtk0}). The general analytic solution of these equations
is given by (\ref{0kab1}) and (\ref{0gab1}).  To obtain this form,
we again adapt a global time coordinate such that the singularity
is at $t=0$.  We label the eigenvalues of $K$, $p_1, \ldots , p_d$,
such that $p_a \leq p_b$ if $a < b$.  The eigenvalues
again satisfy $\sum_{i=1}^d p_i = 1$,
coming from $\tr \, K = 1$.  As in the $D=4$ case, in order
to preserve analyticity even near the points where some of the
eigenvalues coincide, while retaining control of the
solution in terms of the eigenvalues, we introduce a
special construction involving auxiliary exponents and an adapted frame.

In higher dimensions, there are more possibilities to take care of,
but the idea is the same as in the $D=4$ case.
Consider an arbitrary point $x_0 \in \Sigma$.
Let $n$ be the number of distinct eigenvalues of $K$ at $x_0$.
Let $m_i$ be the multiplicity of $p_{A_i}$,
$i \in \{1, \ldots, n \}$, with $p_{A_i}$
such that $p_b$ is strictly less than $p_{A_i}$
if $b < A_i$.  Thus
$p_{A_i}, \ldots, p_{A_i + m_i -1}$ are
equal at $x_0$.  For each integer
$a \in \{ A_i, \ldots, A_i + m_i -1 \}$, define
\begin{displaymath}
q_a = {1 \over m_i} \sum_{j= A_i}^{A_i + m_i -1} p_j
\end{displaymath}
on a neighbourhood of $x_0$, $U_0$.
Note that if $m_i = 1$, then $q_{A_i} = p_{A_i}$.
Shrinking $U_0$ if necessary, choose $\epsilon > 0$ such
that for $x \in U_0$,
for $a \in \{ A_i, \ldots, A_i + m_i -1 \}$ and
for $b \in \{ A_j, \ldots, A_j + m_j -1 \}$,
if $i = j$, then $|p_a - p_b| < \epsilon/2$,
while if $i \neq j$, $|p_a - p_b| > \epsilon/2$.

The adapted frame $\{ e_a \}$ is again required to be analytic and
such that the related frame $\{ \tilde e_a \}$ is orthonormal with
respect to the Kasner-like metric at
some time $t_0 > 0$, with $\tilde e_a = t^{-q_a} e_a$. In
addition, it is required that $e_{A_i},\ldots, e_{A_i + m_i -1}$
span the eigenspace of $K$
corresponding to the eigenvalues $p_{A_i},\ldots, p_{A_i + m_i -1}$.
Note that if $m_i = 1$ then $e_{A_i}$ is an eigenvector of
$K$ corresponding to the eigenvalue $q_{A_i}$.
The auxiliary exponents, $q_a$, are analytic,
satisfy the Kasner relation ($\sum q_a = 1$), are ordered
($q_a \leq q_b$ for $a < b$), and satisfy $q_1 \geq p_1$,
$q_d \leq p_d$ and $\max_a |q_a - p_a| < \epsilon/2$.
If $q_a \neq q_b$, then $^0g_{ab}$, $^0g^{ab}$,
$^0 \tilde g_{ab}$ and $^0 \tilde g^{ab}$ all vanish,
and the same is true with $g$ replaced by $k$.

The velocity-dominated constraints corresponding to the Hamiltonian
and momentum constraints are $^0C=0$ and $^0C_a = 0$, with
$^0C$ and $^0C_a$ as in equations~(\ref{hamiltonian0}) and
(\ref{momentum0}), with the matter terms vanishing.
For the solution (\ref{0kab1}) -- (\ref{0gab1})
the velocity-dominated Hamiltonian constraint
equation is equivalent to $\sum p_a^2 = 1$.
Equations~(\ref{dt0C}) -- (\ref{dt0Ca}) are satisfied, so
if the velocity-dominated constraints are satisfied at some
$t_0$, then they are satisfied for all $t > 0$.

For this class of solutions, the inequality \cite{DHS},
$2 p_1 + p_2 + \cdots + p_{d-2} > 0$, or equivalently,
\begin{equation}
\label{inequalities2}
1 + p_1 - p_d - p_{d-1} > 0,
\end{equation}
defines the set $V$ which was referred to in the introduction.
As shown in \cite{DHS}, this inequality can be realized
when the spacetime dimension $D$ is equal to or greater
than $11$. As in our Maxwell archetypal example above, we expect that
this inequality will be crucial to control the effect of the
source terms (here linked to the spatial curvature) near the singularity.
It is again convenient to introduce
 a number $\sigma > 0$ so that, for all $x \in U_0$,
$4 \sigma < 1 + p_1 - p_d - p_{d-1} $.
Reduce $\epsilon$ if necessary so that
$\epsilon < \sigma/(2 d + 1) $.  If $\epsilon$ is reduced, it
may be necessary to shrink $U_0$ so that the conditions imposed
above remain satisfied.  It is assumed that
$\epsilon$ and $U_0$ are such that the conditions
imposed above and the condition imposed
in this paragraph are all satisfied.

We again recast the evolution equations in
the form (\ref{fuchs0}) and
show, for $D \geq 11$, that (\ref{fuchs0}) is Fuchsian and
equivalent to the vacuum Einstein equation, with
quantities $u$, $\cal A$ and $f$ as follows.  Let
$u = ({\gamma^a}_b, {\lambda^c}_{ef}, {\kappa^h}_i)$ be
related to the Einstein variables by
(\ref{defgamma}) -- (\ref{defkappa}).  For general $d$ define
$\alpha_0=(d+1)\epsilon$ and define ${\alpha^a}_b$ in terms of $\alpha_0$
as in section \ref{scaMax4D}. Let $\cal A$
and $f$, be given by equations~(\ref{fuchgamma}) --
(\ref{fuchkappa}), with ${M^a}_b = 0$.
The argument that $\cal A$ in equation~(\ref{fuchs0})
satisfies the appropriate positivity condition is analogous
to the part of the argument concerning the submatrix of $\cal A$
corresponding to $(\gamma, \kappa)$ in \cite{AR}.  A transformation
to a frame in which $^0g_{ab}$ is diagonal induces a similarity
transformation of $\cal A$.  The eigenvalues of the submatrix
are calculated in this representation in \cite{AR}, and
the generalization of the calculation to integer $d \geq 2$
is straightforward.

To obtain $f = O(t^\delta)$ requires the estimate
$t^{2 - {\alpha^a}_b} \,^S {R^a}_b = O(t^\delta)$.
The strategy used here is different from that used to estimate the
curvature in \cite{AR}. The general problem is one of organization.
There are many terms to be estimated, each of which on its own is
not too difficult to handle. The difficulty is to maintain an 
overview of the different terms. The procedure in \cite{AR}
made essential use of the fact that the indices only take three
distinct values and in the case of higher dimensions, where this
simplification is not available, an alternative approach had to 
be developed.

First $^S{R^a}_b$ is estimated by considering
each of the five terms in the expression (\ref{curvature})
These five terms are expanded
by considering each of the six terms in (\ref{christoffel})
if the indices on $^S\Gamma^c_{ab}$
are distinct, but carrying out the summation before estimating
$^S\Gamma^a_{ab}$.
There are thus 55 terms to estimate.
While many of these terms are actually identical up to
numerical factors, the ease with which each term can be
estimated, using the inequalities~(\ref{makeest1}) --
(\ref{makeest3}), led to estimation of all 55 terms, rather
than first combining terms.  We do however,
take into account that $f^i_{jk} = 0$ if $j = k$ for
obtaining the estimates.

Once an equation such as (\ref{fuchs0}) is shown to be
Fuchsian, then it follows that spatial derivatives of $u$ of any
order are $o(1)$.  At the stage of the argument we are at here,
we cannot assume $u_{xx} = O(1)$.  This means that
$t^{-\zeta} {\lambda^a}_{bc}$ must be used for $e_b({\gamma^a}_c)$
in places where a spatial derivative of $e_b({\gamma^a}_c)$ occurs.
This makes a slight difference, compared to section \ref{stepfour},
in what estimate of the terms in the connection coefficients
is used for the first and second terms of (\ref{curvature})
($t^{-\delta}$ is replaced by $t^{-\zeta}$).  There are additional
differences from (\ref{christoffela}) -- (\ref{christoffelf}),
because there it is assumed that the Kasner-like metric
is diagonal.  The estimates $^0 \tilde g_{ab} = O(t^{-\epsilon})$
and $^0 \tilde g^{ab} = O(t^{-\epsilon})$, obtained in Lemma~2 of
\cite{AR}, hold in the case we are considering, so that
$\tilde g_{(ab)} \preceq t^{|q_a - q_b| -\epsilon}$ and
(see \cite{AR})
$^S \tilde g^{ab} \preceq t^{|q_a - q_b| -\epsilon}$.
This adds factors of $t^{-\epsilon}$ to the estimate
of terms in the connection coefficients.

With these considerations, from (\ref{christoffel}),
\begin{equation}
\label{atracetildegamma2}
^S \tilde \Gamma^a_{ac} =
{1 \over 2} \, ^S \tilde g^{ab} \tilde e_c(\tilde g_{(ab)})
+ \tilde f^a_{ac}
\preceq t^{-q_c - 2 \epsilon - \zeta}.
\end{equation}

Here we do not write out the estimates of all 55 terms,
but instead give some examples, with a number designating
which term of (\ref{curvature}) is being considered (1 -- 5),
and a letter designating which term of (\ref{christoffel})
is being considered (A -- F).  Thus, for example, term 1C is
\begin{eqnarray}
\label{estcurvature12c}
t^{-q_a + q_b} \tilde g^{ac}
\tilde e_h ( \, ^S \tilde g^{hi} \tilde e_i (\tilde g_{(cb)}) ) & \preceq &
\sum_{c,h,i} t^{- q_a + q_b + |q_a - q_c| - q_h + |q_h - q_i|
- q_i + |q_c - q_b| - \delta - 3 \epsilon - \zeta} \nonumber \\
& \preceq & t^{- 2 q_a + 2 q_{\max\{a,b\}} - 2 q_d
- \delta - 3 \epsilon - \zeta}.
\end{eqnarray}
Term 3E is
\begin{eqnarray}
\label{estcurvature32e}
t^{-q_a + q_b} \tilde g^{ac}
\, ^S \tilde g^{ik} \tilde g_{(bj)} \tilde f^j_{ck}
\, ^S \tilde \Gamma^h_{hi} & \preceq &
\sum_{c,i,j,k \neq c} t^{- q_a + q_b + |q_a - q_c| + |q_i - q_k|
+ |q_b - q_j| + q_j - q_c - q_k - q_i - \delta - 5 \epsilon} \nonumber \\
& \preceq & \sum_{c,k \neq c} t^{- 2 q_{\min\{a,c\}} + 2 q_b
- 2 q_{\min\{d,k\}} - \delta - 5 \epsilon}.
\end{eqnarray}
In term 4, $t^{- q_a + q_b} \tilde g^{ac} \, ^S
\tilde \Gamma^h_{ib} \, ^S \tilde \Gamma^i_{ch}$, the
terms resulting from expanding $^S \tilde \Gamma^h_{ib}$
are designated by small letters a -- f, and those from
$^S \tilde \Gamma^i_{ch}$ are designated by capital
letters A -- F.  Term 4dA is
\begin{eqnarray}
\label{estcurvature42dA}
& & t^{-q_a + q_b} \tilde g^{ac}
\, ^S \tilde g^{hl} \tilde g_{(ji)} \tilde f^j_{bl}
\, ^S \tilde g^{ik} \tilde e_c (\tilde g_{(hk)}) \nonumber \\  & \preceq &
\sum_{c,h,i,j,k,l} t^{- q_a + q_b + |q_a - q_c| + |q_h - q_l|
+ |q_j - q_i| + q_j - q_b - q_l + |q_i - q_k| - q_c + |q_h - q_k|
 - \delta - 5 \epsilon} \nonumber \\
& \preceq & t^{- 2 q_a - \delta - 5 \epsilon}.
\end{eqnarray}
Term 4eD is
\begin{eqnarray}
\label{estcurvature42eD}
& & t^{-q_a + q_b} \tilde g^{ac}
\, ^S \tilde g^{hn} \tilde g_{(bj)} \tilde f^j_{in}
\,^ S \tilde g^{il} \tilde g_{(kc)} \tilde f^k_{hl}
\nonumber \\  & \preceq &
\sum_{c,h,i,j,k,l \neq h,n \neq i} t^{- q_a + q_b + |q_a - q_c|
+ |q_h - q_n| + |q_b - q_j| + q_j - q_i - q_n
+ |q_i - q_l| + |q_k - q_c| + q_k - q_h - q_l
 - \delta - 5 \epsilon} \nonumber \\
& \preceq & t^{2 q_b - 2 q_d -2 q_{d-1}
- \delta - 5 \epsilon}
\end{eqnarray}
Term 5D is
\begin{eqnarray}
\label{estcurvature52d}
t^{-q_a + q_b} \tilde g^{ac} \tilde f^i_{cj}
\, ^S \tilde g^{jk} \tilde g_{(hi)} \tilde f^h_{bk}
& \preceq &
\sum_{c,h,i,j \neq c ,k \neq b} t^{- q_a + q_b + |q_a - q_c|
+ q_i - q_c - q_j + |q_j - q_k|
+ |q_h - q_i| + q_h - q_b - q_k  - \delta - 3 \epsilon} \nonumber \\
& \preceq & \sum_{c,j \neq c ,k \neq b}
t^{- 2 q_{\min \{a,c\}} + 2 q_1
-2 q_{\min \{j,k\}} - \delta - 3 \epsilon}.
\end{eqnarray}
The estimates of the remaining terms are obtained as these.
The examples include one of the terms which limits
the estimate for each possible choice of indices $a$ and $b$.
The result is,
\begin{equation}
\label{estrab}
^S {R^a}_b \preceq 
t^{2 q_b - 2 q_d -2 q_{d-1}
- \delta - 5 \epsilon}
+ \sum_{c,j \neq c ,k \neq b}
t^{- 2 q_{\min \{a,c\}} + 2 q_1
-2 q_{\min \{j,k\}} - \delta - 5 \epsilon}.
\end{equation}
And
\begin{eqnarray}
\label{estrabforkappa}
t^{2 - {\alpha^a}_b} \, ^S {R^a}_b
& \preceq & 
\{ t^{2 q_{\min \{a,b\}} - 2 q_d -2 q_{d-1}} +
\sum_{c \geq a,j \neq c ,k \neq b}
t^{-2 q_{\max \{a,b\}} + 2 q_1 -2 q_{\min \{j,k\}}} \}
t^{2 - \alpha_0 - \delta - 5 \epsilon} \nonumber \\
& \preceq & t^{2 - 2q_d - 2q_{d-1} + 2q_1 - (d + 7) \epsilon}
\preceq t^{8 \sigma - (d + 7) \epsilon} = O(t^{\delta}).
\end{eqnarray}

The estimate of the rest of the terms in $f$ is obtained
straightforwardly by checking that the exponent of $t$ in
each case is strictly positive.  The other regularity
conditions that $f$ should satisfy are shown to hold by
equation~(31) in \cite{AR} and the remarks following
equation~(31).  The symmetry of $g_{ab}$ is
shown for all $d$'s in subsection \ref{symmetryofg}.
That the Hamiltonian and momentum constraints are satisfied is shown
by the direct analogue of argument made in section~\ref{stepfour} and 
the estimate $R = o(t^{-2 + \eta_1})$ 
obtained from equation~(\ref{estrab}).
The only change is that equation (\ref{divmom2}) is replaced by
\begin{equation}
\tilde g^{ab} \tilde \Gamma^c_{ab} \, \bar C_c \, t^{-p_c}
\preceq t^{-2 p_d - \delta}.
\end{equation}

To conclude this section we discuss the solution of the velocity-dominated
constraints for the vacuum equations and $D\ge 4$. The case $D=3$ could be
discussed in a similar way but the analogue of the Lichnerowicz equation
has a different form and so for brevity that case will be omitted.
The discussion proceeds in a way which is parallel to that of the last
section. As already indicated there, the essential task is the analysis of
the Lichnerowicz equation. In the present case we start with free data
$\tilde g_{ab}$, $\tilde k_{ab}$ and $H$ where $\tilde k_{ab}$ has zero
divergence. The actual data are defined by  $g_{ab}=\omega^{4/(d-2)}
\tilde g_{ab}$ and $k_{ab}=\omega^{-2}\tilde k_{ab}+Hg_{ab}$. The constraints
will be satisfied is $\omega$ satisfies the following analogue of the 
Lichnerowicz equation:
\begin{equation}\label{lichnod}
\Delta_{\tilde g}\omega+\frac{d-2}{4(d-1)}(-R_{\tilde g}\omega
+\tilde k^{ab}\tilde k_{ab}\omega^{\frac{3d-2}{d-2}})-\frac{d(d-2)}{4}
H^2\omega^{\frac{d+2}{d-2}}=0
\end{equation}
The corresponding equation in the velocity-dominated case is
\begin{equation}\label{vdlichnod}
\frac{d-2}{4(d-1)}\tilde k^{ab}\tilde k_{ab}\omega^{\frac{3d-2}{d-2}}
-\frac{d(d-2)}{4}H^2\omega^{\frac{d+2}{d-2}}=0
\end{equation}
As in the case of (\ref{vdlichno3}) it is trivial to solve 
(\ref{vdlichnod}) provided
$\tilde k_{ab}$ does not vanish at any point. To determine the solvability
of equation (\ref{lichnod}) it is necessary to study the positive zeros of 
the algebraic expressions $x+b x^{\frac{3d-2}{d-2}}-ax^{\frac{d+2}{d-2}}$ 
and $b x^{\frac{3d-2}{d-2}}-ax^{\frac{d+2}{d-2}}$ where $a>0$ and $b>0$.
The second expression is very close to what we had in the
velocity-dominated case and clearly has a unique positive zero for any
values
of $a$ and $b$ satisfying the inequalities assumed. Looking for 
positive zeros of the first algebraic expression is equivalent to looking 
for positive solutions of $x^{-\frac{d+2}{d-2}}+a x^{-2}-b=0$. Note that
the function on the left hand side of this equation is evidently
decreasing for all positive $x$, tends to infinity as $x\to 0$ and
tends to $-b$ as $x\to\infty$. Hence as long as the constant $b$ is
non-zero this function has exactly one positive zero, as desired.
This is what is needed to obtain an existence theorem. It would be
desirable to also obtain a uniqueness theorem for the solution of 
(\ref{lichnod}). To obtain solutions of the velocity-dominated constraints
of the right kind to be consistent with Kasner-like behaviour we can use the
same approach as in the last section, starting with Kasner solutions with
an appropriate set of Kasner exponents.

\section{ Massless scalar field, $D \geq 3$}\label{scalar}
\setcounter{equation}{0}
\setcounter{theorem}{0}
\setcounter{lemma}{0}      
Consider Einstein's equations, $D \geq 3$, with a 
massless scalar field as source, the action given by
$S_E [g_{\alpha \beta}] + S_\phi[g_{\alpha \beta}, \phi]$,
and $d + 1$ decomposition
as in section~\ref{stepzero}.  The stress-energy tensor is
\begin{equation}
\label{stressenergy}
T_{\mu \nu} = \, ^{(D)}\nabla_\mu \phi \,
^{(D)}\nabla_\nu \phi - {1 \over 2} g_{\mu \nu} \,
^{(D)}\nabla_\alpha \phi \, ^{(D)}\nabla^\alpha \phi.
\end{equation}
Thus $\rho = {1 \over 2} \{ (\partial_t \phi)^2 +
g^{ab} e_a(\phi) e_b(\phi)\}$, $j_a = - \partial_t \phi \, e_a(\phi)$,
and ${M^a}_b = g^{ac} e_b(\phi) \, e_c(\phi)$.  A crucial step in
the generalization to arbitrary $D \geq 3$ is that the cancellation
of terms involving $\partial_t \phi$ in the expression for
${M^a}_b$ is not particular to $D = 4$.  The scalar field
satisfies $^{(D)}\nabla_\alpha \, ^{(D)}\nabla^\alpha \phi = 0$,
which has $d + 1$ decomposition
\begin{equation}
\label{eveqphi0}
\partial^2_t \phi - (\tr k) \partial_t \phi =
g^{ab} \nabla_a \nabla_b \, \phi.
\end{equation}

Let the Kasner-like evolution equations be
equations~(\ref{dtg0}) -- (\ref{dtphi0}), with
solution (\ref{0kab1}) -- (\ref{0phi1}) for time
coordinate as in section~\ref{vacuum}.  Given a point
$x_0 \in \Sigma$, let the neighbourhood $U_0$, the (local)
adapted frame and the constant $\epsilon$ be as in
section~\ref{vacuum}.  Define
$^0 \rho = {1 \over 2} (\partial_t \, ^0 \phi)^2$ and
$^0 j_a = - \partial_t \, ^0 \phi \, e_a(\, ^0 \phi)$.
The velocity-dominated constraints corresponding to the
Hamiltonian and momentum constraints are $^0C=0$ and
$^0C_a = 0$, with $^0C$ and $^0C_a$ given by
equations~(\ref{hamiltonian0}) and (\ref{momentum0}).  For
the solution (\ref{0kab1}) -- (\ref{0phi1}) the
velocity-dominated Hamiltonian constraint is equivalent
to $\sum {p_a}^2 + A^2 = 1$.  Equations~(\ref{dt0C}) and
(\ref{dt0Ca}) are satisfied so if the velocity-dominated
constraints are satisfied at some $t_0$, then they are
satisfied for all $t > 0$.  The restriction defining
the set $V$ is the inequality~(\ref{inequalities2}).
(If $D < 11$, then satisfying simultaneously (\ref{inequalities2}),
$\sum p_a = 1$
and $\sum {p_a}^2 + A^2 = 1$ requires $A \neq 0$.  Note that
conversely, for $D=3$, the restrictions defining $V$ are simply
equivalent to $A \not=0$,  since (\ref{inequalities2})
is in this case a consequence 
of $p_1 + p_2 = 1$ and $p_1^2 + p_2^2 <1$.)
The constant $\sigma > 0$
is chosen so that, for all $x \in U_0$,
$4 \sigma < 1 + p_1 - p_d - p_{d-1}$ from which
it follows that
\begin{equation}
\label{inequalities3}
\sigma < 2 - 2 p_d.
\end{equation}
Now reduce $\epsilon$ if necessary so that
$\epsilon < \sigma/(2 d + 1) $.  As before,
this may in turn require shrinking $U_0$.

The unknown $u = ({\gamma^a}_b, {\lambda^a}_{bc},
{\kappa^a}_b, \psi, \omega_a,\chi)$ is related
to the Einstein-matter variables by (\ref{defgamma}) -- (\ref{defchi}).
The quantities $\cal A$ and $f$ appearing in
equation~(\ref{fuchs0}) are given by the evolution
equations (\ref{fuchgamma}) -- (\ref{fuchomega}) and
\begin{equation}
\label{fuchchi0}
t \, \partial_t \chi + \beta \chi = t^{\alpha_0 - \beta}
(\tr \, \kappa) ( A + t ^\beta \chi ) + t^{2 - \beta}
\; ^S g^{ab} \, ^S \nabla_a \, ^S \nabla_b \, ^0 \phi
+ t^{2 - \zeta} \; ^S \nabla^a \omega_a.
\end{equation}
The argument that the matrix $\cal A$ satisfies the appropriate
positivity condition is analogous to the argument in \cite{AR}.
Regarding the estimate $f = O(t^\delta)$, the estimate
$t^{2 - {\alpha^a}_b} \,^S {R^a}_b = O(t^\delta)$ was
obtained in equation~(\ref{estrabforkappa}).  The estimate
$t^{2 - {\alpha^a}_b} {M^a}_b = O(t^\delta)$ follows from
the inequality~(\ref{inequalities3}) and from $q_d < p_d$.
The only other terms in $f$ whose estimates are not immediate
from the estimates made in \cite{AR} are the last two terms
on the right hand side of equation~(\ref{fuchchi0}).  The covariant
derivative compatible with the symmetrized metric is used
in equation~(\ref{fuchchi0}) for convenience.
{}From the estimate
$^S \tilde g^{ab} \preceq t^{|q_a - q_b| -\epsilon}$ \cite{AR},
equations~(\ref{estmetric}) and (\ref{estdgab}),
\begin{displaymath}
^S g^{ab} \preceq t^{-2 q_{\min \{ a,b \}} - \epsilon}
\hspace{20pt} \mbox{and} \hspace{20pt}
e_c(g_{(ab)}) \preceq t^{ 2 q_{\max \{ a,b \}}
- \delta - \epsilon}.
\end{displaymath}
Therefore,
\begin{eqnarray}
^S g^{ab} \, ^S \Gamma^c_{ab} & = &  ^S g^{ab} \;^S g^{ch}
\Big(e_a(g_{(bh)}) - {1 \over 2} e_h(g_{(ab)}) \Big)
- \, ^S g^{ch} f^a_{ah} \nonumber \\       
& \preceq & t^{- 2 q_d - \delta - 3 \epsilon} \nonumber
\end{eqnarray}
and
\begin{eqnarray}
t^{2 - \beta} \; ^S g^{ab} \, ^S \nabla_a
\, ^S \nabla_b \, ^0 \phi & = &
t^{2 - \beta} \; ^S g^{ab}
\{e_a( e_b(\,^0 \phi)) - \, ^S \Gamma^c_{ab}
e_c(\,^0 \phi)\} \nonumber \\
& \preceq & t^{2 - 2 q_d - \beta - \delta
- 3 \epsilon} = O(t^\delta), \nonumber \\
t^{2 - \zeta} \; ^S \nabla^a \omega_a & = &
t^{2 - \zeta} \; ^S g^{ab}
\{e_a( w_b) - \, ^S \Gamma^c_{ab}
w_c \} \nonumber \\
& \preceq & t^{2 - 2 q_d - \zeta - \delta
- 3 \epsilon} = O(t^\delta) \nonumber
\end{eqnarray}

The other regularity conditions that $f$ should satisfy are
again shown to hold by equation (31) in \cite{AR} and the
remarks following equation (31). That $g_{ab}$ is 
symmetric (so that equation~(\ref{fuchchi0})
and equation~(\ref{eveqphi0}) are equivalent) is shown as
in subsection \ref{symmetryofg}.
That the Hamiltonian and momentum constraints are satisfied
is shown by the analogue of the argument made in section \ref{stepfour}
and the estimate $R = o(t^{-2 + \eta_1})$ obtained from
equation~(\ref{estrab}).           

Note that the case $D=3$ of this result has an interesting connection
to the Einstein vacuum equations in $D=4$. As it follows from
standard Kaluza-Klein lines, the solutions of the latter
with polarized $U(1)$ symmetry are equivalent to the Einstein-scalar
field system in $D=3$ (see e.g. \cite{nicolai} and
\cite{andersson99}, section 5). Hence the
result of this section implies that we have constructed the most 
general known class of singular solutions of the Einstein vacuum 
equations in four spacetime dimensions. These spacetimes have one
spacelike Killing vector.  

\section{Matter fields derived from $n$-form
potentials}\label{nform}
\setcounter{equation}{0}
\setcounter{theorem}{0}
\setcounter{lemma}{0}      
\subsection{Equations of motion}
We now turn to the general system (\ref{001}),
but without the interaction terms ``more".  These are
considered in section \ref{couple} below.
The action is the sum of (\ref{action2}), (\ref{action3})
and $k$ additional terms, each of the form 
(\ref{action4}).  The argument is based on that
of section~\ref{scalar}.  It is enough here to note the
differences.  Furthermore, since there is no coupling
between additional matter fields, the differences from the
argument made in section~\ref{scalar} can be noted for
each additional matter field independently of the others.
Therefore consider the $j$th additional matter field,
\begin{displaymath}
F_{\mu_0 \cdots \mu_{n_j}} =
(n_j+1) \nabla_{[\mu_0} A_{\mu_1 \cdots \mu_{n_j}]},
\end{displaymath}
with $A$ an $n_j$-form.  This matter field contributes
the following additional terms to the stress-energy
tensor, equation~(\ref{stressenergy}),
\begin{displaymath}
T_{\mu \nu} = \cdots + \left\{ {1 \over n_j!}
F_{\mu \alpha_1 \cdots \alpha_{n_j}}
{F_\nu}^{\alpha_1 \cdots \alpha_{n_j}}
-{1 \over 2 (n_j + 1)!} g_{\mu \nu}
F_{\alpha_0 \cdots \alpha_{n_j}}
F^{\alpha_0 \cdots \alpha_{n_j}} \right\}
e^{\lambda_j \phi}.
\end{displaymath}
Define ${\cal E}^{a_1 \cdots a_{n_j}} = \sqrt{g} \,
F^{0 a_1 \cdots a_{n_j}} \;
e^{\lambda_j \phi}$.  If $n_j = 0$, $\cal E$ is a spatial
scalar density.  Throughout this section and the next
we use the following conventions.  If $n_j = 0$, then
$P_{a_1 \cdots a_{n_j}}$ is a scalar, $g_{a_1 b_1} \cdots
g_{a_{n_j} b_{n_j}} =1$, etc.
The $d+1$ decomposition of the contribution of
this matter field to the stress-energy tensor is
\begin{eqnarray}
\label{rhonf}
\rho & = & \cdots + { 1 \over 2 \, g \, n_j!}
g_{a_1 b_1} \cdots g_{a_{n_j} b_{n_j}}
{\cal E}^{a_1 \cdots a_{n_j}}
{\cal E}^{b_1 \cdots b_{n_j}} e^{-\lambda_j \phi}
\nonumber \\ & & \; \; \; \; \; + {1 \over 2 (n_j + 1)!}
\, g^{a_0 b_0} \cdots g^{a_{n_j} b_{n_j}}
F_{a_0 \cdots a_{n_j}}
F_{b_0 \cdots b_{n_j}} e^{\lambda_j \phi}, \\
\label{currentnf}
j_a & = & \cdots +
{1 \over \sqrt{g} \, n_j!} \, {\cal E}^{b_1 \cdots b_{n_j}}
\, F_{a b_1 \cdots b_{n_j}}, \\
{M^a}_b & = & \cdots -
{1 \over g} \Big( {n_j \over n_j!} \,
g_{b h_1} g_{c_2 h_2} \cdots g_{c_{n_j} h_{n_j}}
{\cal E}^{a c_2 \cdots c_{n_j}} {\cal E}^{h_1 \cdots h_{n_j}}
\nonumber \\ & & \; \; \; \; \;
- {n_j \over (d - 1) n_j!} \, {\delta^a}_b \, g_{c_1 h_1}
\cdots g_{c_{n_j} h_{n_j}} {\cal E}^{c_1 \cdots c_{n_j}}
{\cal E}^{h_1 \cdots h_{n_j}} \Big) e^{-\lambda_j \phi}
\nonumber \\ & & \; \; \; \; \; + \Big({1 \over n_j!} \,
g^{ac} g^{h_1 i_1} \cdots g^{h_{n_j} i_{n_j}}
F_{c h_1 \cdots h_{n_j}} F_{b i_1 \cdots i_{n_j}}
\nonumber \\ & & \; \; \; \; \;
- {n_j \over (d - 1) (n_j +1)!} \, {\delta^a}_b \,
g^{c_0 h_0} \cdots g^{c_{n_j} h_{n_j}} F_{c_0 \cdots c_{n_j}}
F_{h_0 \cdots h_{n_j}} \Big) e^{\lambda_j \phi}. \nonumber
\end{eqnarray}
The $j$th matter field satisfies
\begin{eqnarray}
\label{matter1nf}
^{(D)}\nabla_\mu (F^{\mu \nu_1 \cdots \nu_{n_j}}
e^{\lambda_j \phi}) & = & 0, \\
\label{matter2nf}
^{(D)} \nabla_{[\mu} F_{\nu_0 \cdots \nu_{n_j}]} & = & 0,
\end{eqnarray}
with $d+1$ decomposition into constraint equations,
\begin{eqnarray}
\label{elconstraintnf}
e_a ( {\cal E}^{a b_2 \cdots b_{n_j}}) +
f^c_{ca} \, {\cal E}^{a b_2 \cdots b_{n_j}} +
{1 \over 2} \sum_{i=2}^{n_j} f^{b_i}_{ac}
{\cal E}^{a b_2 \cdots c \cdots b_{n_j}} & = & 0, \\
\label{magconstraintnf}
e_{[a}(F_{b_0 \cdots b_{n_j}]})
-{(n_j + 1) \over 2} f^c_{[a b_0}
F_{|c| b_1 \cdots b_{n_j}]} & = & 0,
\end{eqnarray}
and evolution equations, 
\begin{eqnarray}
\label{elevnf}
\partial_t {\cal E}^{a_1 \cdots a_{n_j}} & = &
-e_b( \sqrt{g} g^{b c_0}
g^{a_1c_1} \cdots g^{a_{n_j} c_{n_j}}
F_{c_0 \cdots c_{n_j}} e^{\lambda_j \phi})
-\{f^h_{hb} g^{b c_0} g^{a_1 c_1} \cdots g^{a_{n_j} c_{n_j}}
\nonumber \\ & & \; \;
+ {1 \over 2} \sum_{i=1}^{n_j}
f^{a_i}_{bh} g^{b c_0} g^{a_1 c_1} \cdots g^{h c_{i}} \cdots
g^{a_{n_j} c_{n_j}} \}
\sqrt{g} F_{c_0 \cdots c_{n_j}}
e^{\lambda_j \phi}, \\
\label{magevnf}
\partial_t F_{a_0 \cdots a_{n_j}}
& = & -(n_j+1) e_{[a_0} ( {1 \over \sqrt{g}}
g_{a_1 |b_1|} \cdots g_{a_{n_j} ]  b_{n_j}}
{\cal E}^{b_1 \cdots b_{n_j}} e^{-\lambda_j \phi})
\nonumber \\ & & \; \;
+ {(n_j+1) n_j \over 2 \sqrt{g}} f^c_{[a_0 a_1}
g_{|c| |b_1|} g_{a_2  |b_2|} \cdots g_{a_{n_j} ] b_{n_j}}
{\cal E}^{b_1 \cdots b_{n_j}} e^{-\lambda_j \phi}.
\end{eqnarray}
The $j$th matter field contributes the following terms to the evolution
equation~(\ref{eveqphi0}) for $\phi$.
\begin{eqnarray}
\label{phievnf}
\partial^2_t \phi - (\tr k) \partial_t \phi & = & \cdots
+ {\lambda_j \over 2 \, g \, n_j!} g_{a_1 b_1} \cdots g_{a_{n_j} b_{n_j}}
{\cal E}^{a_1 \cdots a_{n_j}}
{\cal E}^{b_1 \cdots b_{n_j}} e^{-\lambda_j \phi} \nonumber \\
& & \; \; \; \; \; - {\lambda_j \over 2 (n_j+1)! }
g^{a_0 b_0} \cdots  g^{a_{n_j} b_{n_j}}
F_{a_0 \cdots a_{n_j}} F_{b_0 \cdots b_{n_j}} e^{\lambda_j \phi}
\end{eqnarray}

\subsection{Velocity-dominated system}
The Kasner-like evolution equations corresponding to this
matter field are
$\partial_t \, ^0 {\cal E}^{a_1 \cdots a_{n_j}} = 0$ and
$\partial_t \, {}^0 F_{a_0 \cdots a_{n_j}} = 0$.
The quantities $^0{\cal E}^{a_1 \cdots a_{n_j}}$ and
$^0F_{a_0 \cdots a_{n_j}}$ are constant in time with
analytic spatial dependence and both are totally antisymmetric.         

The velocity-dominated matter constraint equations
are equations~(\ref{elconstraintnf})
and (\ref{magconstraintnf}) with $^0\cal E$ and $^0F$
substituted for $\cal E$ and $F$.  Since all quantities
in the velocity-dominated matter constraints are independent
of time, if the matter constraints are satisfied at
some time $t_0 > 0$, then they are satisfied for
all $t > 0$.  This matter field does not contribute
to $^0 \rho$.  Its contribution to $^0 j_a$ is
the term shown on the right hand side of equation~(\ref{currentnf})
with $^0 g$, $^0 {\cal E}$ and $^0 F$ substituted for
$g$, ${\cal E}$ and $F$.
The velocity-dominated constraints corresponding to the Hamiltonian
and momentum constraints are $^0C=0$ and $^0C_a = 0$, with
$^0C$ and $^0C_a$ given by equations (\ref{hamiltonian0}) and
(\ref{momentum0}).  Equations~(\ref{dt0C}) and (\ref{dt0Ca})
are satisfied,
so as before, if the velocity-dominated
constraints are satisfied at some $t_0>0$, then they are
satisfied for all $t > 0$.

The presence of the matter field $A^{(j)}$ puts the following
restrictions on the set $V$ \cite{dh1}.
\begin{equation}
\label{inequalitiesnf}
2 p_1 + \cdots + 2p_{n_j} - \lambda_j A > 0
\hspace{20pt} \mbox{and}
\hspace{20pt} 2 p_1 + \cdots + 2 p_{d-n_j-1} + \lambda_j A > 0.
\end{equation}
The restrictions generalize the inequalities (\ref{inequalities1})
found for a Maxwell field in 4 dimensions and, like them, guarantee
that one can asymptotically neglect the $p$-form
$A^{(j)}$ in the Einstein-dilaton dynamical equations.
(For $n_j = 0$, the inequality on the left of (\ref{inequalitiesnf})
is $-\lambda_j A > 0$, while for $n_j = 1$ it is
$2 p_1 - \lambda_j A > 0$.  For $n_j = d-1$, the inequality
on the right is $\lambda_j A > 0$, while for $n_j = d-2$
it is $2 p_1 + \lambda_j A > 0$.)

The constant $\sigma$ is reduced from its value in
section~\ref{scalar}, if necessary, so that, for all $x \in U_0$,
$\sigma < 2 p_1 + \cdots + 2p_{n_j} - \lambda_j A$ and
$\sigma < 2 p_1 + \cdots + 2 p_{d-n_j-1} + \lambda_j A$.
If $\sigma$ is reduced, it may be necessary to reduce
$\epsilon$, and in turn shrink $U_0$, so that the conditions
imposed in section~\ref{scalar} are still all satisfied.

\subsection{Fuchsian property - Estimates}
The $j$th matter field contributes the following components to
the unknown $u$ in the Fuchsian equation~(\ref{fuchs0}).
\begin{eqnarray}
\label{defxinf}
{\cal E}^{a_1 \cdots a_{n_j}}
 & = & \, ^0 {\cal E}^{a_1 \cdots a_{n_j}}
 + t^\beta \xi^{a_1 \cdots a_{n_j}}, \\
\label{defvarphinf}
F_{a_0 \cdots a_{n_j}} 
& = & \, ^0 F_{a_0 \cdots a_{n_j}}
+ t^\beta \varphi_{a_0 \cdots a_{n_j}}.
\end{eqnarray}
Here, $\beta = \epsilon / 100$ as above,
$\xi^{a_1 \cdots a_{n_j}}$ is a totally antisymmetric
spatial tensor density, so contributes $d \choose n_j$
independent components to $u$, and
$\varphi_{a_0 \cdots a_{n_j}}$ is a totally antisymmetric
spatial tensor, so contributes $d \choose n_j + 1$ components
to $u$.  This is consistent with the form of the evolution
equations.  Note that ${\cal E}^{a_1 \cdots a_{n_j}} = O(1)$
and $F_{a_0 \cdots a_{n_j}} = O(1)$.

This matter field contributes additional rows and columns
to the matrix $\cal A$ such that the only non-vanishing new
entries are on the diagonal and strictly positive.  Therefore,
the presence of this matter field does not alter that $\cal A$
satisfies the appropriate positivity condition.

The terms in the source $f$ which must be estimated on account of the
$j$th matter field are the following.  It contributes terms
to the components of $f$ corresponding to $\kappa$ through
its contribution to ${M^a}_b$.
\begin{eqnarray}
\label{estM1nf}
& & t^{2 - {\alpha^a}_b} {1 \over g}
g_{b h_1} g_{c_2 h_2} \cdots g_{c_{n_j} h_{n_j}}
{\cal E}^{a c_2 \cdots c_{n_j}} {\cal E}^{h_1 \cdots h_{n_j}}
e^{-\lambda_j \phi} \nonumber \\ & & \hspace{40pt} \preceq \,
\sum t^{-2 q_{\max\{a,b\}} + 2 q_a + 2 q_{\max\{b,h_1\}} +
\cdots + 2 q_{\max\{c_{n_j},h_{n_j}\}} - \lambda_j A - \alpha_0 -
n_j \epsilon} \nonumber \\
& & \hspace{40pt} \preceq \, t^{2 q_1 + \cdots + 2q_{n_j}
- \lambda_j A - \alpha_0 - n_j \epsilon}
\, = \, O(t^{- \alpha_0 - n_j \epsilon + \sigma})
\, = \, O(t^{\delta}) \\
\label{estM3nf} 
& & t^{2 - {\alpha^a}_b}
g^{ac} g^{h_1 i_1} \cdots g^{h_{n_j} i_{n_j}}
F_{c h_1 \cdots h_{n_j}} F_{b i_1 \cdots i_{n_j}}
e^{\lambda_j \phi} \nonumber \\ & & \hspace{40pt} \preceq \,
\sum t^{2-2 q_{\max\{a,b\}} + 2 q_a - 2 q_{\min\{a,c\}}
- 2 q_{\min\{h_1,i_1\}} - \cdots
- 2 q_{\min\{h_{n_j},i_{n_j}\}} + \lambda_j A - \alpha_0 -
(n_j + 1) \epsilon} \nonumber \\& & \hspace{40pt}  \preceq \,
t^{2 q_1 + \cdots + 2 q_{d - n_j - 1} + \lambda_j A - \alpha_0 -
(n_j + 1) \epsilon}
\, = \, O(t^{- \alpha_0 - (n_j + 1) \epsilon + \sigma})
\, = \, O(t^{\delta})
\end{eqnarray}
Here it is used that both ${\cal E}^{a_1 \cdots a_{n_j}}$
and $F_{a_0 \cdots a_{n_j}}$ are totally antisymmetric,
so that the sums indicated by a summation symbol are not over
all indices.  Note that the inequalities~(\ref{inequalitiesnf})
have been crucially used in getting 
the estimates~(\ref{estM1nf}) and (\ref{estM3nf}).  The
desired estimates for the other two terms are obtained similarly.

The terms contributed to the component of $f$
corresponding to $\chi$ by the $j$th matter
field are obtained by multiplying the right hand side of
equation~(\ref{phievnf}) by $t^{2 - \beta}$.
\begin{eqnarray}
\label{estchinf}
t^{2 - \beta} {1 \over g }
g_{a_1 b_1} \cdots g_{a_{n_j} b_{n_j}}
{\cal E}^{a_1 \cdots a_{n_j}}
{\cal E}^{b_1 \cdots b_{n_j}} e^{-\lambda_j \phi}
& = & O(t^{- \beta - n_j \epsilon + \sigma})=O(t^{\delta}) \\
t^{2-\beta}
g^{a_0 b_0} \cdots  g^{a_{n_j} b_{n_j}}
F_{a_0 \cdots a_{n_j}} F_{b_0 \cdots b_{n_j}} e^{\lambda_j \phi}
& = & O(t^{- \beta - (n_j + 1) \epsilon + \sigma})=O(t^{\delta}).
\end{eqnarray}
The terms in $f$ corresponding to $\xi^{a_1 \cdots a_{n_j}}$
for the $j$th matter field are obtained by multiplying
the right hand side of equation~(\ref{elevnf}) by $t^{1 - \beta}$.
These terms are $O(t^{- \beta - \delta
- (n_j + 1) \epsilon + \sigma})=O(t^{\delta})$.
The terms in $f$ corresponding to
$\varphi_{a_0 \cdots a_{n_j}}$ for the $j$th matter field
are obtained by multiplying
the right hand side of equation~(\ref{magevnf}) by $t^{1 - \beta}$.
These terms are
$O(t^{- \beta - \delta - n_j \epsilon  + \sigma})=O(t^{\delta})$.
Thus the terms which occur in $f$ due to the $j$th matter field
are $O(t^\delta)$.

The time derivative of the matter constraint quantities for the
$j$th field (the left hand side of equations~(\ref{elconstraintnf})
and (\ref{magconstraintnf})) vanishes.  If the
velocity-dominated matter constraints are satisfied, the
matter constraint quantities are $o(1)$.  A quantity which is
both constant in time and $o(1)$ must vanish.  Therefore
the matter constraints for the $j$th field are satisfied.

Next the matter terms due to the $j$th field in the
Einstein constraints are estimated, in order to verify
that they are consistent with equations~(\ref{estconstraints1})
and (\ref{estconstraints2}).  The contribution to the
Hamiltonian constraint is, from equation~(\ref{rhonf}),
\begin{eqnarray}
{ 1 \over g } g_{a_1 b_1} \cdots g_{a_{n_j} b_{n_j}}
{\cal E}^{a_1 \cdots a_{n_j}}
{\cal E}^{b_1 \cdots b_{n_j}} e^{-\lambda_j \phi}
& = & O(t^{-2 - n_j \epsilon + \sigma}) = o(t^{-2 + \eta_1}), \\
g^{a_0 b_0} \cdots g^{a_{n_j} b_{n_j}}
F_{a_0 \cdots a_{n_j}}
F_{b_0 \cdots b_{n_j}} e^{\lambda_j \phi}
& = & O(t^{-2 - (n_j + 1) \epsilon + \sigma})
= o(t^{-2 + \eta_1}).
\end{eqnarray}
The contribution to the momentum constraint is
\begin{eqnarray}
\label{estjbnf}
j_a  - ^0 j_a & = & \cdots +
({1 \over \sqrt{g} } - {1 \over \sqrt{\, ^0 g} })
\, ^0{\cal E}^{b_1 \cdots b_{n_j}}
\, ^0 F_{a b_1 \cdots b_{n_j}} \\ & & \; \; \; \; \; +
{1 \over \sqrt{g} } ( {\cal E}^{b_1 \cdots b_{n_j}}
\, F_{a b_1 \cdots b_{n_j}} -
\, ^0{\cal E}^{b_1 \cdots b_{n_j}}
\, ^0 F_{a b_1 \cdots b_{n_j}} ) = o(t^{-1 + \eta_2}).
\nonumber
\end{eqnarray}
Estimates related to the determinant which are relevant to
(\ref{estjbnf}) are analogues of the estimates for $d=3$
immediately preceding equation~(\ref{estddetg1}). The form of
these estimates for general $d$ will now be presented. These
are $1/\sqrt{g} - 1/\sqrt{\, ^0g} = O(t^{-1 + \alpha_0 - d \epsilon})$,
$e_a(\tilde g) = O(t^{\alpha_0 - \delta - d \epsilon})$,
\begin{equation}\nonumber
e_a(g) = O(t^{2 + \alpha_0 - \delta - d \epsilon}),
\end{equation} 
and
\begin{displaymath}
e_a(g^{-1/2}) = - {e_a(g) \over 2 g^{3/2}}
= O(t^{-1 + \alpha_0 - \delta - d \epsilon}).
\end{displaymath}

\section{Determination of subcritical domain}
\label{subcritical}
\setcounter{equation}{0}
\setcounter{theorem}{0}
\setcounter{lemma}{0} 
 The explicit determination of the subcritical range of the dilaton
couplings for which the inequalities on the
Kasner exponents are consistent so that $V$ exists
may be a complicated matter.  We
consider in this section a few cases and give some general rules.
As in subsection \ref{criticalvalue}, we introduce the metric
\be
dS^2 =  G_{\m \n} dp^\m dp^\n =
\sum {dp_a}^2 - (\sum {dp_a})^2 + (dA)^2 
\label{Kmetricbis}
\ee
in the $D$-dimensional space of the Kasner exponents $(p_a,A)
\equiv (p^\m)$.  
This  metric has again Minkowskian signature
$(-,+,+, \cdots, +)$.  The forward light cone is defined
by
\be
G_{\m \n} p^\m p^\n = 0, \; \; \; \sum {p_a} >0 .
\label{lightbis}
\ee   
The Kasner conditions met in the previous section are equivalent to
the conditions that the Kasner exponents be on the forward
light cone (since $\sum {p_a} = 1$ can always be achieved
by positive rescalings).

The wall chamber ${\cal W}$ is now defined by
\begin{eqnarray}
&{}& p_1 \leq p_2 \leq \cdots \leq p_d \label{symmwalls}\\
&{}& 2 p_1 + p_2 + \cdots + p_{d-2} \geq 0 
\label{gravwalls} 
\end{eqnarray}
and, for each $p$-form,
\begin{eqnarray}
&{}& p_1 + p_2 + \cdots + p_{n_j} - \frac{\lambda_j}{2} A \geq 0
\label{elecwalls} \\
&{}& p_1 + p_2 + \cdots + p_{d-n_j-1} + \frac{\lambda_j}{2} A \geq 0.
\label{magnwalls}
\end{eqnarray} 
These inequalities may not be all independent.  The question
is to determine the ``allowed" values of the dilaton couplings
for which the wall chamber contains in its interior future-directed
lightlike vectors.  It is clear that this set is non-empty 
since the inequalities can be all fulfilled when the
couplings are zero (the $p_a$'s can be chosen to be positive
in the presence of a dilaton).

\subsection{Einstein-dilaton-Maxwell system in $D$ dimensions}
We consider first the case of a single 1-form in $D \geq 4$
dimensions.  This case is simple because the inequalities
(\ref{gravwalls}) are then consequences of (\ref{elecwalls})
and (\ref{magnwalls}), which read
\be
p_1 - \frac{\lambda}{2} A \geq 0 , \; \; \; \; \;
p_1 + p_2 + \cdots + p_{d-2} + \frac{\lambda}{2} A \geq 0.
\ee
Furthermore, the number of faces of the wall chamber (defined
by these inequalities and (\ref{symmwalls})) is exactly $D$
and the edge vectors form a basis.  Thus, the analysis of
subsection \ref{criticalvalue} can be repeated.

A basis of edge vectors can be taken to be
\begin{eqnarray}
&{}& (0,0, \cdots, 0,1,0) \label{first} \\
&{}& \big(-\frac{d-k-2}{k+1}, \cdots, -\frac{d-k-2}{k+1}, 1, \cdots, 1,
-\frac{2(d-k-2)}{\lambda(k+1)} \big), \; \; k = 1,2, \cdots, d-2
\label{second} \\
&{}& (1,1, \cdots, 1, \frac{2}{\lambda}) \label{nexttolast} \\
&{}& (1,1, \cdots, 1, -\frac{2(d-2)}{\lambda}) \label{last}
\end{eqnarray}
In (\ref{second}), the first $k$ components are equal to
$-\frac{d-k-2}{k+1}$ and the next $d-k$ components are equal to $1$.

The first vector is lightlike. The $k$-th vector in the group
(\ref{second})  has squared norm 
\be
-\frac{(d-1)[k^2 -k(d-3)+d]}{(k+1)^2} + \frac{4(d-k-2)^2}
{\lambda^2 (k+1)^2} , \; \; \; \; \; k = 1,2, \cdots, d-2
\label{nc2a}
\ee
while (\ref{nexttolast}) and (\ref{last}) have norm squared equal to
\be
-d(d-1) + \frac{4}{\lambda^2}
\label{nc2b}
\ee
 and 
\be
-d(d-1) +\frac{4(d-2)^2}{\lambda^2},
\label{nc2c}
\ee
respectively.  The subcritical values of $\lambda$ must
(by definition) be such that at least one of the expressions 
(\ref{nc2a}), (\ref{nc2b}) or (\ref{nc2c}) is positive.  To determine
the boundaries $\pm \lambda_c$ of the subcritical interval,
we first note that (\ref{nc2b}) is positive whenever 
$\vert \lambda \vert < \Lambda_1$, with $\Lambda_1 = 2/\sqrt{d(d-1)}$.
Similarly, (\ref{nc2c}) is positive whenever 
$\vert \lambda \vert < \Lambda_2$ with $\Lambda_2
= 2(d-2)/\sqrt{d(d-1)}$.  To analyse the sign of (\ref{nc2a}),
we must consider two cases, according to whether $k^2 -k(d-3)+d$
is positive or negative.

If $d< 9$, the factor $k^2 -k(d-3)+d$ is always positive
(for any choice of $k$, $k = 1, 2, \cdots, d-2$) and the
expression (\ref{nc2a}) is positive provided 
$\vert \lambda \vert < \Pi_k$, with
\be
\Pi_k = \frac{2(d-k-2)}{\sqrt{(d-1)[k^2 -k(d-3)+d]}}. 
\ee
The critical value $\lambda_c$ is equal to the
largest number among $\Lambda_1$, $\Lambda_2$ and $\Pi_k$.
This largest number is $\Lambda_2$ for $d=3,4,5,6$,
$\Pi_1$ for $d=7$ and $\Pi_2$ for $d=8$.
We thus have the following list of critical couplings:
\begin{eqnarray}
&{}& \lambda_c = \sqrt{\frac{2}{3}}, \; \; \; \; \; \; \; d = 3 \nn \\
&{}& \lambda_c = \frac{2}{\sqrt{3}}, \; \; \; \; \; \; \; d = 4 \nn \\
&{}& \lambda_c = \frac{3}{\sqrt{5}} , \; \; \; \;\; \;  \; d = 5 \nn \\ 
&{}& \lambda_c = \frac{4\sqrt{2}}{\sqrt{15}}, \; \; \; \; \; \; \; d = 6 \nn \\
&{}& \lambda_c = \frac{2\sqrt{2}}{\sqrt{3}}, \; \; \; \; \; \; \; d = 7 \nn \\
&{}& \lambda_c = \frac{4\sqrt{2}}{\sqrt{7}}, \; \; \; \; \; \; \; d = 8.
\label{6.16}
\end{eqnarray}
Note that the value of the dilaton coupling that comes from
dimensional reduction of vacuum gravity in one dimension higher
\be
\lambda_{KK} = 2 \sqrt{\frac{d}{d-1}}
\ee
is always strictly greater than the critical value,
except for $d=8$, where $\lambda_{KK} = \lambda_c$.   
(The corresponding values of the Kasner exponents
are those of the point on the Kasner
sphere exhibited in
\cite{DHS} for $D=10$, where all gravitational inequalities are
marginally fulfilled.)

If $d \geq 9$, the factor $k^2 -k(d-3)+d$ is non-positive 
for 
\be
\frac{d-3 - \sqrt{(d-9)(d-1)}}{2}
\leq k \leq
\frac{d-3 + \sqrt{(d-9)(d-1)}}{2}
\ee
(this always occurs for $k=3$).  Thus, the expression (\ref{nc2a})
is positive for such $k$'s no matter what $\lambda$ is.  This
implies that the critical value of $\lambda$ is infinite,
\be
\lambda_c = \infty, \; \; \; \; \; \; \; d \geq 9.
\label{6.19}
\ee
The fact that $D=10$ appears as a critical dimension
for the Einstein-dilaton-Maxwell system,
above which the system is velocity-dominated no matter what the
value of the dilaton coupling is in the line 
of the findings of \cite{DHS}, since the edges (\ref{second}) 
differ from those
of the pure gravity wall chambers only by an additional component
along the spacelike dilaton direction.

\subsection{Einstein-dilaton system with one $p$-form ($p \not=0$,
$p \not= D-2$)}
The same geometrical procedure for determining
the critical values of the dilaton couplings 
can be followed when there
is only one $p$-form in the system
($p \not=0$, $p \not= D-2$), because in that case
the wall chamber has exactly $D$ faces and the edge vectors 
form a basis.  Indeed, the gravitational inequalities
(\ref{gravwalls}) are always consequences of the
symmetry inequalities (\ref{symmwalls}) and the form
inequalities (\ref{elecwalls}) and (\ref{magnwalls})
(for $n_j \not= 0$ and $n_j \not= D-2$),
\be
2 p_1 + p_2 + \cdots + p_{d-2}
= (p_1 + \cdots + p_{n_j} - \frac{\lambda_j}{2}A)
+ (p_1 + p_{n_j + 1} + \cdots + p_{d-2} + \frac{\lambda_j}{2}A).
\ee
So, if there is only one $p$-form
(with $p \not=0$ and $p \not= D-2$), the $D-2$ symmetry 
inequalities (\ref{symmwalls}) together with the two form
inequalities (\ref{elecwalls}) and (\ref{magnwalls}) 
completely define the wall chamber, which has $D$ faces.
We shall not provide an explicit example of a calculation of
$\lambda_c$ for such a system, since it proceeds as for a
1-form. 

When there is more than one exterior form, one can still drop
the gravitational inequalities
(if there is at least one $p$-form with
$p \not=0$ and $p \not= D-2$), but the situation is more
involved because the inequalities corresponding
to different forms are usually independent, so that the
wall chamber has more than $D$ faces (its intersection with
the hyperplane $\sum{p_a} = 1 $ is not a simplex).  The
calculation is then more laborious.  The same feature
arises for a $0$-form, which we now examine.

\subsection{0-form in 4 dimensions}

We consider the case of a $0$-form
in $4$ spacetime dimensions.  As explained
above, we impose the condition $\lambda \not=0$
to the corresponding dilaton coupling\footnote{The case 
$\lambda = 0$ is clearly in the subcritical region but must
be treated separately because there are then two dilatons.  The
Kasner conditions read $p_1 + p_2 + \cdots + p_d = 1$
and $p_1^2 + \cdots p_d^2 + A_1^2 + A_2^2 = 1$, where the scalar
fields behave as $\phi_1 \sim A_1 \ln t$, $\phi_2 \sim
A_2 \ln t$.  This allows positive $p_i$'s, which
enables one to drop spatial derivatives as $t \rightarrow 0$.
The system is velocity-dominated.}.
Without loss of generality (in view of the $\phi \to - \phi$ symmetry),
 we can assume
$\lambda >0$.  The inequalities defining the subcritical
domain relevant to the
$0$-form case can be brought to the form
\begin{eqnarray}
& &p_1 >0 \label{ineq0fa}\\
& & A >0 \label{ineq0fb} \\
& & p_1 + p_2 - \frac{\lambda}{2} A >0 \label{ineq0fc} \\
& & p_2 - p_1 >0 \label{ineq0fd} \\
& & p_3 - p_2 >0 \label{ineq0fe}
\end{eqnarray}
We denote by $\alpha$, $\beta$, $\gamma$, $\delta$
and $\epsilon$ the corresponding border hyperplanes
({\it i.e.,} $\alpha : p_1 = 0$, $\beta : A =0$ etc).
The inequalities (\ref{ineq0fa})--(\ref{ineq0fe}) guarantee
that all potential walls are negligible asymptotically.
They are independent.  The five faces $\alpha$, $\beta$, $\gamma$, $\delta$
and $\epsilon$ intersect along the 7 one-dimensional edges
generated by the vectors:
\begin{eqnarray}
& & e_1 = (0,0,1,0) \in \alpha \cap \beta \cap \gamma =
\alpha \cap \beta \cap \delta = \alpha \cap \gamma \cap \delta
= \beta \cap \gamma \cap \delta
\label{e1e1} \\
& & e_2 = (0,1,1,0) \in \alpha \cap \beta \cap \epsilon
\label{e2e2} \\
& & e_3 = (0,1,1, \frac{2}{\lambda}) \in \alpha \cap \gamma
\cap \epsilon \label{e3e3} \\
& & e_4 = (0,0,0,1) \in \alpha \cap \delta \cap \epsilon
\label{e4e4} \\
& & e_5 = (-1,1,1,0) \in \beta \cap \gamma  \cap \epsilon
\label{e5e5} \\
& & e_6 = (1,1,1,0) \in \beta \cap \delta \cap \epsilon
\label{e6e6} \\
& & e_7 = (1,1,1,\frac{4}{\lambda}) \in \gamma  \cap
\delta \cap \epsilon \label{e7e7}
\end{eqnarray}                  
Among these vectors, neither $e_4$ nor $e_5$ bound
the subcritical domain since $e_4$ is such that
$p_1 + p_2 - (\lambda/2) A <0$ (changing its sign
would make $A<0$), while $e_5$ is such that $p_1 <0$
(changing its sign would make $p_2 - p_1 <0$).

The edge-vectors $\{e_1, e_2 ,e_3, e_6, e_7\}$
form a complete (but not linearly independent) set.
Any vector can be expanded as
\be
v = v_1 e_1 + v_2 e_2 + v_3 e_3 + v_6 e_6 + v_7 e_7
\label{expansion}
\ee
The coefficients $v_1$, $v_2$, $v_3$, $v_6$, $v_7$
are not independent but can be changed as
\be
v_2 \rightarrow v_2 +2 k, \; 
v_3 \rightarrow v_3 - 2 k , \; 
v_6 \rightarrow v_6 - k, \;
v_7 \rightarrow v_7 + k \;
\label{redef1}
\ee
For (\ref{expansion}) to be interior to the wall chamber, the
coefficients $v_1$, $v_2$, $v_3$, $v_6$ and $v_7$ must fulfill
\be
v_1 >0, \; \; v_2+v_3 >0,  \; \; 
\; v_2+ 2 v_6 >0, \;  
\; v_3+ 2 v_7 >0, \; \;
v_6 + v_7 >0. 
\label{wallcone0F}
\ee
Using the above redefinitions, which leave the inequalities
invariant, we can make $v_A \geq 0$, $A = 1,2,3,6,7$, with
at most two $v_A$'s equal to zero.
Indeed, let $s = \min(v_2, v_3, 2v_6, 2v_7)$.  Assume for
definiteness that $s = v_2$ (the other cases are treated
in exactly the same way).  One has then $v_2 \leq 2 v_7$.
Take $2k = -s$
in the redefinitions (\ref{redef1}). This makes $v_2$
 equal to zero and makes
$v_7$ equal to $v_7 -(v_2/2) \geq 0$. 
Because
of (\ref{wallcone0F}), the new $v_3$ and $v_6$ are strictly
positive, as claimed.
Thus, one sees that any vector in the wall chamber can be expanded
as in (\ref{expansion}) with non-negative coefficients.  But
the vectors $e_1$, $e_2$, $e_3$, $e_6$ and $e_7$ are
all future-pointing and timelike or null when
$\lambda \geq \sqrt{8/3}$.  It follows that for such $\lambda $'s, there
is no lightlike direction in the interior of the wall chamber.
Conversely, if $\lambda < \sqrt{8/3}$, the vector $e_7$ is
spacelike and one can find an interior vector
$\a e_1 + \b e_2 + e_7$ ($\a, \b >0$) that is lightlike.  
We can thus conclude:
\be
\lambda_c = \sqrt{\frac{8}{3}} \; \; 
\; \hbox{ for a 0-form in 4 dimensions,}
\ee
i.e., the system is velocity-dominated for $\vert \lambda \vert
< \sqrt{8/3}$.

The action for the matter fields in the case of a 0-form $A$ coupled to
a dilaton $\phi$ is
\be
S_\phi[g_{\alpha \beta}, \phi, A]  = - {1 \over 2} \int
(\partial_\mu \phi \, \partial^\mu \phi  
+e^{\lambda\phi} \partial_\mu A \, \partial^\mu A)\sqrt{-g} \, d^4 x
\ee 
Note that this is the action for a wave map (also known as a nonlinear 
$\sigma$-model or hyperbolic harmonic map) with values in a two-dimensional
Riemannian manifold of constant negative curvature. Its curvature is 
proportional to $\lambda^2$. Thus we obtain an interesting statement on
velocity-dominated behaviour for the Einstein equations coupled to certain
wave maps. 
Note for comparison that wave maps in flat space occurring naturally
in the context of solutions of the vacuum Einstein equations with 
symmetry, for instance in Gowdy spacetimes ({\it cf.} \cite{BG}),
are defined by a Lagrangian of the above type (using the flat metric)
with $\lambda=2$.

\subsection{Collection of 1-forms}

We now turn to a system of several 1-forms.  It is clear
that if these have all the same dilaton coupling, as in
the Yang-Mills action (\ref{actionym}), then, the
critical value of $\lambda$ is just that computed in
(\ref{6.16}) and (\ref{6.19}) since each
form brings in the same walls.
The situation is more complicated if the dilaton couplings
are different.  One could naively think that the subcritical
domain is then just the Cartesian product of the individual
subcritical intervals $[- \lambda_c^{(j)}, \lambda_c^{(j)}]$, but this
is not true because the intersection of the wall chambers
associated with each 1-form may have no interior lightlike direction,
even if each wall chamber has some.

This is best seen on the example of two 1-forms in $D$
spacetime dimensions with opposite dilaton couplings.  The
relevant inequalities, from which all others follow, are
in this case
\begin{eqnarray}
& & p_1 - \frac{\lambda}{2} A >0 , \; \; \; \;
p_1 + \frac{\lambda}{2} A >0 \label{6.37}\\
& & p_1 < p_2 < \cdots <p_d
\end{eqnarray} 
and can be easily analysed because they determine, in
this particular instance,  a simplex
in the hyperplane $\sum p_a = 1$.  It follows from (\ref{6.37})
that $p_1 >0$.  The edge-vectors can be taken to be
$(0, \cdots, 0, 1, \cdots, 1,0)$ ($k$ zeros, $d-k$ ones, $k =
1, \cdots, d-1$) and $(1,1, \cdots, 1,\pm 2/\lambda)$.
The first $d-1$ edge-vectors are timelike or null, while the
last two are spacelike provided $-d(d-1)\lambda^2 + 4 >0$.
This yields
\be
\lambda_c = \frac{2}{\sqrt{d(d-1)}} \; \; \;
\hbox{ for two 1-forms with opposite dilaton couplings}
\ee
Accordingly, $\lambda_c$  is finite for any spacetime dimension
(and in fact, tends to zero as $d \rightarrow \infty$),
even though $\lambda_c = \infty$ for a single 1-form whenever
$d>8$.

\section{Coupling between the matter fields}\label{couple}
\setcounter{equation}{0}
\setcounter{theorem}{0}
\setcounter{lemma}{0}      

The actions for the bosonic sectors of the low-energy limits
of superstring theories or M-theory contain coupling terms
between the $p$-forms, indicated by ``more" in (\ref{001}).
These coupling terms are of the Chern-Simons or the 
Chapline-Manton type.
In this section, we show that these terms
are consistent
with the results obtained in section~\ref{nform}, in
that they are also asymptotically negligible in the
dynamical equations of motion when
the Kasner exponents are subject
to the above inequalities (\ref{symmwalls})--(\ref{magnwalls}).

More precisely, the form of
the velocity-dominated evolution equations and solutions are
in each case exactly as in section~\ref{nform}.
The velocity-dominated matter constraints have additional
terms, but as before, the velocity-dominated matter
variables (besides the dilaton) are constant in time,
so if the constraints are satisfied at some $t > 0$ they
are satisfied for all $t > 0$.  The quantities $^0\rho$
and $^0j_a$ are defined exactly as in section~\ref{nform}.
Since the velocity-dominated evolution equations are
also the same, there is nothing additional to check
concerning the velocity-dominated Hamiltonian and
momentum constraints. 

Turning now to the exact equations, the restrictions defining the
set $V$ are unchanged from section~\ref{nform}.
The form of the evolution equation for the dilaton is
unchanged.  The form of the stress-energy tensor is
also unchanged, and so the form of the Einstein evolution
equations and the Einstein constraints is unchanged.
The additional matter field variables considered in
section~\ref{nform} are still all $O(1)$,
so estimates of terms involving the matter fields do not
need to be reconsidered, as long as their form has not
changed, for instance, in the argument that the
Einstein constraints are satisfied.
That the matter constraints are satisfied follows
as in the other cases, once it is verified that
their time derivative vanishes and that they are $o(1)$.
Since so much of the argument is identical to that of
section~\ref{nform}, we only point out the few places
where there are differences.

\subsection{Chern-Simons terms}

First we consider the coupling of $i$ of the additional matter
fields via a Chern-Simons term in the action.  These additional
matter fields should be such that
\begin{equation}
\label{formscs}
i-1 + \sum_{j=1}^i n_j = D.
\end{equation}
The Chern-Simons term which is added to the action is
\begin{equation}
\label{actioncs}
S_{\rm CS}[A^{(1)}_{\gamma_1 \cdots \gamma_{n_1}},
\cdots,A^{(i)}_{\gamma_1 \cdots \gamma_{n_i}}] =
\int A^{(1)} \wedge \, d A^{(2)}
\wedge \cdots \wedge \, d A^{(i)}.
\end{equation}
The variation of this term with respect to both the
metric and the dilaton field, $\phi$, vanishes.
The matter equation (\ref{matter2nf}) is
unchanged, since it is still the case that
$F^{(j)} = d A^{(j)}$
for all $j$.  But the equation (\ref{matter1nf})
for each of the $i$ coupled matter fields acquires a
non-vanishing right hand side.
\begin{equation}
\label{mattercs}
^{(D)}\nabla_\mu (F^{(j) \mu \nu_1 \cdots
\nu_{n_j}} e^{\lambda_j \phi}) \sqrt{-g} =
C_j \, \epsilon^{\cdots  \nu_1 \cdots \nu_{n_j} \cdots}
F^{(1)}_{\cdots } \cdots F^{(j-1)}_{\cdots }
F^{(j+1)}_{\cdots } \cdots F^{(i)}_{\cdots }
\end{equation}
Here $\epsilon^{0 ... d} = 1$ and $C_j$ is a numerical factor.
Next, considering the
$d+1$ decomposition of equation~(\ref{mattercs}), the constraint
equation~(\ref{elconstraintnf}), for the $j$th coupled matter
field, acquires the following term on its right hand side,
\begin{equation}
\label{elconstraintcs}
-C_j \epsilon^{\cdots 0 b_1 \cdots}
F^{(1)}_{\cdots} \cdots F^{(j-1)}_{\cdots }
F^{(j+1)}_{\cdots } \cdots F^{(i)}_{\cdots }
\end{equation}
Here all indices which are not explicit are spatial.  So,
only magnetic fields appear in (\ref{elconstraintcs}).
The following term is added to the right hand side of the
evolution equation~(\ref{elevnf}) for the $j$th
coupled matter field.
\begin{eqnarray}
\label{elevcs}
& & - C_j \Big\{\sum_{m =1}^{j-1} (n_m + 1)
\epsilon^{\cdots 0 c_1 \cdots c_{n_m}
\cdots a_1 \cdots a_{n_j} \cdots }
F^{(1)}_{\cdots } \cdots \nonumber \\ & &
\times \Big({ 1 \over \sqrt{g} }
 g_{c_1 h_1} \cdots g_{c_{n_m} h_{n_m}}
{\cal{E}}^{(m) h_1 \cdots h_{n_m}} e^{-\lambda_m \phi} \Big)
\cdots F^{(j-1)}_{\cdots } F^{(j+1)}_{\cdots }
\cdots F^{(i)}_{\cdots } \Big\} \nonumber \\ & &
-  C_j \Big\{ \sum_{m =j+1}^{i} (n_m + 1)
\epsilon^{\cdots a_1 \cdots a_{n_j} \cdots
0 c_1 \cdots c_m \cdots} F^{(1)}_{\cdots }
\cdots F^{(j-1)}_{\cdots } F^{(j+1)}_{\cdots }
\cdots \\ & & \times \Big({ 1 \over \sqrt{g} }
 g_{c_1 h_1} \cdots g_{c_{n_m} h_{n_m}}
{\cal E}^{(m) h_1 \cdots h_{n_m}} e^{-\lambda_m \phi} \Big)
\cdots F^{(i)}_{\cdots }\Big\}. \nonumber
\end{eqnarray}
Again, all indices which are not explicit are spatial.
There is in each term only one electric field.

The velocity-dominated matter constraint equations for
the $j$th coupled matter field can be obtained
from the ``full'' matter constraint equations for
the same field by substituting the velocity-dominated
quantities for all variables.

The only additional terms occurring in $f$ are due to
equation~(\ref{elevcs}).  The form of the $m$th
term on the right hand side of equation~(\ref{elevcs})
is just like the form of the terms on the right hand side of
equation~(\ref{magevnf}) for the $m$th coupled field.
The factors which differ, comparing the $m$th term
of (\ref{elevcs}) to equation~(\ref{magevnf})
for the $m$th field, are $0(1)$.  Since in both cases a
factor of $t^{1 - \beta}$ is added in order to obtain
the terms appearing in $f$, the estimate that the
additional terms in $f$ due to the Chern-Simons
coupling are $O(t^\delta)$ is obtained just as the
corresponding previously obtained estimates.

\subsection{Chapline-Manton couplings}

Next we consider Chapline-Manton couplings.
For definiteness, we treat two explicit examples, leaving to the
reader the task of checking that the general case
works in exactly the same way.
The first coupling is between
an $n$-form $A$ and an $(n+1)$-form $B$ and is
equivalent to making $B$ massive.  Let
$F = dA + B$ and $H = dB$.  The gauge transformations
are $ B \rightarrow B + d \eta $, for arbitrary
$n$-form $\eta$, and $A \rightarrow A - \eta + d \gamma$,
for arbitrary $(n-1)$-form $\gamma$.  (If $n=0$, then $d \gamma$ is
replaced by a constant scalar and we require that
the corresponding constant, $\lambda_A$, in the coupling to
the dilaton be nonzero.)  The form of the action is
the same as in section~\ref{nform}, but since $F$ now depends on
$B$ and not just on $A$, the variation of the action with
respect to $B$ acquires an additional term.  Also, it is now
the case that $dF = H$.

The matter equation~(\ref{matter1nf}) is unchanged for $F$
and equation~(\ref{matter2nf}) is unchanged for $H$.
Equation~(\ref{matter1nf}) for $H$
and equation~(\ref{matter2nf}) for $F$ are now as follows.
\begin{eqnarray}
\label{matter1nfcm1}
^{(D)}\nabla_\mu (H^{\mu \nu_0 \cdots \nu_n}
e^{\lambda_B \phi}) & = &
F^{\nu_0 \cdots \nu_n} e^{\lambda_A \phi}, \\
\label{matter2nfcm1}
^{(D)} \nabla_{[\mu} F_{\nu_0 \cdots \nu_n]} & = &
{1 \over (n+2)} H_{\mu \nu_0 \cdots \nu_n}.
\end{eqnarray}
Define ${\cal E}^{a_1 \cdots a_n} = \sqrt{g} \,
F^{0 a_1 \cdots a_n } e^{\lambda_A \phi}$ and
${\cal D}^{a_0 \cdots a_n} = \sqrt{g} \,
H^{0 a_0 \cdots a_n} e^{\lambda_B \phi}$.
The matter constraint equations which are affected are 
\begin{eqnarray}
\label{elconstraintcm1}
e_a ( {\cal D}^{a b_1 \cdots b_n}) +
f^c_{ca} \, {\cal D}^{a b_1 \cdots b_n} +
{1 \over 2} \sum_{i=1}^n f^{b_i}_{ac}
{\cal D}^{a b_1 \cdots c \cdots b_n} & = &
-{\cal E}^{b_1 \cdots b_n}, \\
\label{magconstraintcm1}
e_{[a}(F_{b_0 \cdots b_n]})
-{(n + 1) \over 2} f^c_{[a b_0}
F_{|c| b_1 \cdots b_n]}  & =  &
{1 \over n+2} H_{a b_0 \cdots b_n},
\end{eqnarray}
The additional term which appears on the
right hand side of equation~(\ref{elevnf}) for
${\cal D}^{a_0 \cdots a_n}$ is
\begin{equation}
\label{elevcm1}
\sqrt{g} g^{a_0 b_0} \cdots g^{a_n b_n}
F_{b_0 \cdots b_n} e^{\lambda_A \phi}.
\end{equation}
The additional term which appears on the
right hand side of equation~(\ref{magevnf}) for
$F_{a_0 \cdots a_n}$ is
\begin{equation}
\label{magevcm1}
{-1 \over  \sqrt{g}}
g_{a_0  b_0} \cdots g_{a_n b_n}
{\cal D}^{b_0 \cdots b_n} e^{-\lambda_B \phi}.
\end{equation}

The velocity-dominated matter constraint equations which are
affected can be obtained from equation~(\ref{elconstraintcm1})
and (\ref{magconstraintcm1}) by substituting the corresponding
velocity-dominated quantities for all variables.
The only additional terms occurring in $f$ are due to
equations~(\ref{elevcm1}) for $\cal D$ and
(\ref{magevcm1}) for $F$.  The form of the additional
terms in these equations is just like the form of the
terms which appear in equations~(\ref{elevnf})
for $\cal E$ and in (\ref{magevnf}) for $H$.  Therefore
the estimate that the additional terms are $O(t^\delta)$
is obtained just as the corresponding previously obtained
estimates. 

The second Chapline-Manton type coupling is between an
$n$-form $A$ and a $(2n)$-form $B$.  Let $F = d A $
and $H = dB + A \wedge F$.  The gauge transformations
are $ A \rightarrow A + d \gamma $, for arbitrary
$(n-1)$-form $\gamma$, and $B \rightarrow B + d \eta - \gamma \wedge F$,
for arbitrary $(2n-1)$-form $\eta$.  (If $n=0$ the gauge transformations
are $ A \rightarrow A + C$ and $B \rightarrow B + D - CA$ for constant
scalars $C$ and $D$ and we require both $\lambda_A \neq 0$ and
also $\lambda_B \neq 0$.)  The form of the action is again
the same as in section~\ref{nform}.
Define ${\cal E}^{a_1 \cdots a_n} = \sqrt{g} \,
F^{0 a_1 \cdots a_n } e^{\lambda_A \phi}$ and
${\cal D}^{a_1 \cdots a_{2n}} = \sqrt{g} \,
H^{0 a_1 \cdots a_{2n}} e^{\lambda_B \phi}$.
The matter equations~(\ref{matter1nf}) for $F$
and (\ref{matter2nf}) for $H$ are affected,
only if $n$ is odd.  The equation for $F$ which
is affected (if $n$ is odd)
and its $d + 1$ decomposition are
\begin{equation}
\label{matter1nfcm2}
^{(D)}\nabla_\mu (F^{\mu \nu_1 \cdots \nu_n}
e^{\lambda_A \phi}) =
{2 \over (n + 1)!} H^{\mu \nu_1 \cdots \nu_n
\sigma_1 \cdots \sigma_n} F_{\mu \sigma_1 \cdots \sigma_n}
e^{\lambda_B \phi},
\end{equation}
\begin{equation}
\label{elconstraintcm2}
e_a ( {\cal E}^{a b_2 \cdots b_n}) +
f^c_{ca} \, {\cal E}^{a b_2 \cdots b_n} +
{1 \over 2} \sum_{i=2}^{n} f^{b_i}_{ac}
{\cal E}^{a b_2 \cdots c \cdots b_n} = {2 \over (n+1)!}
{\cal D}^{a b_2 \cdots b_n h_1 \cdots h_n}
F_{a h_1 \cdots h_n },
\end{equation}
\begin{eqnarray}
\label{elevcm2}
\partial_t {\cal E}^{a_1 \cdots a_n} & = & \cdots
- { 2 \over \sqrt{g} \, n!}
{\cal D}^{a_1 \cdots a_n b_1 \cdots b_n}
g_{b_1 c_1} \cdots g_{b_n c_n} {\cal E}^{c_1 \cdots c_n}
 \\ & & \; \; +  {2 \over (n+1)!} \sqrt{g} g^{b h_0}
g^{a_1 h_1} \cdots g^{a_n h_n}
g^{c_1 h_{n +1}} \cdots g^{c_n h_{2n}}
H_{h_0 \cdots h_{2n}}
F_{b c_1 \cdots c_n} e^{\lambda_B \phi}. \nonumber
\end{eqnarray}
The equation for $H$ which is affected (if $n$ is odd)
and its $d+1$ decomposition are
\begin{equation}
\label{matter2nfcm2}
^{(D)} \nabla_{[\mu_0} H_{\mu_1 \cdots \mu_{2n+1}]} =
{(2n + 1)! \over (n+1)! (n+1)!}
F_{[\mu_0 \cdots \mu_n} F_{\mu_{n +1} \cdots \mu_{2n+1}]},
\end{equation}
\begin{equation}
\label{magconstraintcm2}
e_{[a}(H_{b_0 \cdots b_{2n}]})
-{(2n + 1) \over 2} f^c_{[a b_0}
H_{|c| b_1 \cdots b_{2n}]} =
{(2n + 1)! \over (n + 1)! (n + 1)!}
F_{[a b_0 \cdots b_{n-1}} F_{b_n \cdots b_{2n}]},
\end{equation}
\begin{equation}
\label{magevcm2}
\partial_t H_{a_0 \cdots a_{2n}}
= \cdots + {(2n + 2)! \over \sqrt{g} \,
(n + 1)! (n + 1)!}
g_{[a_0  |b_1|} \cdots g_{a_{n-1} |b_n|}
{\cal E}^{b_1 \cdots b_n} e^{-\lambda_A \phi}
F_{a_n \cdots a_{2n}]}. \nonumber
\end{equation}
The velocity-dominated matter constraint equations which are
affected can be obtained from equation~(\ref{elconstraintcm2})
and (\ref{magconstraintcm2}) by substituting
the corresponding
velocity-dominated quantities for all variables.
The only additional terms occurring in $f$ are due to
equations~(\ref{elevcm2}) for $\cal E$ and
(\ref{magevcm2}) for $H$.  Here again,
the estimate that the additional terms in $f$
are $O(t^\delta)$, is just as the estimate of
terms appearing already in section~\ref{nform},
either in equation~(\ref{elevnf}) for $\cal D$
or in equation~(\ref{magevnf}) for $F$.

\section{Yang-Mills}\label{yangmills}
\setcounter{equation}{0}
\setcounter{theorem}{0}
\setcounter{lemma}{0}      

We complete our analysis by proving that Yang-Mills couplings
also enjoy the property of not modifying the conclusions.
The action is 
(\ref{actionym}), with a Yang-Mills field as
source in addition to the scalar field considered
in section~\ref{scalar}
and with $\vert \lambda \vert < \lambda_c$.  The argument is again
based on that of sections~\ref{scaMax4D} --
\ref{nform} and it is enough here
to note differences.  The main one is that
one must work with the vector potential instead
of the fields themselves, because bare $A$'s
appear in the equations.  We could, in fact, have developed
the entire previous analysis in terms of the vector potentials,
thereby reducing the
number of matter constraint equations.
We followed a manifestly gauge-invariant approach for easing
the physical understanding, but this was not
mandatory.
The stress-energy tensor is
\begin{equation}
T_{\mu \nu} = \, ^{(D)}\nabla_\mu \phi \,
^{(D)}\nabla_\nu \phi - {1 \over 2} g_{\mu \nu} \,
^{(D)}\nabla_\alpha \phi \,
^{(D)}\nabla^\alpha \phi + [F_{\mu \alpha} \cdot {F_\nu}^\alpha
-{1 \over 4} g_{\mu \nu} F_{\alpha \beta} \cdot
F^{\alpha \beta} ] e^{\lambda \phi}.
\end{equation}
We work in the temporal gauge, $A_0 = 0$.
The matter fields satisfy the following equations.
\begin{equation}
^{(D)}\nabla_\alpha \, ^{(D)}\nabla^\alpha \phi - {\lambda \over 4}
 F_{\alpha \beta} \cdot F^{\alpha \beta}  e^{\lambda \phi} = 0
\end{equation}
\begin{equation}
^{(D)}\nabla_\mu (F^{\mu \nu} e^{\lambda \phi})
+ [A_\mu, F^{\mu \nu}] e^{\lambda \phi}= 0,
\end{equation}
\begin{equation}
\label{ffroma}
F_{\mu \nu } = \partial_\mu A_\nu - \partial_\nu A_\mu
+ [A_\mu,A_\nu].
\end{equation}
The Lie Bracket has no intrinsic time dependence.
The $d$+1 decomposition of the stress-energy tensor is expressed
in terms of the spatial tensor density
${\cal E}^a = \sqrt{g} \, F^{0a} e^{\lambda \phi}$
and the antisymmetric spatial tensor $F_{ab}$.
\begin{eqnarray}
\rho & = & {1 \over 2} \{ (\partial_t \phi)^2 + g^{ab} e_a(\phi) e_b(\phi)
+{1 \over g} g_{ab} {\cal E}^a \cdot
{\cal E}^b e^{-\lambda \phi} +
{1 \over 2} g^{ab} g^{ch} F_{ac} \cdot
F_{bh} e^{\lambda \phi} \}, \\
j_a & = & - \partial_t \phi \, e_a(\phi) +
{1 \over \sqrt{g} } {\cal E}^b \cdot F_{ab}, \\
\label{calcMym}
{M^a}_b & = & g^{ac} e_b(\phi) \, e_c(\phi)
- {1 \over g} \{ g_{bc} {\cal E}^a \cdot {\cal E}^c
- {1 \over 2} {\delta^a}_b g_{ch} {\cal E}^c \cdot {\cal E}^h \}
e^{-\lambda \phi} \nonumber \\ & & \; \;
+ \{ g^{ac} g^{hi} F_{ch} \cdot F_{bi}
- {1 \over 4} {\delta^a}_b g^{ch} g^{ij} F_{ci} \cdot
F_{hj} \} e^{\lambda \phi}.
\end{eqnarray}
The matter constraint equation is
\begin{equation}
\label{elconstraintym}
e_a ( {\cal E}^a) + f^b_{ba} \, {\cal E}^a + [A_a,{\cal E}^a] = 0.
\end{equation}
The matter evolution equations are
\begin{eqnarray}
\partial^2_t \phi - (\tr k) \partial_t \phi & = & g^{ab} \nabla_a \nabla_b \phi
+ {\lambda \over 2 g } g_{ab} {\cal E}^a \cdot
{\cal E}^b e^{-\lambda \phi}
- {\lambda \over 4 } g^{ab} g^{ch} F_{ac}
\cdot F_{bh} e^{\lambda \phi}, \\
\partial_t {\cal E}^a & = & e_b( \sqrt{g} g^{ac} g^{bh}
F_{ch} e^{\lambda \phi})
+ (f^i_{ib} g^{ac} + {1 \over 2} f^a_{bi} g^{ic})
\sqrt{g} g^{bh} F_{ch} e^{\lambda \phi} \\
\partial_t A_a & = & -{1 \over \sqrt{g}} g_{ab} {\cal E}^b e^{- \lambda \phi}.
\end{eqnarray}
Note that we use as basic matter variables $A_a$ and ${\cal E}^b$ (the
quantity $F_{ab}$ being then defined in terms of $A_a$ as
$F_{ab} = \partial_a A_b - \partial_b A_a
+ [A_a,A_b]$).

The Kasner-like evolution equations are
equations~(\ref{dtg0}) -- (\ref{dtE0}) and $\partial_t \, ^0 A_a = 0$.
We consider analytic solutions of the Kasner-like evolution equations
of the form (\ref{0kab1}) -- (\ref{0E1}) along with the
quantity $\, ^0A_a$ which is constant
in time.  Given a point $x_0 \in \Sigma$, we use an adapted
spatial frame on a neighbourhood of $x_0$, $U_0$, as in
section~\ref{vacuum}.  Thus, $^0g_{ab}(t_0)$ and ${K^a}_b$
are specialized as in that section.
There is one velocity-dominated matter constraint
equation, obtained from equation~(\ref{elconstraintym})
by replacing ${\cal E}^a$ and $A_a$ with
$^0 {\cal E}^a$ and $^0 A_a$.  If the velocity-dominated
matter constraint is satisfied at some time
$t_0 > 0$, then it is satisfied for all $t > 0$.  Define
\begin{eqnarray}
^0 \rho & = & {1 \over 2} (\partial_t \, ^0 \phi)^2, \\
^0 j_a & = & - \partial_t \, ^0 \phi \, e_a(\, ^0 \phi) +
{1 \over \sqrt{ \, ^0 g} } \, ^0 {\cal E}^b \,
\cdot \, ^0 F_{ab}.
\end{eqnarray}
The velocity-dominated Einstein constraints are defined
as in the other cases.  Equations~(\ref{dt0C}) and
(\ref{dt0Ca}) are again satisfied, so if the velocity-dominated
constraints are satisfied at some $t_0$, then they
are satisfied for all $t > 0$.  The restrictions defining
the set $V$ are as in section~\ref{nform}, with $n_j = 1$.
The relation of the unknown, $u$, in equation~(\ref{fuchs0})
to the Einstein-matter variables is given by
equations~(\ref{defgamma}) -- (\ref{defxi}) and
\begin{equation}
\label{defvarphiym}
A_a = \, ^0 A_a + t^\beta \varphi_a.
\end{equation}
The quantities $\cal A$ and $f$ in equation~(\ref{fuchs0})
are given by equations~(\ref{fuchgamma}) -- (\ref{fuchomega}) and
\begin{eqnarray}
\label{fuchchiym}
t \, \partial_t \chi + \beta \chi & = & t^{\alpha_0 - \beta}
(\tr \, \kappa) ( A + t ^\beta \chi ) + t^{2 - \beta}
\, ^S g^{ab} \, ^S \nabla_a \, ^S \nabla_b \, ^0 \phi
+ t^{2 - \zeta} \, ^S \nabla^a \omega_a
\nonumber \\ & & \; \;
+t^{2 - \beta} \{ {\lambda \over 2 g}
g_{ab} {\cal E}^a \cdot {\cal E}^b e^{-\lambda \phi}
- {\lambda \over 4} g^{ab} g^{ch} F_{ac} \cdot F_{bh}
e^{\lambda \phi} \}, \\
\label{fuchxiym}
t \, \partial_t \xi^a + \beta \xi^a & = &
t^{1 - \beta} \{
e_b( \sqrt{g} g^{ac} g^{bh} F_{ch} e^{\lambda \phi}) \nonumber \\
& & \; \; + (f^i_{ib} g^{ac} + {1 \over 2} f^a_{bi} g^{ic})
\sqrt{g} g^{bh} F_{ch} e^{\lambda \phi} \}, \\
\label{fuchvarphiym}
t \, \partial_t \varphi_a + \beta \varphi_a & = & - t^{1 - \beta}
{1 \over \sqrt{g}}  g_{ab} {\cal E}^b e^{-\lambda \phi}.
\end{eqnarray}
The estimate that $f = O(t^\delta)$ is obtained as before,
using ${\cal E}^a = O(1)$ and $F_{ab} = O(1)$.
The matter constraint quantity,
the left hand side of equation~(\ref{elconstraintym}), is $o(1)$
and its time derivative vanishes,
so the matter constraint is satisfied.
The estimate of the matter terms in the Einstein
constraints is obtained as in section~\ref{nform}
for $n_j = 1$.

To conclude: the whole analysis goes through even
in the presence of the Yang-Mills coupling terms and the
system is asymptotically Kasner-like provided 
$\vert \lambda \vert < \lambda_c$, where $\lambda_c$
is the same as in the abelian case and explicitly given by
(\ref{6.16}) and (\ref{6.19}).

\section{Self-interacting scalar field}
\setcounter{equation}{0}
\setcounter{theorem}{0}
\setcounter{lemma}{0}      
\label{nonlinear}
Consider Einstein's equations, $D \geq 3$, with sources as in
sections~\ref{scalar}, \ref{nform},
\ref{couple} or \ref{yangmills}, except that the
 massless scalar field, $\phi$, is replaced by a
self-interacting scalar field.  That is, the expression~(\ref{actionnl})
is added to the action. Solutions with a monotone singularity
can be constructed as in sections~\ref{scalar} -- \ref{yangmills},
with assumptions regarding the function $V(\phi)$ which appears
in (\ref{actionnl}) given below. There is no change in the
velocity-dominated evolution equations and solutions, nor in
the velocity-dominated constraints.  The only change to
equation~(\ref{fuchs0}) is that two new terms
appear in $f$.  There is a new term,
$t^{2 - \alpha_0} {\delta^a}_b \, 2 \, V(\phi) / (D-2)$,
on the right hand side of the evolution equation for ${\kappa^a}_b$ 
(through ${M^a}_b$).  There is also a new term,
$ - t^{2 - \beta} V'(\phi)$, on the right hand side of the evolution
equation for $\chi$.  For equation~(\ref{fuchs0})
to be Fuchsian, it must be the case that $f = O(t^\delta)$
and, in addition, that $f$ satisfy other regularity
conditions \cite{AR,KR}.  

Some examples were considered in \cite{R00b}. A trivial example is
obtained by taking $V$ to be a constant. Then the equation for the
scalar field is not changed by the potential while its effect on
the Einstein equations is equivalent to the addition of a 
cosmological constant. Thus we see that the analysis of \cite{AR}
generalizes directly to the case of the Einstein-scalar field
system with non-zero cosmological constant. Of course the analogous
statement applies to the other dimensions and matter fields 
considered in previous sections. To get another simple example
take $V(\phi)=\lambda\phi^p$ for a constant $\lambda$ and an integer 
$p\ge 2$. Showing that the equation is Fuchsian involves examining 
the expression
\be
V(A\ln t+B+t^\beta\psi)=\lambda(A\ln t+B+t^\beta\psi)^p
\ee
and corresponding expressions for the first and second derivatives
of $V$. Of course in this particular case these are given by 
multiples of smaller powers of $t$. The aim is to estimate these
quantities by suitable powers of $t$. In this case a Fuchsian system
is always obtained. A linear massive scalar field is obtained by 
choosing $p=2$. Another interesting possibility is to choose 
$V(\phi)=e^{\lambda\phi}$ for a constant $\lambda$, in which 
case the derivatives of $V$ are also exponentials. Then
\be
V(A\ln t+B+t^\beta\psi)=e^{\lambda B} t^{\lambda A}
\exp (\lambda t^\beta \psi) 
\ee
Note that such an exponential potential can be (formally) 
generated by adding,
as matter field, a $d-$form $A_{\mu_1 \cdots \mu_d}$ with dilaton
coupling $\lambda_d = - \lambda$. Indeed, eliminating the field-strength
$ F = d A$ (which satisfies $e^{\lambda_d \phi} F = C \eta $, where 
$C$ is a constant and $\eta$ the volume form),
leads to a term in the action proportional to 
$e^{- \lambda_d \phi} C^2$. 
A Fuchsian system is obtained provided the general ``electric'' 
$p-$form condition (\ref{inequalitiesnf}) (with $n_j=d$),
$2 p_1 + \cdots + 2p_d - \lambda_d A > 0 $ is satisfied,
 i.e. (after using 
$p_1 + \cdots + p_d = 1$ and $ \lambda_d = -\lambda $)
 provided $ \lambda A>-2$. This 
therefore yields a restriction on the data.

More generally, it is enough to have a function $V$ on the real line which
has an analytic continuation to the whole complex plane and which satisfies
estimates of the form
\begin{eqnarray}
\label{restrictV}
t^{2 - c_1} \tilde V(\tilde A \ln t + \tilde B
+t^\beta \tilde \psi) & = & O(1), \nonumber \\
t^{2 - c_2} \tilde V'(\tilde A \ln t + \tilde B
+t^\beta \tilde \psi) & = & O(1), \\
t^{2 - c_3} \tilde V''(\tilde A \ln t + \tilde B
+t^\beta \tilde \psi) & = & O(1), \nonumber
\end{eqnarray}
for some positive numbers $c_1$, $c_2$ and $c_3$.
Here $\tilde A$ and $\tilde B$ are the analytic continuations
of $A(x)$ and $B(x)$, to some (small, simply connected) complex 
neighbourhood of the range of a coordinate chart.  And
$\tilde \psi$ lies in some region of the complex plane
containing the origin. For $f$ to be regular, it must be the case 
that $c_1 \geq \alpha_0$ and $c_2 \geq \beta$, which can
be achieved by reducing $\epsilon$, if necessary,
and also possibly $U_0$, so that previous assumptions
are satisfied. By taking suitable account of the domains of the
functions involved it is also possible to obtain an analogue of
this result when the functions $V$ and $\tilde V$ are only defined 
on some open subsets of $\bf R$ and $\bf C$.

The only other change to the construction given in
sections~\ref{scalar} -- \ref{yangmills}
is that $\rho \rightarrow \rho + V(\phi)$.
It is still the case that $^{(D)} \nabla_\mu T^{\mu \nu} = 0$,
so equations~(\ref{fuchsianC}) and (\ref{fuchsianCa}) are
satisfied.  Equation (\ref{estconstraints1}) is satisfied
due to the assumptions concerning $V(\phi)$,
so the Einstein constraints are satisfied.

\section{Conclusions}
\setcounter{equation}{0}
\setcounter{theorem}{0}
\setcounter{lemma}{0}      
\label{conclusions}

Our paper establishes the Kasner-like behaviour for
vacuum gravity in spacetime dimensions greater than
or equal to $11$, as well as the Kasner-like behaviour
for the Einstein-dilaton-matter systems with
subcritical dilaton couplings.  Our results can be summarized
as follows

\begin{theorem}\label{thvacuum}
Let $\Sigma$ be a d-dimensional analytic manifold, $d \geq 10$
and let $(\,^0g_{ab}, \, ^0k_{ab})$ be a $C^\omega$
solution of the Kasner-like vacuum Einstein equations on
$(0,\infty) \times \Sigma$ such that $t \, \tr \,^0k=-1$ and such
that the ordered eigenvalues of $-t \, ^0k_{ab}$
satisfy $1 + p_1 - p_d - p_{d-1} > 0$.

Then there exists an open neighbourhood $U$ of
$\{0\} \times \Sigma$ in $[0,\infty) \times \Sigma$ and
a $C^\omega$ solution
$(g_{ab},k_{ab})$ of the Einstein vacuum
field equations on $U\cap ((0,\infty) \times \Sigma)$ such
that for each compact subset $K\subset \Sigma$
there are positive real numbers $\alpha^a_{\ b}$
for which the following estimates hold uniformly on $K$:

\begin{enumerate}
\item
$^0g^{ac}g_{cb}=\delta^a_{\ b}+o(t^{\alpha^a_{\ b}})$
\item
$k^a_{\ b}=\,^0k^a_{\ b}+o(t^{-1+\alpha^a_{\ b}})$
\end{enumerate}
\end{theorem}

\begin{theorem}\label{thscalarmaxwell}
Let $\Sigma$ be a d-dimensional analytic manifold,
$d \geq 2$ and let \hfil\break
$X=(\,^0g_{ab}, \, ^0k_{ab}, \, ^0 \phi ,
\, ^0 {\cal E}^{(j) a_1 \cdots a_{n_j}},
\, ^0 F^{(j)}_{a_0 \cdots a_{n_j}})$,
with $j$ taking on values 1 through $k$ for some
nonnegative integer $k$ (possibly 0, in which case
$j$ takes on no values), $0 \leq n_j \leq d-1$.
Let $\lambda_j$ be constants in the subcritical range.
Let $X$ be a $C^\omega$ solution of the Kasner-like
Einstein-matter equations on $(0,\infty) \times \Sigma$
such that $t \, \tr \,^0k=-1$ and such that
the ordered eigenvalues of $-t \, ^0k_{ab}$
satisfy $1 + p_1 - p_d - p_{d-1} > 0$
and, for each $j$, $2p_1 + \cdots + 2p_{n_j}
- \lambda_j \, t \, \partial_t \, ^0 \phi > 0$
and $2p_1 + \cdots + 2p_{d-n_j-1} +
\lambda_j \, t \, \partial_t \, ^0 \phi > 0$.

Then there exists an open neighbourhood $U$ of
$\{0\} \times \Sigma$ in $[0,\infty) \times \Sigma$ and
a $C^\omega$ solution
$(g_{ab}, k_{ab}, \phi,
{\cal E}^{(j) a_1 \cdots a_{n_j}},
F^{(j)}_{a_0 \cdots a_{n_j}})$ of the Einstein-matter
field equations on $U\cap ((0,\infty) \times \Sigma)$ such
that for each compact subset $K\subset \Sigma$
there are positive real numbers  $\beta, \alpha^a_{\ b}$,
with $\beta < \alpha^a_{\ b}$,
for which the following estimates hold uniformly on $K$:

\begin{enumerate}
\item
$^0g^{ac}g_{cb}=\delta^a_{\ b}+o(t^{\alpha^a_{\ b}})$
\item
$k^a_{\ b}=\,^0k^a_{\ b}+o(t^{-1+\alpha^a_{\ b}})$ 
\item
$\phi=\, ^0\phi+o(t^\beta)$
\item
${\cal E}^{(j) a_1 \cdots a_{n_j}} =
\, ^0{\cal E}^{(j) a_1 \cdots a_{n_j}} + o(t^\beta)$
\item
$F^{(j)}_{a_0 \cdots a_{n_j}} =
\, ^0 F^{(j)}_{a_0 \cdots a_{n_j}} + o(t^\beta)$
\end{enumerate}
\end{theorem}

\noindent
{\bf Remarks} 
\begin{enumerate}
\item Corresponding estimates hold for certain first order
derivatives of the basic unknowns in Theorems \ref{thvacuum} and 
\ref{thscalarmaxwell} ({\it cf.} Theorem 2.1 in \cite{AR}). These are
the derivatives which arise in the definition of new unknowns when 
second order equations are reduced to first order so as to
produce a first order Fuchsian system.

\item Our analysis shows that a solution of the full subcritical 
Einstein-matter equations satisfying the estimates given in the theorems
and the corresponding estimates for first order derivatives just
mentioned is uniquely determined by the solution of the 
velocity-dominated equations (the integration functions are included in 
the zeroth order, Kasner-like solutions; the deviation from them is 
uniquely determined).

\item The Einstein-matter field equations may include
interaction terms of Chern-Simons, Chapline-Manton
and Yang-Mills type, and the scalar field may
be self-interacting,
with assumptions on $V(\phi)$ as
stated in section~\ref{nonlinear}.
If the $j$th field is a Yang-Mills field, then
$F^{(j)}_{a b}$ is obtained from $A^{(j)}_a$ and
$^0F^{(j)}_{a b}$ is obtained from $^0A^{(j)}_a$
through equation~(\ref{ffroma}).
Note that the condition on $\tr \, ^0k$ which is assumed in both
theorems can always be arranged by means of a time translation.

\item The spacetimes of the class whose existence 
is established by these theorems
have the desirable property that it is possible to determine the detailed
nature of their singularities by algebraic calculations. This allows them
to be checked for consistency with the cosmic censorship hypothesis. What
should be done from this point of view is to check that some invariantly
defined physical quantity is unbounded as the singularity at $t=0$ is 
approached. This shows that $t=0$ is a genuine spacetime singularity
beyond which no regular extension of the spacetime is possible. For this
purpose it is common to examine curvature invariants but in fact it is just
as good if an invariant of the matter fields can be found which is unbounded
in the approach to $t=0$. This is particularly convenient in the cases where
a dilaton is present. Then $\nabla^\alpha\phi\nabla_\alpha\phi$ is equal in
leading order to the corresponding velocity-dominated quantity and the 
latter is easily seen to diverge like $t^{-4}$ for $t\to 0$. The vacuum
case is more difficult. It will be shown below that the approximation of 
the full solution by the velocity-dominated solution is sufficiently good 
that it is enough to do the calculation for the velocity-dominated metric.
This means that it is enough to do the calculation for the Kasner metric 
in $D$ dimensions. Note that the Kasner metric is invariant under 
reflection in each of the spatial coordinates. Hence curvature components 
of the form $R_{0abc}$ vanish, as do components of the form $R_{0a0b}$ with
$a\ne b$. Hence the Kretschmann scalar 
$R^{\alpha\beta\gamma\delta}R_{\alpha\beta\gamma\delta}$ is a sum of
non-negative terms of the form $R^{abcd}R_{abcd}$ and $(R^a{}_{0a0})^2$.   
In order to show that the Kretschmann scalar is unbounded it is enough
to show that one of these terms is unbounded. A simple calculation
shows that $(R^a{}_{0a0})^2=p_a^2(1-p_a)^2 t^{-4}$ in a Kasner spacetime.
Thus the curvature invariant under consideration can only be bounded
as $t\to 0$ if all Kasner exponents are zero or one, which
does not occur for the solutions we construct.
To see that the approximation of the full solution by the
velocity-dominated solution is valid for determining the asymptotics
of the Kretschmann scalar it is enough to note that all terms
appearing in the Kretschmann scalar which were not just considered
are $o(t^{-4})$.  Only two estimates additional to those
already obtained are needed -- for these, the estimates
$\tilde R_{abc}^{\; \; \; \; \; \; h} = O(t^{-2 + \epsilon})$
and $\tilde \nabla_a {\tilde k^b}_c = O(t^{-2 + \epsilon})$
are sufficient.  Both of these estimates are straightforward to obtain.
The main input is $\tilde \Gamma^c_{ab} = O(t^{-1 + 4 \sigma
-2 \epsilon - \delta})$ ({\it i.e.,} the connection coefficients
{\it do not} need to be expanded).
The expression for the Kretschmann scalar is
\begin{eqnarray}
&&4 ((\tr \, k) {k^a}_b - {k^a}_c {\k^c}_b)
((\tr \, k) {k^b}_a - {k^b}_h {k^h}_a)
+ ({k^a}_b {k^c}_h - {k^a}_h {k^c}_b)
({k^b}_a {k^h}_c - {k^h}_a {k^b}_c)
\nonumber \\ &&
+ 4 \{ ({R^a}_b - {M^a}_b) ({R^b}_a - {M^b}_a)
+ 2 ({R^a}_b - {M^a}_b) ((\tr \, k) {k^b}_a - {k^b}_h {k^h}_a)
\nonumber \\ &&
-2 (\tilde \nabla_a \tilde k^b_{\; \; c})
(\tilde \nabla_h \tilde k^c_{\; \; b}) \tilde g^{ah}
-2 (\tilde \nabla_a \tilde k^b_{\; \; c})
(\tilde \nabla_b \tilde k^a_{\; \; h})
\tilde g^{ch} \} - \tilde g^{ab} \tilde g^{ch} 
\tilde R_{aci}^{\; \; \; \; \; \; j}
\tilde R_{bhj}^{\; \; \; \; \; \; i} \nonumber \\ &&
+ 2 \tilde R_{abc}^{\; \; \; \; \; \; h}
( \tilde k^a_{\; \; i} \tilde k^b_{\; \; h} -
\tilde k^a_{\; \; h} \tilde k^b_{\; \; i}) \tilde g^{ci}. \nonumber
\end{eqnarray}
Apart from the Kasner terms (which can each be written as
two factors, with each factor $O(t^{-2})$), the remaining
terms can each be written as two factors,
with each factor $O(t^{-2}$) and at least one of the two
factors $o(t^{-2})$.

\item We have constructed large classes of 
solutions of the Einstein-matter equations
with velocity-dominated singularities for matter models
defined by those field theories where the BKL picture predicts that solutions
of this kind should exist. No symmetry assumptions were made. When symmetry
assumptions are made there are more possibilities of finding specialized 
classes of spacetimes with velocity-dominated singularities. See for
instance
\cite{narita00}, where there are results for the Einstein-Maxwell-dilaton
and other systems under symmetry assumptions. There are also results for the
case where the Einstein equations are coupled to phenomenological matter 
models such as a perfect fluid and certain symmetry assumptions are made. 
For one of the most general results of this kind so far see \cite{anguige00}.

\item When solutions are constructed by Fuchsian methods as is done is
this paper there is the possibility of algorithmically constructing an
expansion of the solution about the singularity to all orders which is 
convergent when the input data are analytic, as in this paper. (If the
input data are only $C^\infty$ the expansion is asymptotic in a rigorous
sense when Fuchsian techniques can be applied.) At the same time, there
is the possibility of providing a rigorous confirmation of the
reliability of existing expansions such as those of \cite{GM93} and 
\cite{BDV}. This is worked out for the case of \cite{GM93} in \cite{KR}.

\end{enumerate}

\section*{Acknowledgements}
We thank Mme Choquet-Bruhat for comments
which led to clarifications in the exposition.
The work of MH and MW is supported in part by the ``Actions
de Recherche Concert{\'e}es" of the ``Direction de
la Recherche Scientifique - Communaut{\'e} Fran{\c c}aise
de Belgique", by a ``P\^ole d'Attraction Interuniversitaire"
(Belgium) and by IISN-Belgium (convention 4.4505.86).
The research of MH is also supported by
Proyectos FONDECYT 1970151 and 7960001 (Chile) and
by the European Commission RTN programme HPRN-CT-00131,
in which he is associated to K. U. Leuven.
MW would also like to thank the organizers of the
Mathematical Cosmology Program at the Erwin Schr\"{o}dinger
Institute, Summer 2001, where a portion of this work was
completed.

\end{document}